%% file: main.tex
\documentclass[sigconf]{acmart}

\usepackage{caption}
\usepackage[ruled,linesnumbered]{algorithm2e} 
\usepackage{graphicx} 

\usepackage{enumerate} 
\usepackage{enumitem}
\usepackage{multicol}  
\usepackage{multirow}
\usepackage{booktabs} 
\usepackage{setspace} 

\usepackage{tikz,float}
\usepackage{tabularx}
\usepackage{xparse}

\usepackage{etoolbox}
\usepackage{tkz-euclide}

\usepackage{makecell}
\usepackage{subcaption}

\usepackage{subfiles}
\usepackage{hyperref}
\hypersetup{colorlinks = true, allcolors = black}

\tikzset{%
	pics/sema/.style args={#1/#2/#3}{code={%
			\ifstrequal{#2}{0}{%
				\draw[fill=#1] (0,0) circle (0.34em){};
			}{%
				\tkzDefPoint(0,0){O}
				\tkzDrawSector[R,fill=#3](O,0.34em)(90,90-#2)
				\tkzDrawSector[R,fill=#1](O,0.34em)(90-#2,90-360)
			}
	}},
}
\newcommand{\ballnumber}[1]{\tikz[baseline=(myanchor.base)] \node[circle,fill=.,inner sep=1pt] (myanchor) {\color{-.}\bfseries\footnotesize #1};}

\copyrightyear{2021}
\acmYear{2021}
\setcopyright{acmcopyright}
\acmConference[CCS '21]{Proceedings of the 2021 ACM
SIGSAC Conference on Computer and Communications Security}{November 15--19,
2021}{Virtual Event, Republic of Korea}
\acmBooktitle{Proceedings of the 2021 ACM SIGSAC Conference on Computer and
Communications Security (CCS '21), November 15--19, 2021, Virtual Event, Republic
of Korea}
\acmPrice{15.00}
\acmDOI{10.1145/3460120.3484589}
\acmISBN{978-1-4503-8454-4/21/11}

\begin{document}

\fancyhead{}

\title{DeepAID: Interpreting and Improving Deep Learning-based Anomaly Detection in Security Applications} 

\author{Dongqi~Han$^1$,
        Zhiliang~Wang$^1$,
        Wenqi~Chen$^1$,
        Ying~Zhong$^1$,
        Su~Wang$^2$,
        Han~Zhang$^1$,
        Jiahai~Yang$^1$,
        Xingang~Shi$^1$, 
        and~Xia~Yin$^2$}
\affiliation{%
  \institution{$^1$Institute for Network Sciences and Cyberspace, BNRist, Tsinghua University, Beijing, China}
  \country{}
  }
 \affiliation{%
  \institution{$^2$Department of Computer Science and Technology, BNRist, Tsinghua University, Beijing, China} 
  \country{}
  }
\affiliation{
   \textsf{\normalsize{ \{handq19, chenwq19, zhongy18, wangsu17\}@mails.tsinghua.edu.cn,
 \{wzl, yang, shixg\}@cernet.edu.cn,
 \{zhhan,yxia\}@tsinghua.edu.cn } }
 \country{}
}

\renewcommand{\authors}{Dongqi~Han, 
        Zhiliang~Wang,
        Wenqi~Chen,
        Ying~Zhong,
        Su~Wang,
        Han~Zhang,
        Jiahai~Yang,
        Xingang~Shi, 
        and~Xia~Yin}
\renewcommand{\shortauthors}{Dongqi~Han,
        Zhiliang~Wang,
        Wenqi~Chen,
        Ying~Zhong,
        Su~Wang,
        Han~Zhang,
        Jiahai~Yang,
        Xingang~Shi, 
        and~Xia~Yin}


 
  




\begin{abstract}
Unsupervised Deep Learning (DL) techniques have been widely used in various security-related anomaly detection applications, owing to the great promise of being able to detect unforeseen threats and the superior performance provided by Deep Neural Networks (DNN). However, the lack of interpretability creates key barriers to the adoption of DL models in practice. Unfortunately, existing interpretation approaches are proposed for supervised learning models and/or non-security domains, which are unadaptable for unsupervised DL models and fail to satisfy special requirements in security domains.

In this paper, we propose \textsf{DeepAID}, a general framework aiming to 
(1) interpret DL-based anomaly detection systems in security domains,
and (2) improve the practicality of these systems based on the interpretations. We first propose a novel interpretation method for unsupervised DNNs by formulating and solving well-designed optimization problems with special constraints for security domains.
Then, we provide several applications based on our \emph{Interpreter} as well as a model-based extension \emph{Distiller} to improve security systems by solving domain-specific problems.
We apply \textsf{DeepAID} over three types of security-related anomaly detection systems and extensively evaluate our Interpreter with representative prior works. Experimental results show that \textsf{DeepAID} can provide high-quality interpretations for unsupervised DL models while meeting several special requirements in security domains.
We also provide several use cases to show that \textsf{DeepAID} can help security operators to understand model decisions, diagnose system mistakes, give feedback to models, and reduce false positives.  

\end{abstract}



\maketitle

\input{intro}

\input{method}
\input{eval}
\input{end}

\bibliographystyle{ACM-Reference-Format}
\bibliography{REF}

\input{supp}

\end{document}

%% file: intro.tex
\section{Introduction}
Anomaly detection has been widely used in diverse security applications \cite{chandola2009anomaly}.
Security-related anomaly detection systems are desired to be capable of detecting \emph{unforeseen} threats such as zero-day attacks.
To achieve this goal, \emph{unsupervised} learning with only normal data, also known as ``zero-positive'' learning \cite{du2019lifelong}, becomes more promising since it does not need to fit any known threats (i.e., abnormal data) during training, compared with supervised methods.

Recently, deep learning (DL) has completely surpassed traditional methods in various domains, owing to the strong ability to extract high-quality patterns and fit complex functions \cite{wang2020deep}.
Consequently, \emph{unsupervised deep learning} models are more desirable in security domains for owing both the ability to detect unforeseen anomalies and high detection accuracy.	
So far, unsupervised deep learning models have been applied in various security-related anomaly detection applications, such as network intrusion detection \cite{mirsky2018kitsune}, system log anomaly detection \cite{du2017deeplog}, advanced persistent threat (APT) detection \cite{bowman2020detecting}, domain generation algorithm (DGA) detection \cite{sidi2020helix} and web attack detection \cite{tang2020zerowall}.

While demonstrated great promise and superior performance, DL models, specifically Deep Neural Networks (DNN) are lack of transparency and interpretability of their decisions. The black-box reputation creates key barriers to the adoption of DL models in practice, especially in security-related domains.
Firstly, it is hard to establish trust on the system decision from simple binary (abnormal or normal) results without sufficient reasons and credible evidence.
Secondly, black-box DL-based systems are difficult to incorporate with expert knowledge, troubleshoot and debug decision mistakes or system errors.
Thirdly, reducing false positives (FP) is the most  challenging issue for anomaly detection systems in practice. It is impossible to update and adjust DL models to reduce FPs without understanding how the models work.
As a result, security operators are confused with over-simplified model feedback, hesitated to trust the model decisions, shriveled towards model mistakes, and overwhelmed by tons of meaningless false positives.

In recent years, several studies have attempted to develop techniques for interpreting decisions of DL models by pinpointing a small subset of features which are most influential for the final decision \cite{montavon2018methods}.
However, they are not directly applicable to DL-based anomaly detection in security applications for two reasons.
Firstly, existing studies such as \cite{guo2018lemna, meng2020interpreting} mostly focus on interpreting DL models for supervised classification rather than unsupervised anomaly detection. 
Since the mechanisms of these two types of learning methods are fundamentally different, directly applying prior techniques is improper and ineffective, as validated by our experiments in \S \ref{sec6.2}.
Secondly, most prior methods are designed for non-security domains such as computer vision \cite{zhou2016learning, ZintgrafCAW17, shrikumar2017learning,fong2017interpretable} to understand the  mechanism of DNNs.
Instead, security practitioners are more concerned about how to design more reliable and practical systems based on interpretations. 
Besides, studies have shown that prior methods failed to be adopted in security domains due to their poor performance \cite{warnecke2020evaluating,fan2020can}. 

\noindent \textbf{Our Work.} In this paper, we develop \textbf{\textsf{DeepAID}}, a general framework to interpret and improve DL-based anomaly detection in security applications. The high-level design goals of \textsf{DeepAID} include (1) developing a novel interpretation approach for unsupervised DL models that meets several special requirements in security domains
(such as high-fidelity, human-readable, stable, robust, and high-speed),
as well as (2) solving several domain-specific problems of security systems (such as decision understanding, model diagnosing and adjusting, and reducing FPs).
To this end, we develop two techniques in \textsf{DeepAID} referred to as \emph{Interpreter} and \emph {Distiller}.

\noindent \textbf{\textsf{DeepAID} Designs.} 
\emph{Interpreter} provides interpretations of certain anomalies in unsupervised DL models to help security practitioners understand why anomalies happen.
Specifically, we formulate the interpretation of an anomaly as solving an optimization problem of searching a normal ``\emph{reference}'', and pinpoint the most effective \emph{deference} between it and the anomaly.
We propose several techniques in formulating and solving the optimization problem to ensure that our interpretation can meet the special concerns of security domains.  
Based on the interpretations from Interpreter, we additionally propose a model-based extension \emph{Distiller} to improve security systems. 
We ``distill'' high-level heuristics of black-box DL models and expert feedback of the interpretations into a rather simple model (specifically, finite-state machines (FSM)). 
Compared with the original black-box DL model, security practitioners can more easily understand and modify the simple FSM-based Distiller.

\noindent \textbf{Implementation and Evaluations.} 
We separate security-related anomaly detection systems into three types according to the structure of source data: \emph{tabular}, \emph{time-series}, and \emph{graph data}.
Then we provide prototype implementations of \textsf{DeepAID} Interpreter over three representative security systems (\texttt{Kitsune}\cite{mirsky2018kitsune}, \texttt{DeepLog}\cite{du2017deeplog} and \texttt{GLGV}\cite{bowman2020detecting}), 
and Distiller over tabular data based systems (\texttt{Kitsune}).
We extensively evaluate our Interpreter with representative prior methods. 
Experimental results show that \textsf{DeepAID} can provide high-quality interpretations for unsupervised DL models while meeting the special requirements of security domains.
We also provide several use cases to show that \textsf{DeepAID} can help security operators to understand model decisions, diagnose system mistakes, give feedback to models, and reduce false positives.

\noindent \textbf{Contributions.} This study makes the following contributions:
\begin{itemize}[leftmargin=*]
	\item We propose \textsf{DeepAID}, a general framework for interpreting and improving DL-based anomaly detection in security applications, which consists of two key techniques: Interpreter provides high-fidelity, human-readable, stable, robust, and fast interpretations for unsupervised DNNs. Distiller serves as an extension of Interpreter to further improve the practicality of security systems.
	
	\item We provide prototype implementations of \textsf{DeepAID} Interpreter over three types of DL-based anomaly detection applications in security domains using tabular, time-series, and graph data, as well as Distiller for tabular data based security applications\footnotemark[1].
	
	\item We conduct several experiments to demonstrate that our interpretation outperforms existing methods with respect to fidelity, stability, robustness, and efficiency.  
	
	\item We introduce several use cases and improvements of security applications based on \textsf{DeepAID} such as understanding model decisions, model diagnosing and feedback, and reducing FPs.
	
\end{itemize}

The rest of the paper is organized as follows: 
In \S \ref{sec2}, we provide backgrounds on DL interpretability and unsupervised DL for anomaly detection in security applications, as well as the overview of \textsf{DeepAID} design.
We introduce the motivation of this work in \S \ref{sec3}.
\S \ref{sec4} introduces the problem formulation for interpreting unsupervised DL models and its instantiations on three types of systems.
\S \ref{sec5} introduces the model-based extension Distiller.
In \S \ref{sec6}, we extensively evaluate the performance of interpreters and showcase how \textsf{DeepAID} solves some critical problems of security systems.  
We make several discussions in \S \ref{sec7} and introduce related work in \S \ref{sec8}.
\S \ref{sec9} concludes this study.

\footnotetext[1]{The implementation of \textsf{DeepAID} is available at: \url{https://github.com/dongtsi/DeepAID}}

\begin{figure*}
	\begin{minipage}{.68\textwidth}
		\includegraphics[width=\textwidth,trim=0 1 0 8,clip]{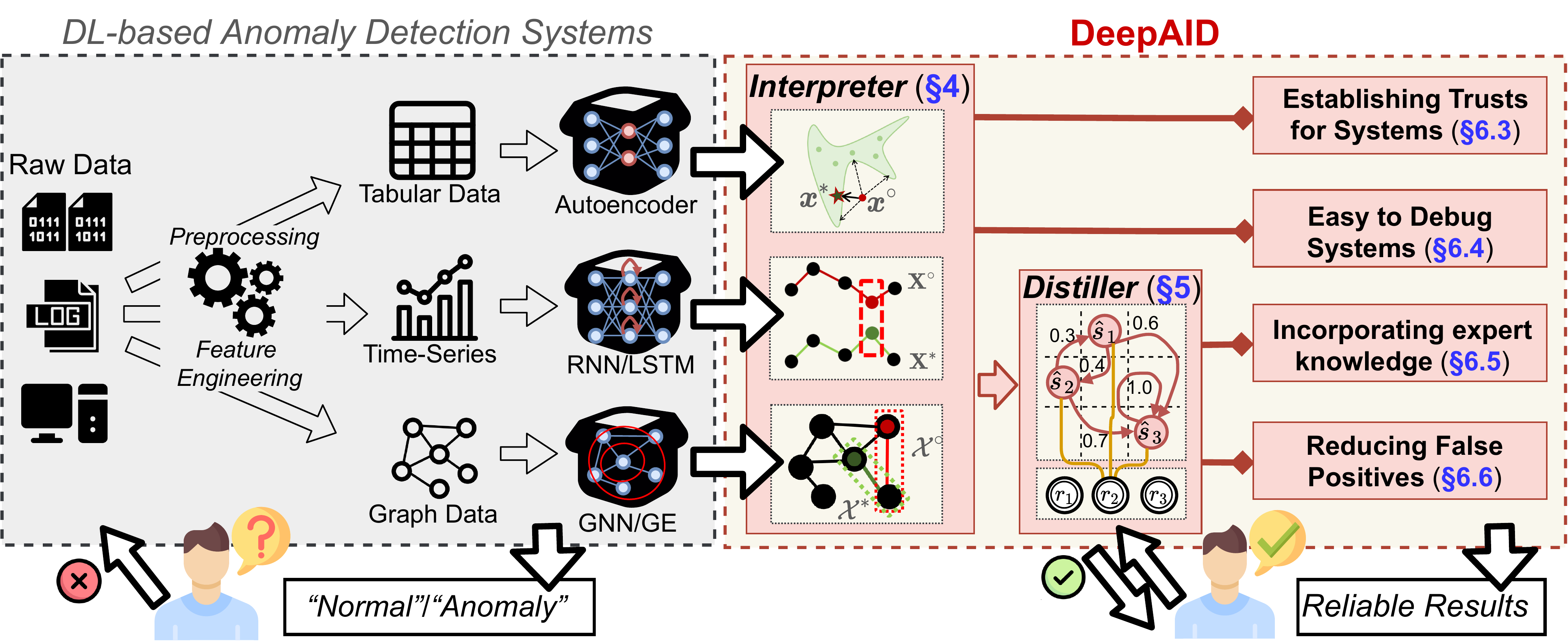}
		\captionof{figure}{The overview of our work.}
		\label{fig1}
	\end{minipage}
	\hfill
	\begin{minipage}{.28\textwidth}
		\centering
		\includegraphics[width=\textwidth]{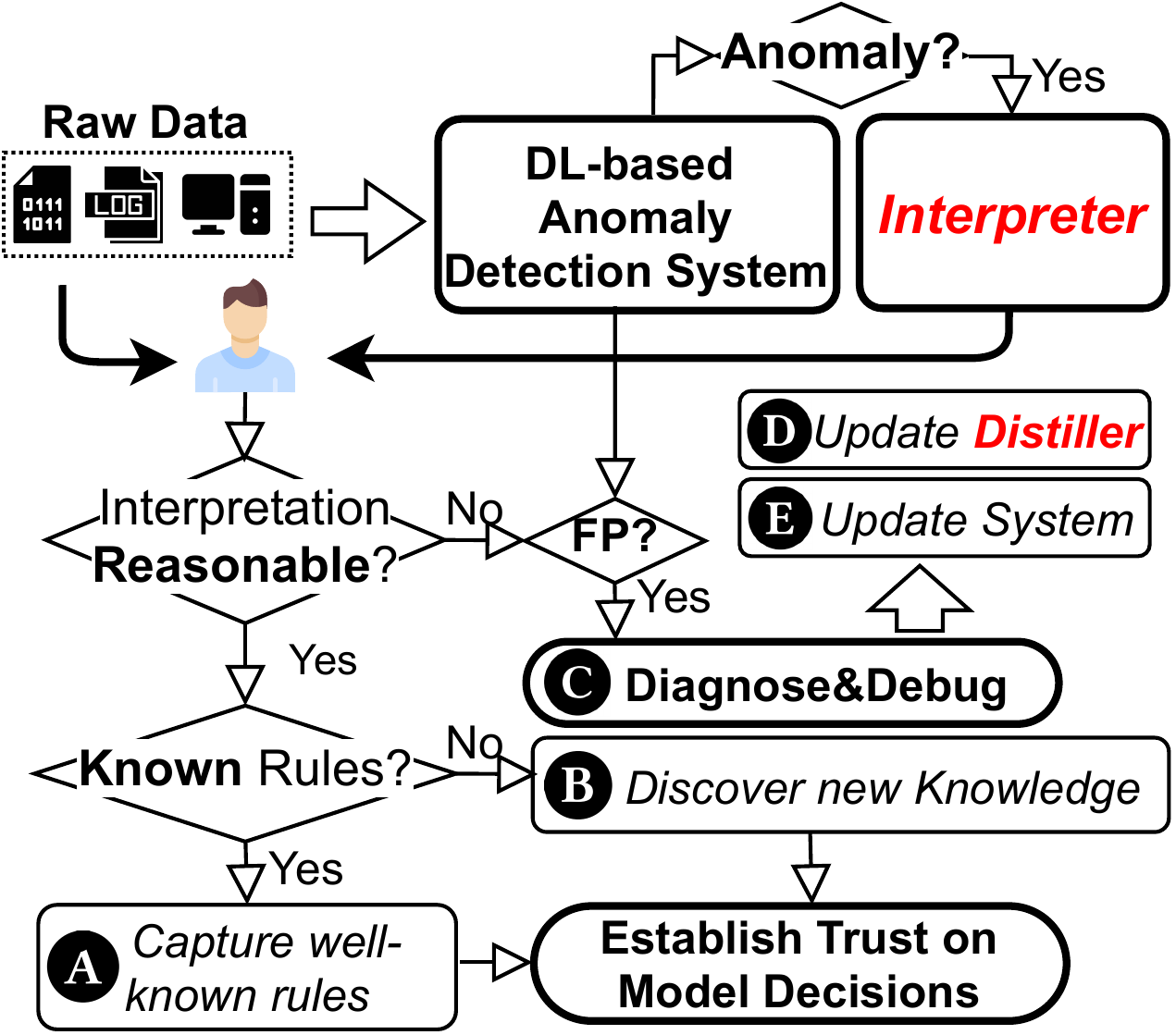}
		\captionof{figure}{The workflow of applying \textsf{DeepAID} on security systems.}
		\label{fig4}
	\end{minipage}
	\vspace{-2ex}
\end{figure*}

\section{Background and Overview}
\label{sec2}
In this section, we first introduce prior techniques for DL interpretability (\S \ref{sec2.1}).
Then, two kinds of unsupervised DL for anomaly detection are introduced (\S \ref{sec2.2}).
Next, we introduce the pipeline and three types of security-related anomaly detection systems (\S \ref{sec2.3}).
Finally, the overview of \textsf{DeepAID} is introduced (\S\ref{sec2.4}).

\subsection{Local Interpretations of DNNs}
\label{sec2.1}
\emph{Local} interpretation refers to interpreting certain decisions/outputs of DL models, which is the most prevailing form in the broad domain of DL Interpretability. 
We introduce three major categories of local interpretations and representative interpreters as follows.

\noindent \textbf{Approximation-based Interpretations.} 
These approaches use a rather simple and interpretable model to locally approximate original DNNs with respect to certain decisions. They commonly make the assumption that although the mapping functions in DNNs are extremely complex, the decision boundary of a specific sample can be simply approximated (e.g., by linear models).
Then, the interpretation of the sample is provided by the simple model.
For example, \texttt{LIME}\cite{ribeiro2016should} uses a set of neighbors of the interpreted sample to train an interpretable linear regression model to fit the local boundary of the original DNN.
\texttt{LEMNA} \cite{guo2018lemna} is another approach dedicated to Recurrent Neural Networks (RNN) in security domains. 
Unlike \texttt{LIME}, the surrogate interpretable model used in \texttt{LEMNA} is a non-linear mixture regression model. In addition, \texttt{LEMNA} leverages fused Lasso to cope with the feature dependence problem in RNNs. 
\texttt{COIN} \cite{liu2018contextual} supports interpreting unsupervised models by training a surrogate linear SVM from interpreted anomalies and normal data.

\noindent \textcolor{black}{\textbf{Perturbation-based Interpretations.} These approaches make perturbations on the input neurons and observe corresponding changes in model prediction.
Impactful input features with small perturbations are more likely to induce major change in the outputs of DNNs \cite{fong2017interpretable,ZintgrafCAW17,FongPV19,ChangCGD19}.
\texttt{CADE} \cite{yang2021cade} provides a more rational method for unsupervised interpretation of concept drift samples by perturbing the inputs and observing the distance changes to the nearest centroid in the latent space.}

\noindent \textbf{Back Propagation-based Interpretations.} The core idea of back propagation-based approaches is to propagate the decision-related information from the output layer to the input layer of DNNs. 
The simple approach is to compute gradients of the output relative to the input based on the back propagation mechanism of DNNs \cite{simonyan2013deep}. However, gradients-based approaches have proven to be unreliable for corner cases and noisy gradients.
Several improvements have been studied to solve the problem \cite{bach2015pixel,sundararajan2017axiomatic,smilkov2017smoothgrad,shrikumar2017learning}. Among them, \texttt{DeepLIFT} \cite{shrikumar2017learning} outperforms others by defining a reference in the input space and computing the back propagate changes in neuron activation to compute the contribution of change between the reference and interpreted sample.

\subsection{Unsupervised DL for Anomaly Detection}
\label{sec2.2}
Unsupervised deep learning for anomaly detection is trained with purely normal data, thus are also called ``zero-positive'' learning \cite{du2019lifelong}. Existing approaches enable DNNs to learn the distribution of normal data and find anomalies that are deviated from the distribution, which can be divided into the following two types:

\noindent \textbf{Reconstruction-based Learning.} 
This kind of learning approach-es are achieved by generative DL models, such as Autoencoder \cite{zhou2017anomaly, mirsky2018kitsune, xu2018unsupervised} and Generative Adversarial Networks (GAN) \cite{zenati2018adversarially}. 
Let $f_R:\mathbb{R}^{N} \rightarrow \mathbb{R}^{N}$ represent generative DL models mapping the input to the output space with the same dimensionality of $N$. The training process of these models is to minimize the distance between normal data $x$ and its output through DNNs denoted with $f(x)$. Then, reconstruction-based models can detect anomalies by compute the reconstruction error (such as mean square error (MSE)) between given inputs and outputs.
The intuition is that DNNs can learn the distribution of normal data after training, thus are more likely to reconstruct normal data with small errors, while failing to reconstruct anomalies that differ from the normal data.

\noindent \textbf{Prediction-based Learning.} This kind of learning approaches are usually used for time-series or sequential data and achieved by RNNs or Long short-term memory (LSTM) \cite{du2017deeplog, du2019lifelong}. 
Given a \emph{normal} time-series $\mathbf{x} = x_1, x_2, ..., x_t$ with the length of $t$, prediction-based approaches force RNNs to increase the probability of predicting $x_t$ through the sequence with first $t-1$ samples. In other words, the loss function of RNNs is to maximize $f(\mathbf{x}) = {\rm Pr}(x_t|x_1x_2...x_{t-1})$ during training. 
Similar to reconstruction-based models, such RNNs can predict the next (i.e., $t$-th) sample in normal time-series with higher probability, while wrongly predict $t$-th sample for anomaly (i.e., with lower probability).

\subsection{Anomaly Detection Systems for Security}
\label{sec2.3}
We start by introducing the general pipeline of unsupervised DL-based anomaly detection systems in security domains. 
Then, we briefly introduce three types of systems with several state-of-the-art works in diverse security domains.

\noindent \textbf{Pipeline of DL-based Anomaly Detection Systems.}
As shown on the left of Figure \ref{fig1}, security applications often start with unstructured raw data, such as network traffic and system logs. These data cannot be directly fed into DNNs, thus pre-processing and feature engineering based on domain-specific knowledge are required.
For example, network traffic can be processed by aggregating related packets into flows through Netflow \cite{umer2017flow}, and free-text system log entries can be parsed into the sequence of log keys indicating message type for entries \cite{xu2009detecting,du2017deeplog}.
After preprocessing, well-structured data can be fed into different kinds of DNNs,
performing the aforementioned reconstruction/prediction based learning to detect anomalies.
In this study, we separate security-related anomaly detection systems into three types according to different structures of source data: \emph{tabular data}, \emph{time-series} and \emph{graph data}.

\noindent \textbf{Tabular Data based Systems.} Tabular data is the most common type that is structured into rows (also called feature vectors), each of which contains informative features about one sample. 
In general, most types of DNNs are designed to process tabular data.
Tabular data based anomaly detection systems have been developed for network intrusion detection \cite{mirsky2018kitsune,zenati2018adversarially} and key performance indicators (KPI) anomaly detection \cite{xu2018unsupervised}. 
For example, \texttt{Kitsune}\cite{mirsky2018kitsune} is a state-of-the-art network intrusion detection systems based on unsupervised DL. It collects raw packets from the monitoring network and then retrieves feature vector from the meta information of each packet which contains over 100 statistics. Finally, tabular features are fed into ensemble Autoencoders to perform reconstruction-based anomaly detection.

\noindent \textbf{Time-Series based Systems.} Time-series is a sequence of data samples indexed in time order. Any data with temporal information can essentially be represented as time-series, such as network traffic \cite{zhong2020helad} and system logs \cite{du2017deeplog}.
Specifically, \texttt{DeepLog} \cite{du2017deeplog} first abstracts each kind of system log as a discrete key and parse Hadoop file system (HDFS) logs to sequences of discrete log key indexes. Then, RNN/LSTM is leveraged for prediction-based anomaly detection.

\noindent \textbf{Graph Data based Systems.} 
Graph data structure is useful for modeling a set of objects (nodes) and their relationships (links) at the same time \cite{zhou2018graph}.
Graph data based anomaly detection is receiving more attention in security domains. Especially for advanced persistent threat (APT) detection, researchers have leveraged Graph Neural Networks (GNN) or Graph Embedding (GE) for unsupervised detection \cite{han2020unicorn,bowman2020detecting}. 
For example, \texttt{GLGV} \cite{bowman2020detecting} is developed to detect lateral movement in APT campaigns.
\texttt{GLGV} constructs authentication graphs from authentication logs within enterprise networks. The training data is collected from the purely benign authentication graphs through DeepWalk \cite{perozzi2014deepwalk}, which is a NN-based GE method. Thus, \texttt{GLGV} can use the reconstruction-based approach to detect abnormal authentications indicating lateral movement in APT.

\begin{table}[]
	\caption{Comparison of representative DL interpreters.}
	
	\resizebox{\linewidth}{!}{%
		\begin{tabular}{@{ }c@{ }|@{ }c@{ }c@{ }c@{ }c@{ }c@{ }|@{ }c@{ }}
			\hline
			\textbf{\begin{tabular}[c]{@{}c@{}} Interpreters\end{tabular}} & \texttt{LIME}\cite{ribeiro2016should} & \texttt{LEMNA}\cite{guo2018lemna} & \texttt{COIN}\cite{liu2018contextual}  & \texttt{CADE}\cite{yang2021cade}    & \texttt{DeepLIFT}\cite{shrikumar2017learning} &  \textbf{\textsf{DeepAID}} \\ \hline
			\textbf{Method (\S \ref{sec2.1})} & Approx. & Approx. & Approx. & Perturb. & B.P. & B.P. \\ \hline
			\textbf{\begin{tabular}[c]{@{}c@{}}Support Unsupervised$^*$\end{tabular}} & \tikz\pic{sema=white/0/}; & \tikz\pic{sema=white/0/}; & \tikz\pic{sema=black/0/}; & \tikz\pic{sema=black/0/};  & \tikz\pic{sema=white/180/black}; & \tikz\pic{sema=black/0/}; \\ \hline
			\textbf{\begin{tabular}[c]{@{}c@{}}Support Tabular Data\end{tabular}} & \tikz\pic{sema=black/0/}; & \tikz\pic{sema=white/180/black}; & \tikz\pic{sema=black/0/}; & \tikz\pic{sema=black/0/}; & \tikz\pic{sema=black/0/}; & \tikz\pic{sema=black/0/}; \\
			\textbf{\begin{tabular}[c]{@{}c@{}}Support Time Seires\end{tabular}} & \tikz\pic{sema=white/180/black}; & \tikz\pic{sema=black/0/}; & \tikz\pic{sema=white/180/black}; & \tikz\pic{sema=white/0/}; & \tikz\pic{sema=black/0/}; & \tikz\pic{sema=black/0/}; \\
			\textbf{\begin{tabular}[c]{@{}c@{}}Support Graph Data\end{tabular}} & \tikz\pic{sema=white/0/}; & \tikz\pic{sema=white/0/}; & \tikz\pic{sema=white/0/}; & \tikz\pic{sema=white/0/}; & \tikz\pic{sema=white/0/}; & \tikz\pic{sema=black/0/}; \\ \hline
			\textbf{Stable$^\dag$} & \tikz\pic{sema=white/0/}; & \tikz\pic{sema=white/0/}; & \tikz\pic{sema=white/180/black}; & \tikz\pic{sema=white/0/}; & \tikz\pic{sema=black/0/}; & \tikz\pic{sema=black/0/}; \\
			\textbf{Efficient$^\dag$} & \tikz\pic{sema=white/0/}; & \tikz\pic{sema=white/0/}; & \tikz\pic{sema=white/0/}; & \tikz\pic{sema=white/0/};& \tikz\pic{sema=black/0/}; & \tikz\pic{sema=black/0/}; \\
			\textbf{Robust$^\dag$} & \tikz\pic{sema=white/0/}; & \tikz\pic{sema=white/0/}; & \tikz\pic{sema=white/180/black}; & \tikz\pic{sema=white/180/black}; & \tikz\pic{sema=white/180/black}; & \tikz\pic{sema=black/0/}; \\ \hline
		\end{tabular}%
	}
	
	\begin{flushleft}
		\setstretch{0.4}
		\small
		\noindent (\textbf{\tikz\pic{sema=black/0/}; = true, \  \tikz\pic{sema=white/180/black}; = partially true, \  \tikz\pic{sema=white/0/}; = false}); \\
		\noindent $*$ \emph{Here we mean whether an interpreter can directly support unsupervised models without abnormal training data;} \\
		\noindent $\dag$ \emph{These three indicators are measured through our experiments in} \S \ref{sec6}.
	\end{flushleft}

	\label{tab2}
\end{table}

\subsection{Overview of \textsf{DeepAID} Design}
\label{sec2.4}
The overview of \textsf{DeepAID} is shown on the right of Figure \ref{fig1} including two core parts: \emph{Interpreter} and \emph {Distiller}.
Interpreter provides high-quality interpretations dedicated to DL-based anomaly detection systems for security, 
which provides a general model-independent framework by formulating the interpretation to a unified optimization problem with security-related constraints.
We instantiate Interpreter on aforementioned three types of security systems.
Based on the interpretations from \textsf{DeepAID} Interpreter, we additionally propose a model-based extension called Distiller to facilitate human interaction with DL models.
Distiller can store informative interpretations and expert feedback and into its FSM-based model.
With Interpreter, security operators can better understand system behaviors and trust system decisions.
With Distiller, DL-based security system becomes debuggable, transparent, and human-in-the-loop.
In Figure \ref{fig4}, we provide the workflow of some use cases to improve the practicality of such security systems with the help of \textsf{DeepAID}, which will be discussed in detail in \S \ref{sec6}.

\section{Motivation}
\label{sec3}
In this section, we provide the key motivations of this work from three aspects. 
Firstly, why unsupervised DL-based security systems need interpretations and improvements? (\S \ref{sec3.1})
Secondly, why existing interpretation approaches are insufficient for unsupervised DNNs? (\S \ref{sec3.2})
Thirdly, why we need Distiller? (\S \ref{sec3.3})

\subsection{Why Security Systems Need \textsf{DeepAID}?}
\label{sec3.1}
In general, there are four major drawbacks in unsupervised DL-based security systems.

\noindent \textbf{Hard to Establish Trusts for System Decisions.}
Security operators need sufficient reasons about the anomaly to take further actions since misoperation will induce a very high cost.
Compared with traditional rule-based systems which contain detailed reasons in their alert logs, DL-based anomaly detection algorithms can only return ``abnormal'' or ``normal''.
As a result, such systems become unreliable with over-simplified outputs.

\noindent \textbf{Difficult to Diagnose System Mistakes.}
Without understanding the black-box models, it is almost impossible to diagnose and repair mistakes in DL-based systems.
For example, it is difficult to tell whether a mistake is caused by implementation bugs, or over-fitting/under-fitting, or redundant features, or other reasons.

\noindent \textbf{Failed to be Human-in-the-Loop.}
Unlike tasks in other domains such as image analysis, security-related systems should be human-in-the-loop instead of fully automated, owing to high cost of errors and reliance on expert knowledge.
That is to say, security systems require to incorporate experts' knowledge and be adjusted according to expert feedback.
However, the black-box nature makes it difficult for experts to interact with DL models or use expert knowledge/experience to improve systems.

\noindent \textbf{Fatigue on tons of False Positives.} 
One of the most critical challenges of developing practical DL-based anomaly detection systems in security domains is to save operators from overwhelming FPs.
There are many reasons for FPs such as insufficient learning of distribution of normal training data, and the appearance of \emph{concept drift} (i.e., distribution of test data is changing and deviates from training data).

We attribute the above drawbacks to the lack of interpretability in DNNs. Therefore, we develop \textsf{DeepAID} to interpret DL-based security systems and further solve the above drawbacks.

\subsection{Why Not Existing Interpretations?}
\label{sec3.2}
Although DL interpretability has been extensively studied \cite{montavon2018methods,guidotti2018survey}, we argue that existing interpretations are unadaptable for unsupervised DL-based security systems for the following two reasons. 


\noindent \textbf{Ill-suited for Unsupervised Learning.}
Most existing interpretation methods serve supervised models.
However, there are clear differences between supervised and
unsupervised models with respect to the mechanism of training and detection.
For supervised DNNs, the interpretation goal is to answer \emph{why a certain input is classified as an anomaly}.
However, this is not the case for unsupervised DL models since they are not supposed to learn any knowledge from abnormal data. Here we introduce a more reasonable goal, which is to answer \emph{why a certain anomaly is not considered as normal}.
That is to say, for unsupervised DL models, what we need to interpret is \emph{deviation}, not \emph{classification}.
We need to seek the reason why anomalies \emph{deviate} from normal data. 
  


\noindent \textbf{Failed to Deploy in Security Domains.}
Firstly, the intention of interpreting security applications differs: 
Security practitioners are more concerned about how to design more reliable systems based on interpretations and solve the aforementioned domain-specific problems, rather than excessive pursuit of understanding the details of DNN. Consequently, Some interpretations used to understand the operating mechanism of DNN are not applicable to security applications \cite{zhou2016learning,nguyen2016multifaceted, du2018towards}.
Secondly, different from other domains, security applications have a low tolerance for errors, which requires the interpretations to be high-quality and robust. 
However, recent studies have shown that existing interpretations failed to be adopted in security domains due to their poor performance \cite{warnecke2020evaluating,fan2020can}.
For one thing, approximation-based interpretation approaches are inaccurate, unstable and time-consuming due to the stochastic sampling for training \emph{surrogate} simple models.
For another, existing interpretations have shown to be vulnerable to adversarial attacks \cite{ghorbani2019interpretation,slack2020fooling, zhang2020interpretable} or even random noise \cite{warnecke2020evaluating,fan2020can}.

\subsection{Why \textsf{DeepAID} Needs Distiller?}
\label{sec3.3}

\noindent \textbf{Why Distiller?} With interpretations from \textsf{DeepAID} Interpreter, model decisions become interpretable, but it is still difficult for security participants to get involved in the decision-making process.
In this case, Distiller provides a bridge to make the entire system become human-in-the-loop.

\noindent \textbf{Why Not Existing Distillation Approaches?}
Knowledge distillation has been proposed to provide \emph{global} interpretations by replacing the entire DL models with interpretable models \cite{wu2018beyond,meng2020interpreting}
(Note that they differ from approximation-based methods in \S \ref{sec2.1} since the latter uses surrogate models for \emph{local} approximation).
However, global substitute model will inevitably induce performance reduction.
By contrast, Distiller in \textsf{DeepAID} is just a back-end extension to DL-based systems, instead of replacing them.

%% file: method.tex
\section{Interpreting Anomalies}
\label{sec4}

In this section, we first give high-level ideas of \textsf{DeepAID} Interpreter and provide a unified problem formulation for interpreting unsupervised DL models, followed by detailing its instantiations on three types of unsupervised DL-based security systems. Table \ref{tab1} lists some important notations used in this paper.

\begin{table}
	\caption{Notations}
	\label{tab1}
	\centering
	\resizebox{\linewidth}{!}{
		\begin{tabular}{@{ }c@{ }|l@{ }}
			\toprule
			\textbf{Notation} & \textbf{Description}\\
			\midrule
			$\mathbf{x}^\circ$; ($\boldsymbol{x}^\circ$,  $\mathbf{X}^\circ$, $\mathcal{X}^\circ$) & Interpreted
			anomaly; (tabular, time-series, graph form) \\
			
			$\mathbf{x}^\ast$; ($\boldsymbol{x}^\ast$,  $\mathbf{X}^\ast$, $\mathcal{X}^\ast$) & Reference for interpretation; (tabular, time-series, graph form) \\
			
			$N,K$ &  \# dimension in features and interpretation vectors (non-zero) \\
			
			$f_R(\cdot),\mathcal{E}_R(\cdot,\cdot),t_R$ & Reconstruction-based DNN, error function and threshold \\
			$f_P(\cdot),t_P$ & Prediction-based DNN and threshold \\
			$\|\cdot\|_p,(\cdot)_i,(\cdot)_{(t)}$ & $L_{p}$-norm, $i$-th element, $t$-th iteration\\
			
			$\mathsf{E}_{G}(\cdot)$ & Graph embedding function\\  
			$\mathcal{D}_{tab}(\cdot),\mathcal{D}_{ts}(\cdot),\mathcal{D}_{gra}(\cdot)$ & Objective function (for tabular, time-series, graph form)\\ 
			$\sigma_{n}$ & Neighborhood scale for initialization of tabular Interpreter\\  
			
			\bottomrule
		\end{tabular}
	}
\end{table}

\subsection{Unified Formulation of Interpretations}
\label{sec4.1}

\noindent \textbf{High-level Ideas for Interpreting Anomalies.} 
In this study, we primarily focus on interpreting \emph{anomalies} since they are more concerned than normal data.
As mentioned in \S \ref{sec3.2}, motivated by the different learning mechanism of unsupervised DNNs from supervised ones, we redefine interpretations of unsupervised models as seeking the reason why anomalies deviate from normal data in the DL model.
From a high level, the proposed Interpreter uses ``\emph{deference} from \emph{reference}'' to interpret the \emph{deviation} of anomalies.
Specifically, we formulate the interpretation of an anomaly as solving an optimization problem which searches a most suitable reference which is considered as normal by the DL model. Then, the interpretation of this anomaly is derived by pinpointing the \emph{deference} between this anomaly and its reference.

\noindent \textbf{Special Concerns for Security Domains.}
As mentioned before, there are special security requirements for interpreters in security domains \cite{warnecke2020evaluating,fan2020can}.
\begin{enumerate}[leftmargin=*]
	\item \textbf{Fidelity}: should be high-fidelity for imitating original DNNs; 
	\item \textbf{Conciseness}: results should be concise and human-readable;
	\item \textbf{Stability}: results on the same sample should be consistent;
	\item \textbf{Robustness}: robust against adversarial attacks or noises;
	\item \textbf{Efficiency}: interpretations should be available without delaying the workflow of high-speed online systems.
\end{enumerate}

 
\noindent \textbf{Unified Problem Formulation.}  Given an anomaly $\mathbf{x}^\circ$, our goal is to find a reference $\mathbf{x}^\ast$, then the interpretation of $\mathbf{x}^\circ$ is provided by the difference between $\mathbf{x}^\circ$ and $\mathbf{x}^\ast$. In light of the above requirements, we formulate interpretations of $\mathbf{x}^\circ$ in unsupervised DNNs (denoted with $f$) for security domains as the following optimization problem:
\begin{align}
\operatorname{argmin}_{\mathbf{x}^\ast} \ & \ 
\mathcal{L}_{fid}(\mathbf{x}^\ast;f)
+ \lambda_{1} \mathcal{L}_{sta}(\mathbf{x}^\circ;f, \mathbf{x}^\ast) 
+ \lambda_{2} \mathcal{L}_{con}(\mathbf{x}^\circ,\mathbf{x}^\ast)
\\
{\rm s.t.} \ & \  \mathbf{x}^\ast \in \mathcal{R}(\mathbf{x}^\circ), \label{eq2}
\end{align}
where $\mathcal{L}_{fid}$, $\mathcal{L}_{sta}$ and $\mathcal{L}_{con}$ respectively measures the loss of fidelity, stability, and conciseness, balanced by weight coefficients $\lambda_{1}$ and $\lambda_{2}$. 
\textcolor{black}{The constraint (\ref{eq2}) means that $\mathbf{x}^\ast$ must be searched in the same feature space as the anomaly $\mathbf{x}^\circ$ (measured by $\mathcal{R}$).}
Below, we will detail instantiations of the formulation and technical solution for three types of systems (according to different data types of $\mathbf{x}^\circ$).



\subsection{Tabular Data Interpreter}
\label{sec4.2}
As mentioned in \S \ref{sec2}, tabular data needs reconstruction-based learning models. Following the unified framework in \S \ref{sec4.1}, the fidelity loss of tabular anomaly data $\boldsymbol{x}^\circ$ can be transformed by adding a constraint to ensure searched reference $\boldsymbol{x}^\ast$ is decided to be normal. 
To define stability loss, we ensure $\boldsymbol{x}^\ast$ to be as close to $\boldsymbol{x}^\circ$ as possible while satisfying other constraints. We use Euclidean distance for measuring this term.
To define conciseness loss, we leverage $L_0$-norm to measure the difference between $\boldsymbol{x}^\ast$ and $\boldsymbol{x}^\circ$, which measures total number of non-zero elements in $\boldsymbol{x}^\ast-\boldsymbol{x}^\circ$. 
Consequently, the interpretation of tabular anomaly can be formulated as:
\begin{align}
\operatorname{argmin}_{\boldsymbol{x}^\ast} \ & \ 
\| \boldsymbol{x}^\ast - \boldsymbol{x}^\circ \|_{2} + \lambda \| \boldsymbol{x}^\ast - \boldsymbol{x}^\circ \|_{0}
\label{eq3} \\
{\rm s.t.}
\ & \ \mathcal{E}_R					(\boldsymbol{x}^\ast,f_R(\boldsymbol{x}^\ast)) < t_R \label{eq4} \\
\ & \ {\boldsymbol{x}^\ast \in[0,1]^{N}}, \label{eq5}
\end{align}
where the first term in objective function (\ref{eq3}) is the stability loss, and the second term is the conciseness loss (weighted by $\lambda$).
The fidelity loss is constrained by (\ref{eq4}), which means $\boldsymbol{x}^\ast$ is decided to be normal. 
Constraint (\ref{eq5}) limits $\boldsymbol{x}^\ast$ within the valid range. Note that, $\boldsymbol{x}^\ast$ here is after normalization to $[0,1]$. Normalizations to other ranges are also supported, which will be explained later.



\noindent \textbf{Challenges.} The above formulation is difficult for solving directly due to three challenges. Firstly, $L_0$-norm minimization problem is proven to be NP-hard \cite{hyder2009approximate}. Secondly, constraint (\ref{eq4}) is highly non-linear. Thirdly, constraint (\ref{eq5}) is a box constraint which cannot be directly solved by most algorithms. Below, several techniques are introduced to address these challenges.

\noindent \textbf{Iteratively Optimizing.} To address the first challenge, we transform the $L_0$-norm term in the objective function into iteratively optimizing another terms. That is, in each iteration, we identify some dimensions in $\boldsymbol{x}^\ast$ that don not have much effect on minimizing the objective function, then their value will be replaced by corresponding value in $\boldsymbol{x}^\circ$. 
How to select \emph{ineffective dimensions} will be discussed later.
In this way, $\| \boldsymbol{x}^\ast - \boldsymbol{x}^\circ \|_{0}$ can be effectively limited through changing a small number of influential dimensions.

\noindent \textbf{Bounding Loss.} To address the second challenge, we transform the highly non-linear constraint (\ref{eq4}) by adding a term into the objective function to minimize $\mathcal{E}_R		(\boldsymbol{x}^\ast,f_R(\boldsymbol{x}^\ast))$. However, this term should not be minimized indefinitely since the goal of $\boldsymbol{x}^\ast$ is to probe the decision boundary of $f_R$. Thus, we limited $\mathcal{E}_R		(\boldsymbol{x}^\ast,f_R(\boldsymbol{x}^\ast))$ to be close to $t_R$ by employing $\mathsf{ReLU}$ function ($\mathsf{ReLU}(x) = \max(0,x)$). 
To ensure $\boldsymbol{x}^\ast$ is ``\emph{within}'' the side of normal data with respect to the decision boundary, $t_R$ is subtracted by a small $\epsilon$. Therefore, constraint (\ref{eq4}) is replaced by the following term:
\begin{equation}
	\mathsf{ReLU}\big(\mathcal{E}_R	(\boldsymbol{x}^\ast,f_R(\boldsymbol{x}^\ast))-(t_R-\epsilon)\big).
\end{equation}

\noindent \textbf{Variable Changing.} We eliminate the box constraint (\ref{eq5}) through the idea of change-of-variables \cite{carlini2017towards} with $\tanh$ function. We introduce a new vector $\boldsymbol{u}$ with the same shape of $\boldsymbol{x}^\ast$. Since the range of  $\tanh(\boldsymbol{u})$ is $(-1,1)$, $\boldsymbol{x}^\ast$ can be replaced by simple linear transformations of $\tanh(\boldsymbol{u})$.
For $\boldsymbol{x}^\ast \in [0,1]^{N}$, $\boldsymbol{x}^\ast = \frac{1}{2}\left(\tanh \left(\boldsymbol{u}\right)+1\right)$.
For $\boldsymbol{x}^\ast \in [a,b]^{N}$($a,b$ is arbitrary as long as $a<b$), $\boldsymbol{x}^\ast = \frac{b-a}{2}\tanh \left(\boldsymbol{u}\right)+\frac{b+a}{2}$.
Thus, the constraint (\ref{eq5}) is eliminated.
\textcolor{black}{As for categorical features, we simply relax their encoding (used in original DL models) to continuous variables and convert them to discrete values after solving the reference $\boldsymbol{x}^\ast$.}

With above techniques, the optimization problem is converted into the following unconstrained one \textbf{in each iteration}:
\begin{align}
\operatorname{argmin}_{\boldsymbol{u}} \ & \ 
	\mathsf{ReLU}\big(\mathcal{E}_R	(\boldsymbol{x}^\ast,f_R(\boldsymbol{x}^\ast))-(t_R-\epsilon)\big)+\lambda \| \boldsymbol{x}^\ast - \boldsymbol{x}^\circ \|_{2} 
\notag \\
{\rm where} \ & \  \boldsymbol{x}^\ast=
\left\{
\begin{array}{cc}
\frac{1}{2}\left(\tanh \left(\boldsymbol{u}\right)+1\right), & {\boldsymbol{x}^\ast \in[0,1]^{n}},\\
\frac{b-a}{2}\tanh \left(\boldsymbol{u}\right)+\frac{b+a}{2}, & {\boldsymbol{x}^\ast \in[a,b]^{n}}.
\end{array}\right.
\label{eq7}
\end{align} 


For illustration purposes, we abbreviate the
objective function in (\ref{eq7}) as $\mathcal{D}_{tab}(\boldsymbol{x}^\ast;\mathcal{E}_R,f_R,t_R,\epsilon,\lambda,\boldsymbol{x}^\circ)$, more concisely $\mathcal{D}_{tab}(\boldsymbol{x}^\ast;\boldsymbol{x}^\circ)$ henceforth. 
This problem can be solved by gradient-based optimization approaches (we use Adam optimizer \cite{kingma2014adam} in this study) since the objective function is fully differentiable.

We now introduce how to select \emph{ineffective dimensions} during iterative optimizations. In each iteration, the best $\boldsymbol{x}^\ast$ is computed by solving (\ref{eq7}), denoted with $\boldsymbol{x}^\ast_{(t)}$ for the $t$-th iteration.
We first compute the gradient of $\mathcal{D}_{tab}$ evaluated at $\boldsymbol{x}^{\ast}_{(t)}$, denoted by $\nabla \mathcal{D}_{tab}(\boldsymbol{x}^{\ast}_{(t)};\boldsymbol{x}^\circ)$.
Motivated by the prior work \cite{bach2015pixel}, we use \emph{gradient} $\times$ \emph{input} together instead of only gradients for accurately measuring the effectiveness of $(\boldsymbol{x}^{\ast}_{(t)})_i$.
Thus, we can select \emph{effective} dimension(s) by solving:
\begin{equation*}
i^\ast = \operatorname{argmax}_{i} \big(\nabla \mathcal{D}_{tab}(\boldsymbol{x}^{\ast}_{(t)};\boldsymbol{x}^\circ)\big)_{i} \cdot (\boldsymbol{x}^{\ast}_{(t)})_i \ {\rm where} \ i \in \{1,2,...,N\} 
\end{equation*}
by calculating and sorting $N$ objective function values. Then, for each $i \neq i^\ast$, $(\boldsymbol{x}^{\ast}_{(t)})_i$ is replaced by $(\boldsymbol{x}^\circ)_i$, ending this iteration.

\noindent \textcolor{black}{\textbf{Adversarial Robustness.}} Gradient-based interpretations are vulnerable against adversarial attacks \cite{ghorbani2019interpretation, zhang2020interpretable} since attackers can also use the gradients for adversarial optimization. However, our method has two natural advantages to defend against adversarial attacks compared with previous interpretations.
First, our method is indifferentiable from overall perspective as we use iterative optimization to cope with $L_0$-norm. Therefore, the attacker cannot directly obtain the overall gradients.
\textcolor{black}{Second, our method is \emph{search}-based. \textsf{DeepAID} interprets anomalies through searching reference $\boldsymbol{x}^\ast$, rather than directly based on the gradient of interpreted anomalies $\boldsymbol{x}^\circ$.
Therefore, it is more difficult for attackers to affect the entire search process by only perturbing the searching entrance.
}


\noindent \textcolor{black}{\textbf{Initialization of $\boldsymbol{x}^\ast$.}}
\textcolor{black}{A straightforward initialization method of $\boldsymbol{x}^\ast$ is directly searching from anomaly $\boldsymbol{x}^\circ$.
To further improve robustness, we initialize $\boldsymbol{x}^\ast$ from the neighborhood of anomaly $\boldsymbol{x}^\circ$. 
Formally, $\boldsymbol{x}^{\ast}_{(1)}$$=\boldsymbol{x}^\circ+\mathcal{N}(0,\sigma_{n}^2)$ where $\mathcal{N}$ is Gaussian distribution and $\sigma_{n}$ reflects the scale of neighborhood.
Such initialization can mitigate the attack effect on the entire search process for reference by mitigating the effect on the searching entrance $\boldsymbol{x}^{\ast}_{(1)}$.}

With aforementioned techniques, the whole procedure for interpreting tabular anomalies is available at Algorithm \ref{alg1}.

\noindent \textbf{Improving Efficiency.} There are several skills to improve efficiency of our interpretation in Algorithm \ref{alg1}. First, we can use the loss change of two iterations to ``early stop'' the iterative optimization (instead of a fixed number of iterations $max_{iter}$ on line 7). Second, increasing the learning rate $\alpha$ gracefully can also speed up the solution (on line 8). Third, we can computer multiple $i^\ast$ (on line 9) to speed up the solution. Fourth, We can update only a few effective dimension(s) $i^\ast$  instead of replacing the majority of ineffective dimensions (on lines 10-13).


\noindent \textbf{Improving Conciseness.} We can fix the \emph{non-zero} dimension of interpretations to $K$ by slightly modifying Algorithm \ref{alg1}: We accumulate the importance of each dimension (measured on line 9) over iterations, then only 
preserve top-$K$ dimensions in $\boldsymbol{x}^\circ-\boldsymbol{x}^\ast$ 
and clip others to $0$ at the end of algorithm.


\begin{algorithm}[!t]
	\small
	\setstretch{0.85}
	\SetKwComment{Comment}{$\triangleright$\ }{}
	\SetKwComment{Commentld}{$\triangledown$\ }{} 
	\SetCommentSty{em}
	\SetKwFunction{Adam}{Adam}
	
	\caption{Interpreting tabular anomalies}
	\label{alg1}
	
	\KwIn{Interpreted anomaly $\boldsymbol{x}^\circ$; $max_{iter}$,  $\alpha$ (learning rate); $\sigma_{n}$}
	\KwOut{Interpretation result $\boldsymbol{x}^\circ-\boldsymbol{x}^\ast$ with the reference $\boldsymbol{x}^\ast$ }
	
	
	
	
	
	\textcolor{black}{$\boldsymbol{x}^\ast_{(1)} \leftarrow \boldsymbol{x}^\circ+\mathcal{N}(0,\sigma_{n}^2)$}; \ \ $t \leftarrow 1$
	\Comment*{\textcolor{black}{initialization of reference}}
	
	\While(\Comment*[f]{iterative optimization}){$t \leq max_{iter}$}
	{	
		$\boldsymbol{x}^{\ast}_{(t)} \leftarrow$\Adam{$\boldsymbol{x}^{\ast}_{(t)};\mathcal{D}_{tab},\alpha$} \Comment*{update $\boldsymbol{x}^{\ast}_{(t)}$ by Adam optimizer}
		
		$i^\ast \leftarrow \operatorname{argmax}_{i} \big(\nabla \mathcal{D}_{tab}(\boldsymbol{x}^{\ast}_{(t)};\boldsymbol{x}^\circ)\big)_{i} \cdot (\boldsymbol{x}^{\ast}_{(t)})_i $;
		
		\For(\Comment*[f]{replacing ineffective dimensions}){$i = 1$ to $N$}{
			\lIf{$i \neq i^\ast$}{
				$(\boldsymbol{x}^{\ast}_{(t)})_i \leftarrow (\boldsymbol{x}^\circ)_i$
			}
		}
		
		$\boldsymbol{x}^\ast_{(t+1)} \leftarrow \boldsymbol{x}^{\ast}_{(t)}$; \ \ 
		$t \leftarrow t+1$;
	}
	
	
	\Return{$\boldsymbol{x}^\circ-\boldsymbol{x}^\ast_{(max_{iter})}$}
	
\end{algorithm}


\subsection{Time-Series Interpreter}
\label{sec4.3}
For interpretations of the remaining two data structures,
we primarily introduce techniques that are different from the tabular case.
As introduced in \S \ref{sec2.2}, anomaly detection for time-series data prefers prediction-based learning with RNN/LSTM. We denote the interpreted time-series anomaly as $\mathbf{X}^\circ$ consists of 
$\boldsymbol{x}^\circ_1,\boldsymbol{x}^\circ_2,...,\boldsymbol{x}^\circ_t$. Time-series data can be generally divided into \emph{univariate} and \emph{multivariate}. Each $\boldsymbol{x}^\circ_i$ ($i\in\{1,2,...,t\}$) in $\mathbf{X}^\circ$ is a one-hot vector for univariate time series, while is a tabular feature vector for multivariate time series. Below, we primarily introduce how to interpret univariate time-series since the demonstrative system (\texttt{DeepLog} \cite{du2017deeplog}) in our study uses univariate data, and then briefly introduce how to extend it to multivariate time series.

As before, our interpretation goal for univariate time-series anomaly is to find a normal time-series reference $\mathbf{X}^\ast$. Intuitively, this is like to \emph{correct} a small number of influential abnormal points in the time-series. Unlike tabular feature vectors, the changes of one-hot vectors in univariate time-series are discrete (i.e., $0$ or $1$). Thus, the $L_2$-norm stability loss and $L_0$-norm conciseness loss (previously in (\ref{eq3})) become equally indifferentiable. The fidelity constraint (\ref{eq4}) becomes $f_P(\mathbf{X}^\ast)>t_P$. The validity constraint (\ref{eq5}) turns into constraining each $\boldsymbol{x}^\ast_i$ in $\mathbf{X}^\ast$ is still one-hot ($i\in\{1,2,...,t\}$). We again use the idea of \emph{iteratively optimizing} and \emph{bounding loss} to solve similar challenges as tabular data. Thus, we have the objective function \textbf{in each iteration} denoted with $\mathcal{D}_{ts}(\mathbf{X}^\ast;\boldsymbol{x}^\ast_t)$ as follows: 
\begin{align}
\mathcal{D}_{ts}(\mathbf{X}^\ast;\boldsymbol{x}^\ast_t) & =
\mathsf{ReLU}\big((t_P+\epsilon) -f_P(\mathbf{X}^\ast)\big) \notag \\
& = \mathsf{ReLU}\big((t_P+\epsilon) -{\rm Pr}(\boldsymbol{x}^\ast_t|\boldsymbol{x}^\ast_1\boldsymbol{x}^\ast_2...\boldsymbol{x}^\ast_{t-1})\big). \label{eq8}
\end{align}

\noindent \textbf{Locating Anomaly.} Before solving $\mathcal{D}_{ts}(\mathbf{X}^\ast)$, there is a special challenge for time-series interpretation. That is, we need to figure out whether an anomaly $\mathbf{X}^\circ$ is caused by $\boldsymbol{x}^\circ_t$ or $\boldsymbol{x}^\circ_1\boldsymbol{x}^\circ_2...\boldsymbol{x}^\circ_{t-1}$. 
\textcolor{black}{This is because $\mathbf{x}_t$ serves as the ``label'' of $\mathbf{x}_1\mathbf{x}_2...\mathbf{x}_{t-1}$ (recall the prediction-based learning introduced in \S \ref{sec2.2}). If the label itself is abnormal, modifying other parts will be useless.}
If the anomaly is caused by $\boldsymbol{x}^\circ_t$, then we can simply replace $\boldsymbol{x}^\circ_t$ with the following $\boldsymbol{x}^\ast_t$ to turn the time-series into normal:
\begin{equation}
\boldsymbol{x}^\ast_t = \boldsymbol{x}^c = \operatorname{argmax}_{\boldsymbol{x}^c} \ {\rm Pr}(\boldsymbol{x}^c|\boldsymbol{x}^\circ_1\boldsymbol{x}^\circ_2...\boldsymbol{x}^\circ_{t-1}). \label{eq9}
\end{equation}
Such $\boldsymbol{x}^\ast_t$ can be obtained directly by observing the index of the maximum probability in the last layer (usually softmax) of the RNN/LSTM. Once $\boldsymbol{x}^\circ_t$ is replaced, the whole process of solving $\mathbf{X}^\ast$ ends immediately. One the other hand, if the anomaly is not caused by $\boldsymbol{x}^\circ_t$, we need to solve $\boldsymbol{x}^\ast_1\boldsymbol{x}^\ast_2...\boldsymbol{x}^\ast_{t-1}$ with $\boldsymbol{x}^\circ_t$ through (\ref{eq8}). Formally,
\begin{equation}
\boldsymbol{x}^\ast_1\boldsymbol{x}^\ast_2...\boldsymbol{x}^\ast_{t-1} = \big( \operatorname{argmax}_{\mathbf{X}^\ast} \ \mathcal{D}_{ts}(\mathbf{X}^\ast;\boldsymbol{x}^\circ_t) \big)_{1,2,...,t-1}  \label{eq10} . 
\end{equation}

To locate anomaly, we introduce \textbf{Saliency Testing}. The intuition is that, if (1) it is hard to make $\mathbf{X}^\circ$ normal by changing $\boldsymbol{x}^\circ_1\boldsymbol{x}^\circ_2...\boldsymbol{x}^\circ_{t-1}$, and (2) RNN/LSTM originally has great confidence of $\boldsymbol{x}^c$ ($\boldsymbol{x}^c \neq \boldsymbol{x}^\circ_t$), then we decide the anomaly is caused by $\boldsymbol{x}^\circ_t$. Formally, saliency testing denoted with $ST(\mathbf{X}^\circ;\mu_1,\mu_2)$ is conducted by:
\begin{equation}
ST(\mathbf{X}^\circ;\mu_1,\mu_2) = (\max(\nabla \mathcal{D}_{ts}(\mathbf{X}^\circ;\boldsymbol{x}^\circ_t))<\mu_1) \wedge (\boldsymbol{x}^c>\mu_2), \label{eq11}
\end{equation}
which respectively represent the above two conditions.
\textcolor{black}{The detailed configuration method of $\mu_1$ and $\mu_2$ is in Appendix \ref{app5}.} 
To conclude, time-series reference $\mathbf{X}^\ast$ is solved as:
\begin{equation*}
\mathbf{X}^\ast=
\left\{
\begin{array}{cc}
	\boldsymbol{x}^\circ_1\boldsymbol{x}^\circ_2...\boldsymbol{x}^\circ_{t-1}\underline{\boldsymbol{x}^\ast_t \leftrightarrow {\rm Eq.}(\ref{eq9})}, & ST(\mathbf{X}^\circ;\mu_1,\mu_2) = {\textbf{True}}, \\
		\underline{\boldsymbol{x}^\ast_1\boldsymbol{x}^\ast_2...\boldsymbol{x}^\ast_{t-1} \leftrightarrow {\rm Eq.}(\ref{eq10})}\boldsymbol{x}^\circ_t, & {\rm otherwise \ (iteratively)}.
\end{array}\right.
\end{equation*}

The whole procedure for interpreting univariate time-series anomalies is available at Algorithm \ref{alg2} in Appendix \ref{app1}. We omit the discussion of robustness for this kind of interpreters since there is no study on adversarial attacks against such kind of interpreter to the best of our knowledge. Besides, the robustness is naturally stronger due to the discreteness of changes in time-series data.

\noindent \textbf{Interpreting Multivariate Time-Series.} We briefly introduce how to extend our interpretation to multivariate time-series, which can be viewed as a combination of tabular and univariate time-series interpretation methods. We first locate a small number of influential data points in an abnormal time-series. Then, for each data point (essentially a feature vector), we can use the tabular interpretation method (in \S \ref{sec4.2}) to search its reference.


\subsection{Graph Data Interpreter}
\label{sec4.4}
There are many types of graphs (such as \emph{attributed} and \emph{unattributed}) and learning tasks (such as \emph{node classification} and \emph{link prediction}). For attributed graph (i.e., each node or link has a feature vector), interpretations can be obtained by the same idea as the interpretation of multivariate time series. That is, first locating a small number of abnormal nodes/links in the graph, and then using tabular interpretation method to modify important feature dimensions to find the reference. Therefore, below we primarily introduce how to interpret unattributed graph anomalies. Since the learning task in our demonstrative system (\texttt{GLGV} \cite{bowman2020detecting}) is link prediction, we will introduce how to interpret abnormal links in graph anomaly.

The common method of processing graph data is to embed it into a continuous feature space, or directly leveraging  end-to-end graph neural networks (GNN). Unattributed graphs usually require the first method, namely graph embedding (GE), such as DeepWalk \cite{perozzi2014deepwalk}. Let $\mathsf{E}_{G}$ denotes GE, then the general workflow of unsupervised DL with graph $\mathcal{G}$ is to first compute the embedding vector $\boldsymbol{e} = \mathsf{E}_{G}(\mathcal{G})$.
For link-level interpretation, an abnormal link $\mathcal{X}^\circ=(x_a^\circ,x_b^\circ)$ ($x_a^\circ$ and $x_b^\circ$ are two nodes identify this link)
is embedded as $\boldsymbol{e}^\circ = \mathsf{E}_{G}(\mathcal{X}^\circ)$.
Therefore, the interpretation of graph link anomaly is derived through two steps.
(1) \textbf{Step 1}: Interpreting $\boldsymbol{e}^\circ$ by finding its reference $\boldsymbol{e}^\ast$. (2) \textbf{Step 2}: Finding the original representation of $\boldsymbol{e}^\ast$ in the graph.

Details of step 1 are omitted since it is the same as the tabular interpretation.
As for step 2, we find the original representation $\mathcal{X}^\ast=(x_a^\ast,x_b^\ast)$  of $\boldsymbol{e}^\ast$ by solving the following optimization problem:
\begin{align}
\setstretch{0.8}
\operatorname{argmin}_{\mathcal{X}^\ast=(x_a^\ast,x_b^\ast)} \ & \
\mathcal{F}_1(\mathcal{X}^\ast) + \lambda \mathcal{F}_2(\mathcal{X}^\ast,\boldsymbol{e}^\ast) \ \ \ {\rm s.t.} \  x_1^\ast, x_2^\ast \in \mathcal{V}, \label{eq12}
\\
{\rm where} \ \ \ \mathcal{F}_1(\mathcal{X}^\ast)= & \  \mathsf{ReLU}\big(\mathcal{E}_R	\big(\mathsf{E}_{G}(\mathcal{X}^\ast),f_R(\mathsf{E}_{G}(\mathcal{X}^\ast))\big)-(t_R-\epsilon)\big) \notag \\
{\rm and } \ \ \mathcal{F}_2(\mathcal{X}^\ast,\boldsymbol{e}^\ast)= & \ \| \mathsf{E}_{G}(\mathcal{X}^\ast) - \boldsymbol{e}^\ast \|_{2}.
\label{eq13}
\end{align} 
Here the first term in objective function (\ref{eq12}) means $\mathsf{E}_{G}(\mathcal{X}^\ast)$ is decided to be normal by the interpreted system, and the second term measures the difference of $\mathsf{E}_{G}(\mathcal{X}^\ast)$ and $\boldsymbol{e}^\ast$ obtained in step 1. 
The intuition of (\ref{eq12}) is to force $\mathsf{E}_{G}(\mathcal{X}^\ast)$ to close to  $\boldsymbol{e}^\ast$, and ensure $\mathsf{E}_{G}(\mathcal{X}^\ast)$ is a normal embedding vector. The constraint in (\ref{eq12}) means $x_a^\circ$ and $x_b^\circ$ are still in the node set denoted with $\mathcal{V}$.


	


Problem (\ref{eq12}) can be simply solved by gradient-based optimizer if $\mathsf{E}_{G}$ is differentiable. 
However, embeddings of inputs are derived by an indifferentiable \emph{look-up} operation \cite{mikolov2013distributed} in many cases. For example, most DL frameworks (e.g., PyTorch \cite{paszke2019pytorch}) set the embedding layer to be indifferentiable. 
A simple method is to use a fully connected network (FCN) to achieve the same embedding function, but this requires additional work to modify the interpreted model. In Appendix \ref{suppA.2}, we propose an alternative greedy solution to (\ref{eq12}) when $ \mathsf{E}_{G}$ is indifferentiable, and also provide the whole algorithm.
 




\setlist{nosep}
\section{Distilling Interpretations}
\label{sec5}
As introduced in \S \ref{sec3.3}, to facilitate human interaction with DL-based anomaly detection systems, we propose a model-based extension based on our interpretations, referred to as \emph{Distiller}. \textcolor{black}{In this study, we primarily design Distiller for tabular data and will extend it to other data types in future work.} Below, we introduce overview, details, superiority, and application of Distiller.

\noindent \textbf{Distiller Overview.} In a nutshell, Distiller allows security analysts to give feedback on interpretation results thus integrating expert knowledge into the security system. 
Specifically, \textcolor{black}{analysts can give ``\emph{feedback}'' (denoted with $r_{i}$) on a set of anomalies after checking their interpretations ($\boldsymbol{x}^\circ-\boldsymbol{x}^\ast$) from Interpreter.}
The number and content of feedback is highly free for analysts.
We refer to ``feedback on interpretations'' as ``\emph{rule}'' denoted as $(\boldsymbol{x}^\circ-\boldsymbol{x}^\ast) \rightarrow r_{i}$.
The intuition is that, after adding rules based on expert knowledge, Distiller is able to automatically recognize similar anomalies in the future to output more reliable results.
The design goal of Distiller is to store rules when analysts need to add new ones (called \emph{Update} mode), and to match ``ruled'' anomalies and report previous expert feedback (called \emph{Test} mode). 
To this end, Distiller essentially consists of two finite-state machines (FSM). The first FSM is used to model $K$-dimension interpretation ($\boldsymbol{x}^\circ-\boldsymbol{x}^\ast$), and \textcolor{black}{the second FSM is used to model the transition from interpretation to feedback $r_{i}$.} Below, we will introduce the design of two FSMs and two modes.

\begin{figure}
    	\includegraphics[width=.89\linewidth]{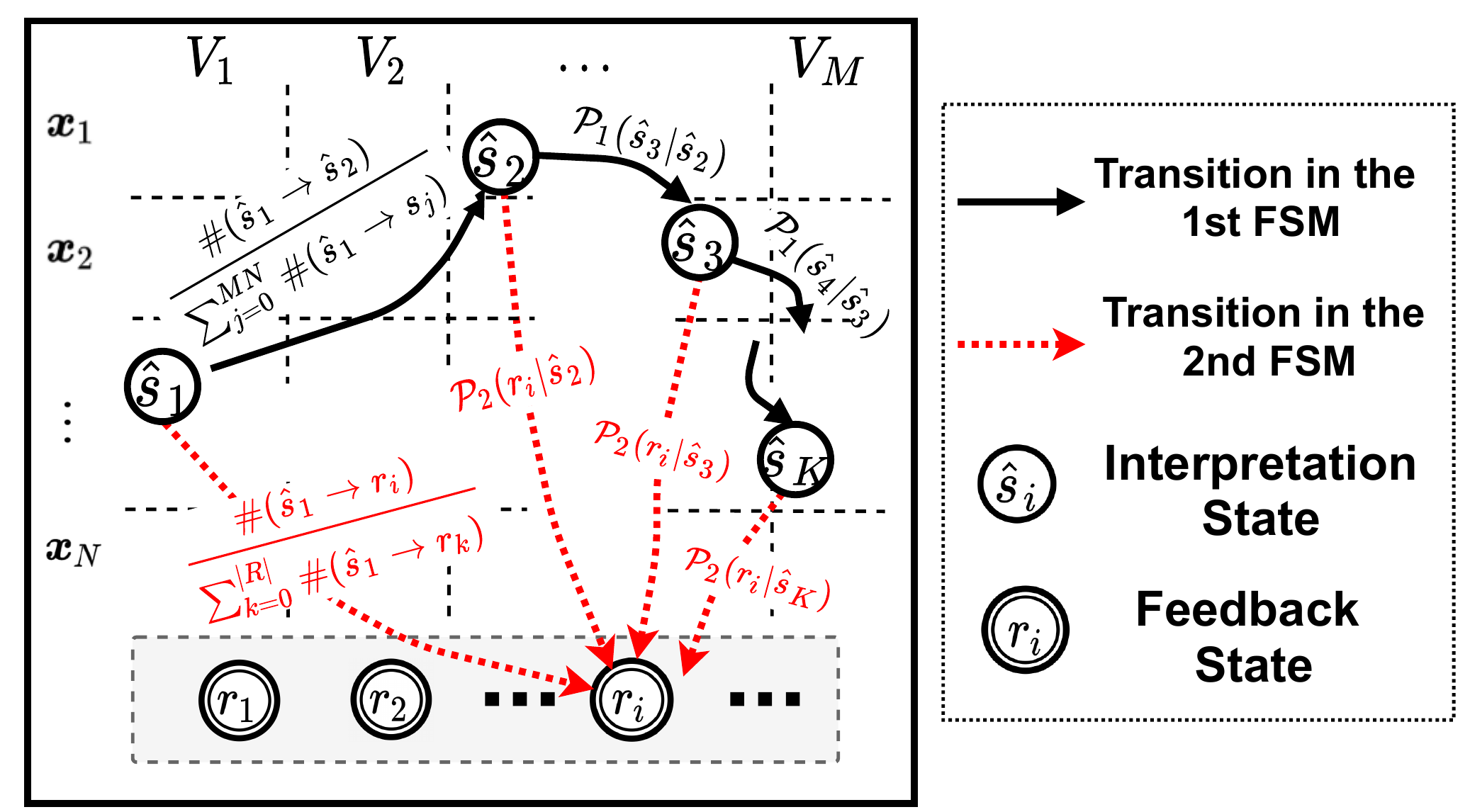}
		\vspace{-1ex}
		\caption{States and transitions of \textsf{DeepAID} Distiller.}
	\label{fig2a}
\end{figure}

\noindent \textbf{State and Transition Abstraction of FSMs
} (Figure \ref{fig2a}).
In the first FSM, we split the value range of each dimension in the interpretation vector $\boldsymbol{x}^\circ-\boldsymbol{x}^\ast$ into $M$ equal-length intervals, then we have totally $M \times N$ states. 
Each of $K$ dimensions in $\boldsymbol{x}^\circ-\boldsymbol{x}^\ast$ is mapped into one of states $\hat{s}_i$ according to its dimension and value.
Subsequently, transitions of these $K$ states are arranged according to the order of decreasing effectiveness. That is, the state with the largest effectiveness becomes the initial state, and the smallest one is the final state.
\textcolor{black}{The second FSM contains all $MN$ states in the first FSM and variable states indicating feedback ${r}_i$.
Transitions in the second FSM are all single-step from an interpretation $\hat{s}_i$ (initial state) to a feedback state ${r}_i$ (final state).}
The intuition of two FSMs is that all dimensions of interpretations ``contribute'' to the feedback (the second FSM), and their order matters (the first FSM).

\begin{figure*}
	\includegraphics[width=.925\textwidth]{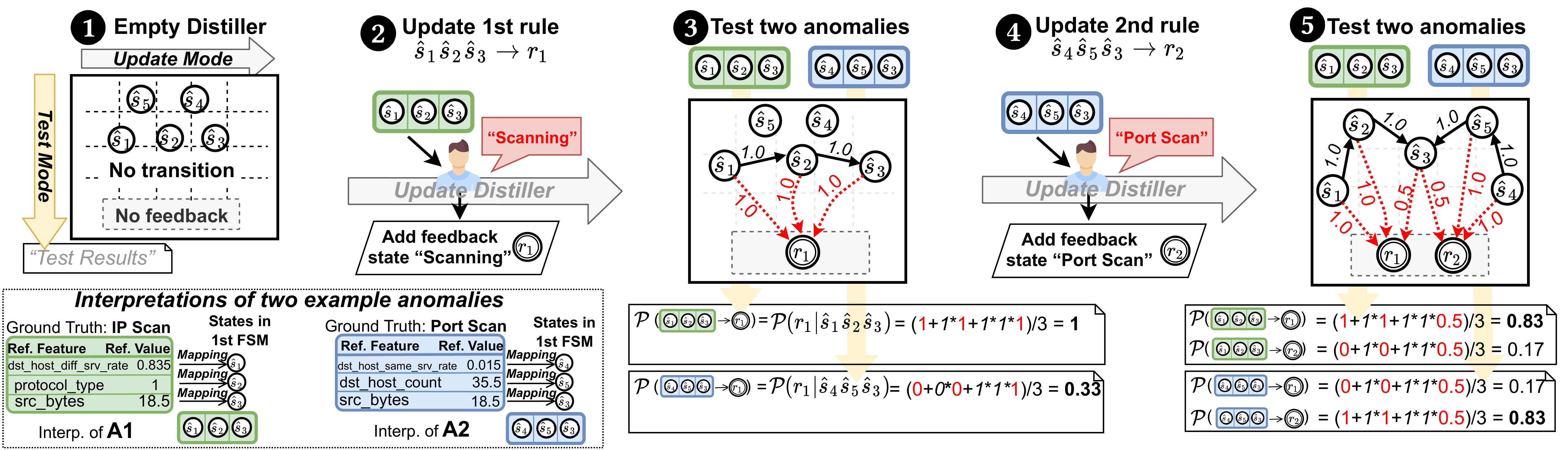}
		\vspace{-1ex}
		\caption{Toy example of updating and testing two anomalies \textbf{A1} (green) and \textbf{A2} (blue) in \textsf{DeepAID} Distiller.}
	\label{fig2b}

\vspace{-1ex}
\end{figure*}

\noindent \textbf{Update and Test Mode.} Distiller works on two modes: (1) \emph{Update} mode indicates adding new rules proposed by analysts.
\textcolor{black}{This can be implemented by separately updating two  \emph{transition matrices} of two FSMs according to the aforementioned transitions of the new rule in two FSMs.}  
(2) \emph{Test} mode matches anomaly interpretations to existing feedback. 
Interpretation $\boldsymbol{x}^\circ-\boldsymbol{x}^\ast$ raised by our Interpreter is first mapped into a sequence of states $\hat{s}_{1} \hat{s}_{2} ... \hat{s}_{K}$. 
\textcolor{black}{Then, the matching probability of the interpretation to each feedback $\hat{r}$ is calculated as:}
\begin{equation}
 \mathcal{P}\left(\hat{r} \mid \hat{s}_{1} \hat{s}_{2} ... \hat{s}_{K}\right)  =  \frac{1}{K} \sum_{i=1}^{K}\big(\mathcal{P}_2(\hat{r} \mid \hat{s}_{i}) \prod_{j=1}^{i-1}\mathcal{P}_1(\hat{s}_{j+1} \mid \hat{s}_{j})\big), \label{eq14}
\end{equation}
where $\mathcal{P}_1$ and $\mathcal{P}_2$ is transition probability in two FSMs/transition matrices. The intuition is to calculate the mean of the product of transition probability from $\hat{s}_{1}$ to $\hat{s}_{i}$ and from  $\hat{s}_{i}$ to $\hat{r}$ ($i\in\{1,2,...,K\}$).

\noindent \textcolor{black}{\textbf{Toy Example.} In Figure \ref{fig2b}, we provide a toy example of updating and testing two anomalies (called \textbf{A1}/\textbf{A2}) from an empty Distiller. The ground truth (unknown to analysts) of \textbf{A1} and \textbf{A2} is IP and Port Scan.
Suppose the analysts mark \textbf{A1} as ``Scanning'' after reading the interpretation ($K=3$ here) and then updates Distiller with the new rule} (\ballnumber{2}),
\textcolor{black}{which essentially adds a new feedback state $r_1$ indicating ``Scanning'' and updates the corresponding transitions in the two FSMs. 
At this time, if we test \textbf{A1} and \textbf{A2}} (\ballnumber{3}),
\textcolor{black}{Distiller will return strong confidence of matching \textbf{A1} to ``Scanning'' ($\mathcal{P}(r_{1}|\hat{s}_1\hat{s}_2\hat{s}_3)=1
$) and weak confidence of matching \textbf{A2} to ``Scanning'' ($\mathcal{P}(r_{1}|\hat{s}_4\hat{s}_5\hat{s}_3)=0.33
$).
Then after updating the second rule indicating \textbf{A2} to feedback ``Port Scan''} (\ballnumber{4}), 
\textcolor{black}{Distiller}  (\ballnumber{5}) \textcolor{black}{will return both strong confidences of matching \textbf{A1} to ``Scanning'' and matching \textbf{A2} to ``Port Scan'' ($\mathcal{P}(r_{1}|\hat{s}_1\hat{s}_2\hat{s}_3)=\mathcal{P}(r_{2}|\hat{s}_4\hat{s}_5\hat{s}_3)=0.83$).}


\noindent \textcolor{black}{\textbf{Distiller vs. ML/DL Classifier}.
	Distiller shares some similarities with ML/DL classifiers, e.g., update/test mode is similar to training/prediction on multi-class tasks, and feedback can be viewed as labels. However, Distiller is more competent for human-in-the-loop detection for three reasons. First and most important, Distiller is more transparent and easy to modify. Security analysts can clearly and simply modify transitions and states in Distiller.
	Second, the number of classes/feedback in Distiller is adaptable. Third, Distiller can preserve the ability of anomaly detection systems to detect unknown threats by initially transiting all states in the first FSM to a feedback state representing unforeseen threats. We conduct experiments in \S \ref{sec6.5} to demonstrate the superiority of Distiller.}

\noindent \textcolor{black}{\textbf{Distiller vs. Rule Matching}. Another potential baseline of Distiller is rule-matching methods. Compared with them, Distiller has the generalization ability to report feedback for \emph{similar} anomalies, e.g., at} \ballnumber{3} \textcolor{black}{(Figure \ref{fig2b}), \textbf{A2} is also reported as ``Scanning'' due to its similarity to the first rule (i.e., $\mathcal{P}(r_{1}|\hat{s}_4\hat{s}_5\hat{s}_3)=0.33$ but not $0$).}

\noindent \textbf{Reducing FPs via Distiller}. We showcase how Distiller may help to reduce two types of FPs in original anomaly detection models: (1) Low confidence positives. Such FPs are caused by normal data nearly decision boundary/threshold (e.g., $t_R,t_P$). To address them, we can modify Distiller by forcing some selected FP interpretations to transit to feedback states indicating normal (or FP).
(2) Concept drift of normal data. We can locate and retrain FPs with very low transition probability in the first FSM,  which indicates such FPs are neither normal (judged by DL model) nor known anomalies.


\noindent \textbf{Theoretical Analysis.} We provide proofs of the correctness, accuracy bound, and complexity of Distiller in Appendix \ref{app2}. 

%% file: eval.tex
\section{Experiments}
\label{sec6}


In this section, we conduct several experiments with baseline interpreters, as well as provide several case studies to introduce how to interpret and improve security-related anomaly detection systems with \textsf{DeepAID}. 
Due to space limit, we primarily evaluate tabular and time-series data based systems (as none of baseline supports graph data). Experiments on graph data based systems as well as several details and supplements of our experiments are left in Appendix \ref{app3} and \ref{app4}. Overall, we provide a roadmap of this section:
\begin{itemize}[leftmargin=*]
	\item \textbf{Experimental Setup.} We introduce implementation and setup of security systems, \textsf{DeepAID}, and baselines in this study (\S \ref{sec6.1}).
	
	\item \textbf{Interpreter Performance.} We demonstrate the performance of \textsf{DeepAID} Interpreter with respect to fidelity, stability, efficiency, and robustness, compared with existing works (\S \ref{sec6.2}).
	
	\item \textbf{Understanding Model Decisions.} We provide case studies on leveraging \textsf{DeepAID} interpretations to establish trusts in model decisions and discovering implicit model knowledge (\S \ref{sec6.3}). 
	
	\item \textbf{Enabling Debuggability.} We provide a case on diagnosing and debugging system mistakes based on \textsf{DeepAID} (\S \ref{sec6.4}).
	

	\item \textbf{Human-in-the-Loop Detection.} For tabular security systems, we conduct reliable detection with expert feedbacks and knowledge based on \textsf{DeepAID} Interpreter and Distiller (\S \ref{sec6.5}).

	\item \textbf{Reducing FPs.} For tabular security systems, we showcase that \textsf{DeepAID} can reduce FPs based on Distiller (\S \ref{sec6.6}).
\end{itemize}


\subsection{Experimental Setup}
\label{sec6.1}
Here we briefly introduce the implementation and experimental setup, 
and we refer readers to Appendix \ref{app3} for details, as they involve many domain-specific backgrounds.

\noindent \textbf{Security Systems.} 
We use three demonstrative DL-based anomaly detection systems in security domains using three data types: 
\texttt{Kitsune}\cite{mirsky2018kitsune}, \texttt{DeepLog}\cite{du2017deeplog}, and \texttt{GLGV}\cite{bowman2020detecting}, which are introduced in \S \ref{sec2.3}. The data used in this work is primarily from their released datasets 
(see Appendix \ref{app3} for more details).


\noindent \textbf{\textsf{DeepAID} Implementation.}
For Interpreter, we fix the interpretation results into $K$-dimension.
By default, we set $\lambda=0.001$ and $\epsilon=0.01$ in  objective functions (\ref{eq7}), (\ref{eq8}), (\ref{eq12}), learning rate $\alpha=0.5$ in Adam optimizer, and $max_{iter}=20$.
For tabular Interpreter, we set neighborhood scale $\sigma_{n}=0$ by default for initialization and evaluate its effect in \S \ref{sec6.2}. For time-series Interpreter, we set $\mu_1=0.01$ and $\mu_2=0.3$ by default.
For Distiller, we set $M=20$ (number of intervals) by default.
\textcolor{black}{The sensitivity evaluation and configuration guideline of these hyper-parameters are in Appendix \ref{app5}}.

\noindent \textbf{Baseline Interpreters and Implementation.} In \S \ref{sec6.2}, we evaluate \textsf{DeepAID} Interpreter with several baseline interpreters listed in Table \ref{tab2} and introduced in \S \ref{sec2.1}. 
\textcolor{black}{(1) For supervised interpretations (\texttt{LIME}, \texttt{LEMNA}, and \texttt{DeepLIFT}), we approximate the decision boundary of the anomaly detection with a supervised DNN trained with additionally sampled anomalies.
(2) \texttt{COIN} and \texttt{CADE} are unsupervised baselines. 
\texttt{CADE} needs a supervised encoding model before interpreting \cite{yang2021cade} (See Appendix \ref{app3} for details).
(3) We also directly \textbf{S}elect the \textbf{R}eference from \textbf{T}raining \textbf{D}ata (abbreviated as \textbf{S.R.T.D.}). 
} 

\subsection{Interpreter Performance}
\label{sec6.2}
We evaluate the performance of \textsf{DeepAID} and baseline interpreters from four aspects: fidelity (with conciseness), stability, robustness, and efficiency, which are of particular concern when deployed in security domains. Their definitions are introduced in \S \ref{sec4.1}.

\begin{figure*}
		\begin{subfigure}{.386\textwidth}
			\includegraphics[width=\textwidth,trim=10 10 5 5,clip]{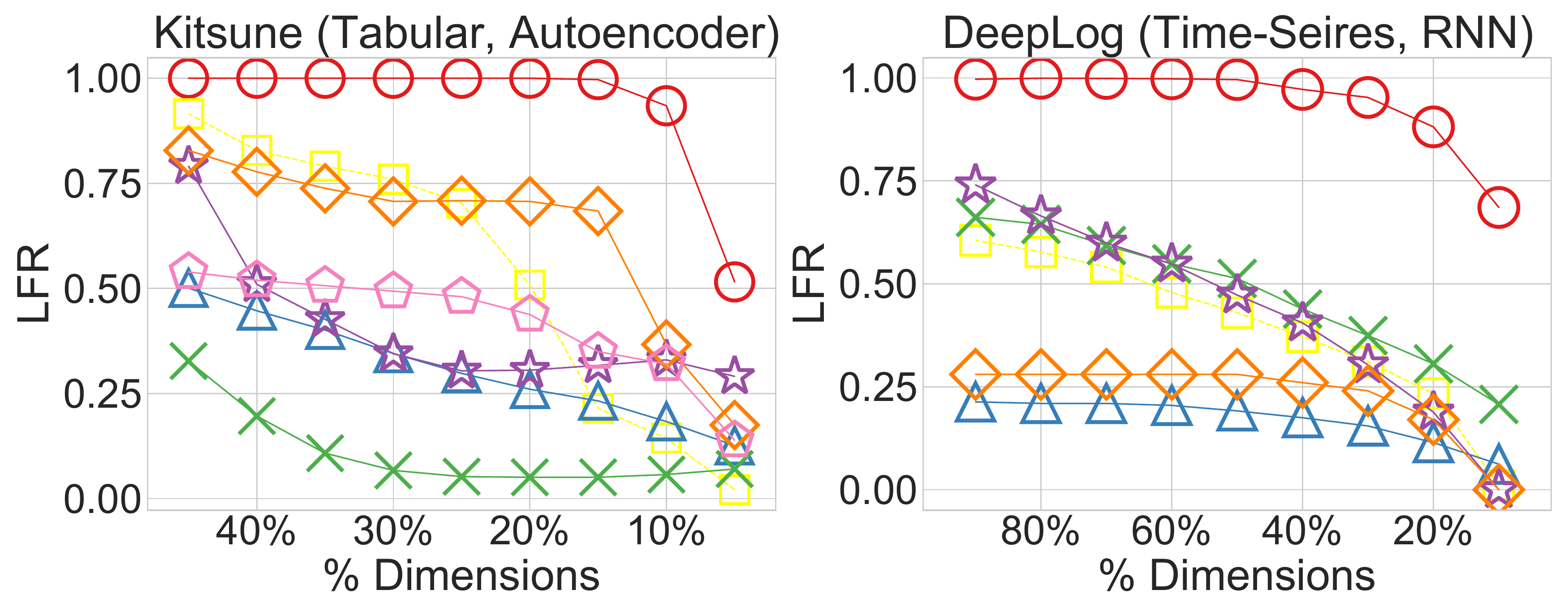}
			\vspace{-4ex}
			\caption{Fidelity evaluation. }
			\label{fig3a}
		\end{subfigure} %
		\hfill
		\begin{subfigure}{.386\textwidth}
			\includegraphics[width=\textwidth,trim=10 10 5 5,clip]{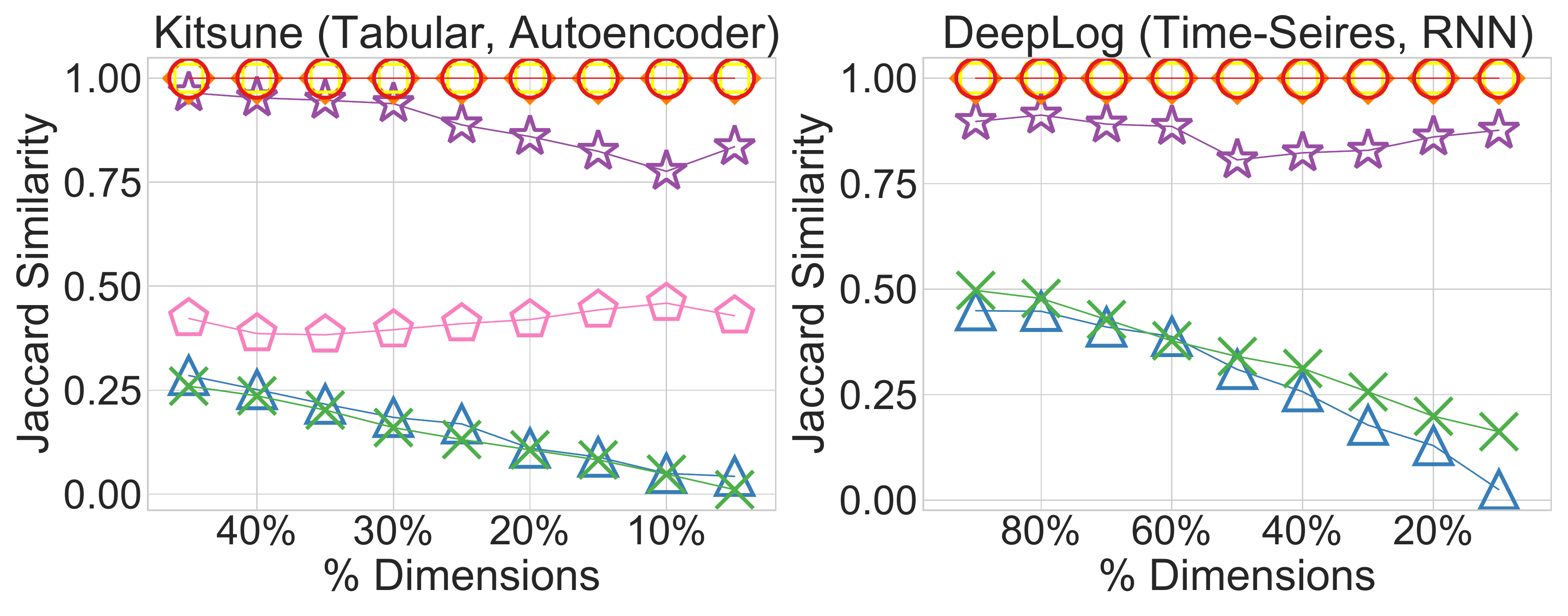}
			\vspace{-4ex}
			\caption{Stability evaluation. }
			\label{fig3b}
		\end{subfigure}
		\hfill
		\begin{subfigure}{.202\textwidth}
			\includegraphics[width=\textwidth,trim=10 10 5 5,clip]{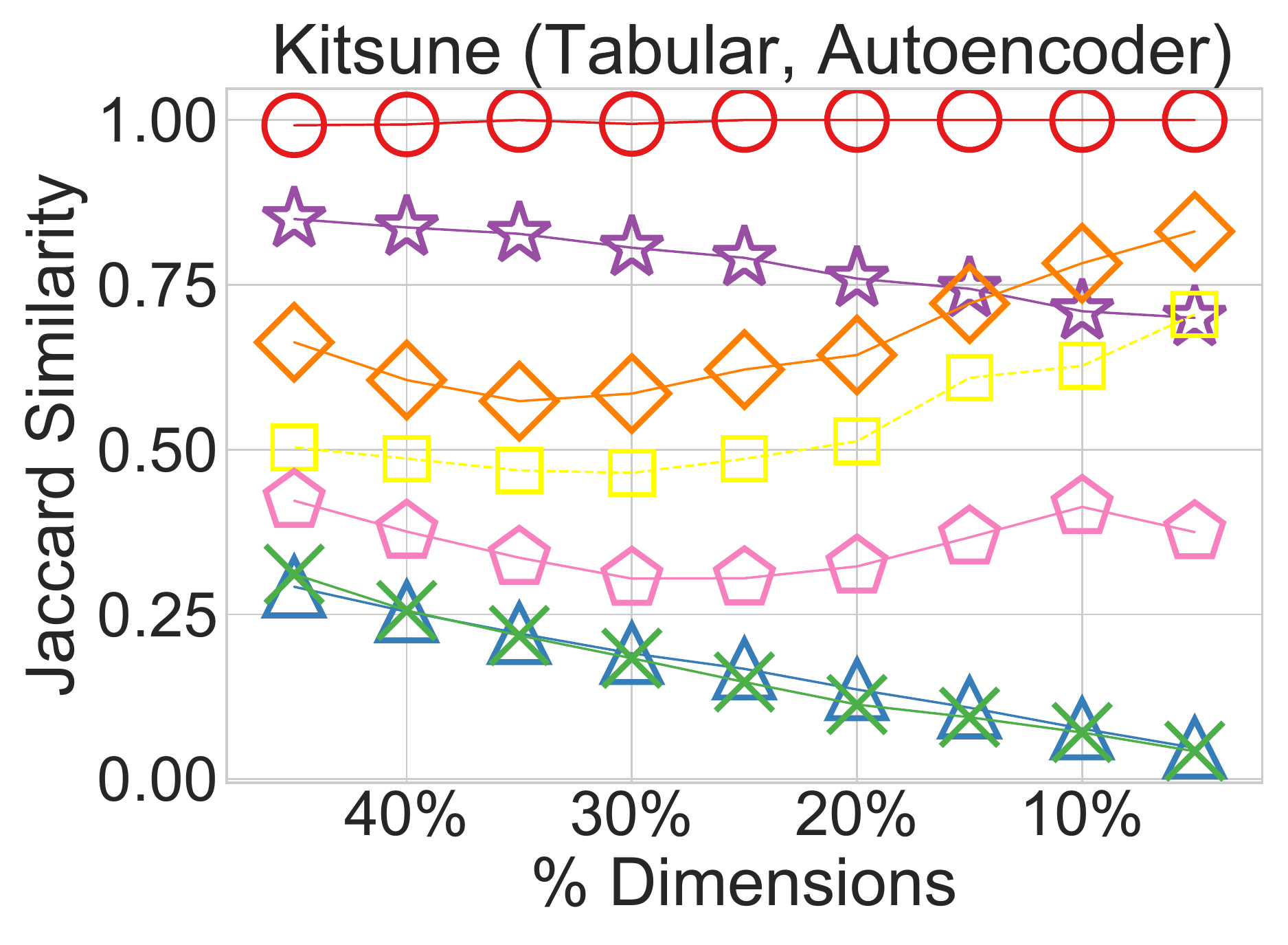}
			\vspace{-4ex}
			\caption{Robustness (to noise).}
			\label{fig3c}
		\end{subfigure}

		\vspace{-0.25ex}
		\begin{subfigure}{0.9\textwidth}
			\centering
			\includegraphics[width=\textwidth]{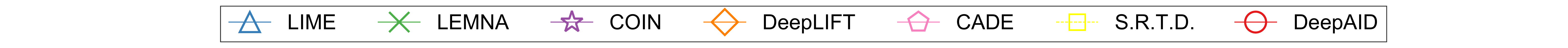}
		\end{subfigure}
        \vspace{-1.5ex}
		\setcounter{figure}{3}
		\captionof{figure}{\textcolor{black}{Fidelity, stability, and robustness (to noise) evaluation of interpreters (\emph{higher is better}).}}
		\label{fig3}
		
\end{figure*}

\begin{figure*}
	\begin{minipage}{.737\textwidth}
		\begin{subfigure}{.325\textwidth}
			\includegraphics[width=\textwidth,trim=10 10 5 5,clip]{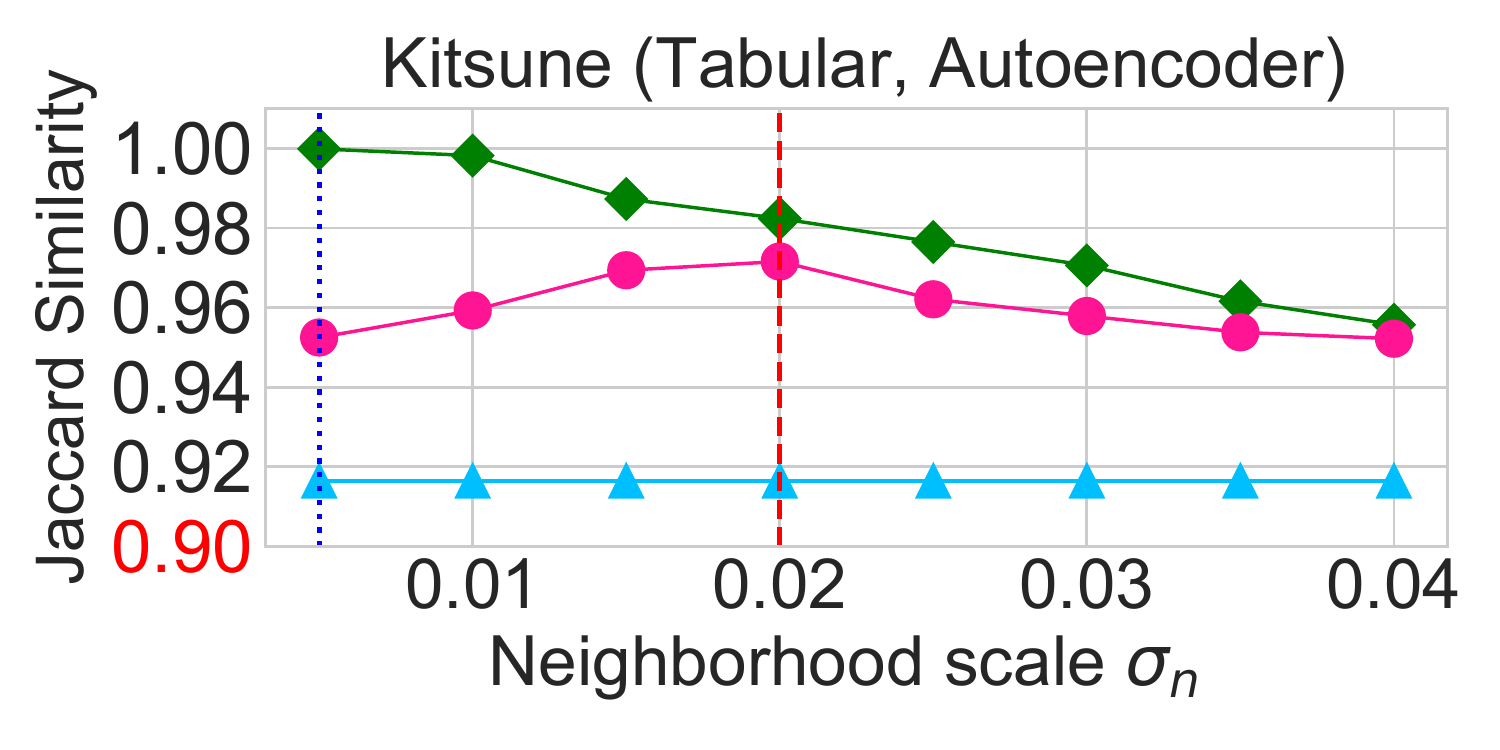}
			\vspace{-4.5ex}
			\caption{attack scale $\delta_{a}=0.2$ }
			\label{fig9a}
		\end{subfigure} %
		\hfill
		\begin{subfigure}{.325\textwidth}
			\includegraphics[width=\textwidth,trim=10 10 5 5,clip]{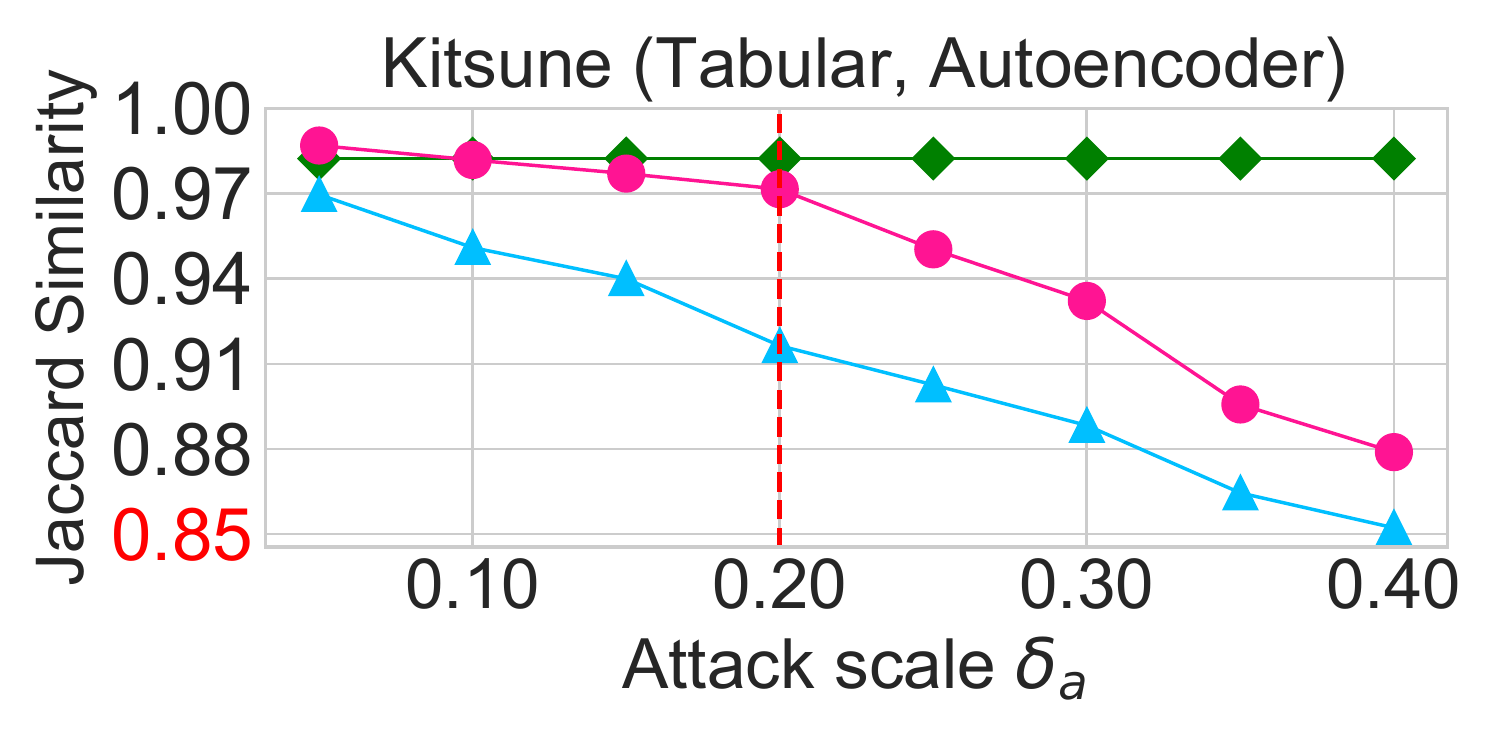}
			\vspace{-4.5ex}
			\caption{neighborhood scale $\sigma_{n}=0.02$ }
			\label{fig9b}
		\end{subfigure}		
		\hfill
		\begin{subfigure}{.325\textwidth}
			\includegraphics[width=\textwidth,trim=10 10 5 5,clip]{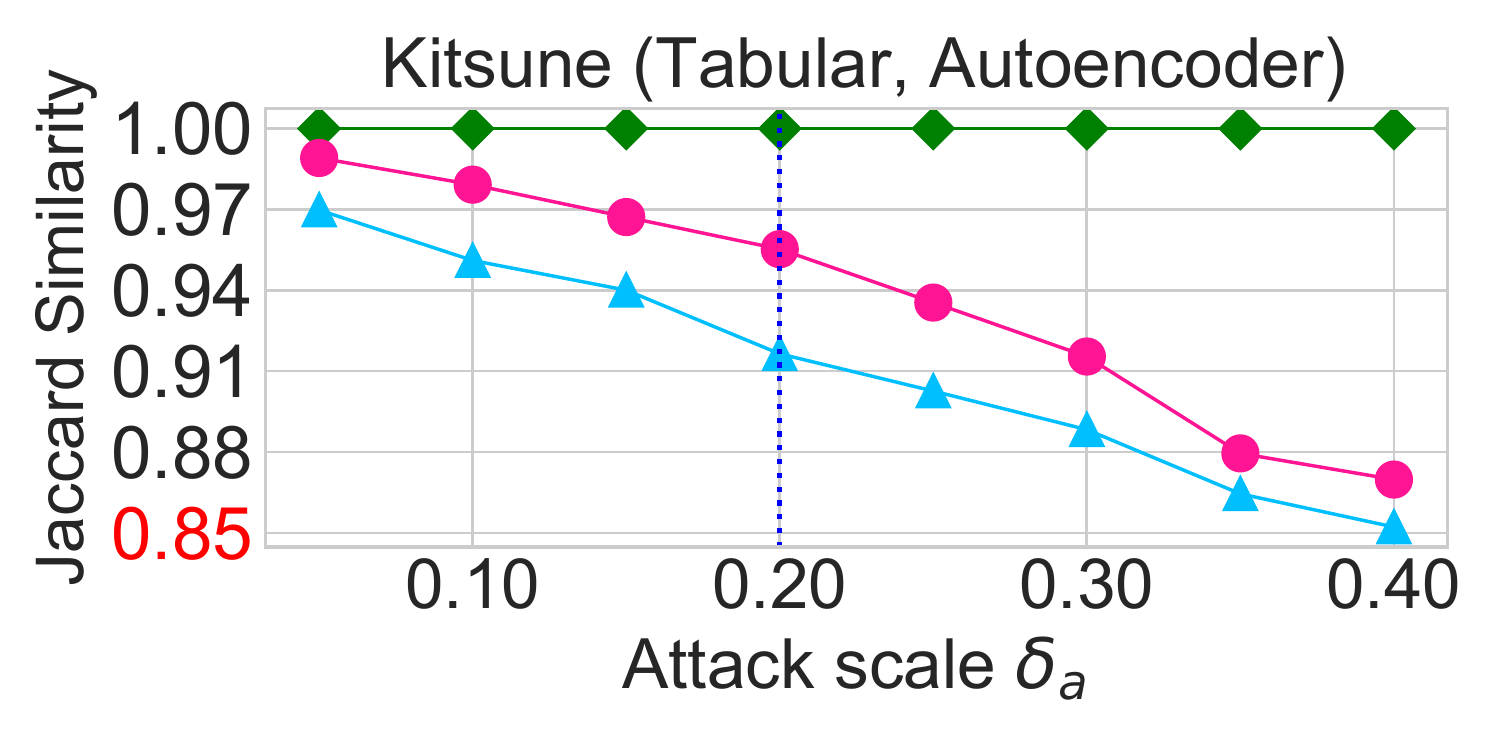}
			\vspace{-4.5ex}
			\caption{neighborhood scale $\sigma_{n}=0.005$ }
			\label{fig9c}
		\end{subfigure}
	
		\begin{subfigure}{\textwidth}
			\centering
			\includegraphics[width=\textwidth]{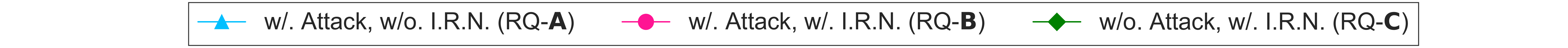}
		\end{subfigure}
        \vspace{-4ex}
		\setcounter{figure}{4}
		\captionof{figure}{\small Adversarial robustness (to optimization-based attack) evaluation of \textsf{DeepAID} (\emph{higher is better}).}
		\label{fig9}
	\end{minipage}
	\hfill
	\begin{minipage}{.25\textwidth}
		\centering
		\includegraphics[width=\textwidth]{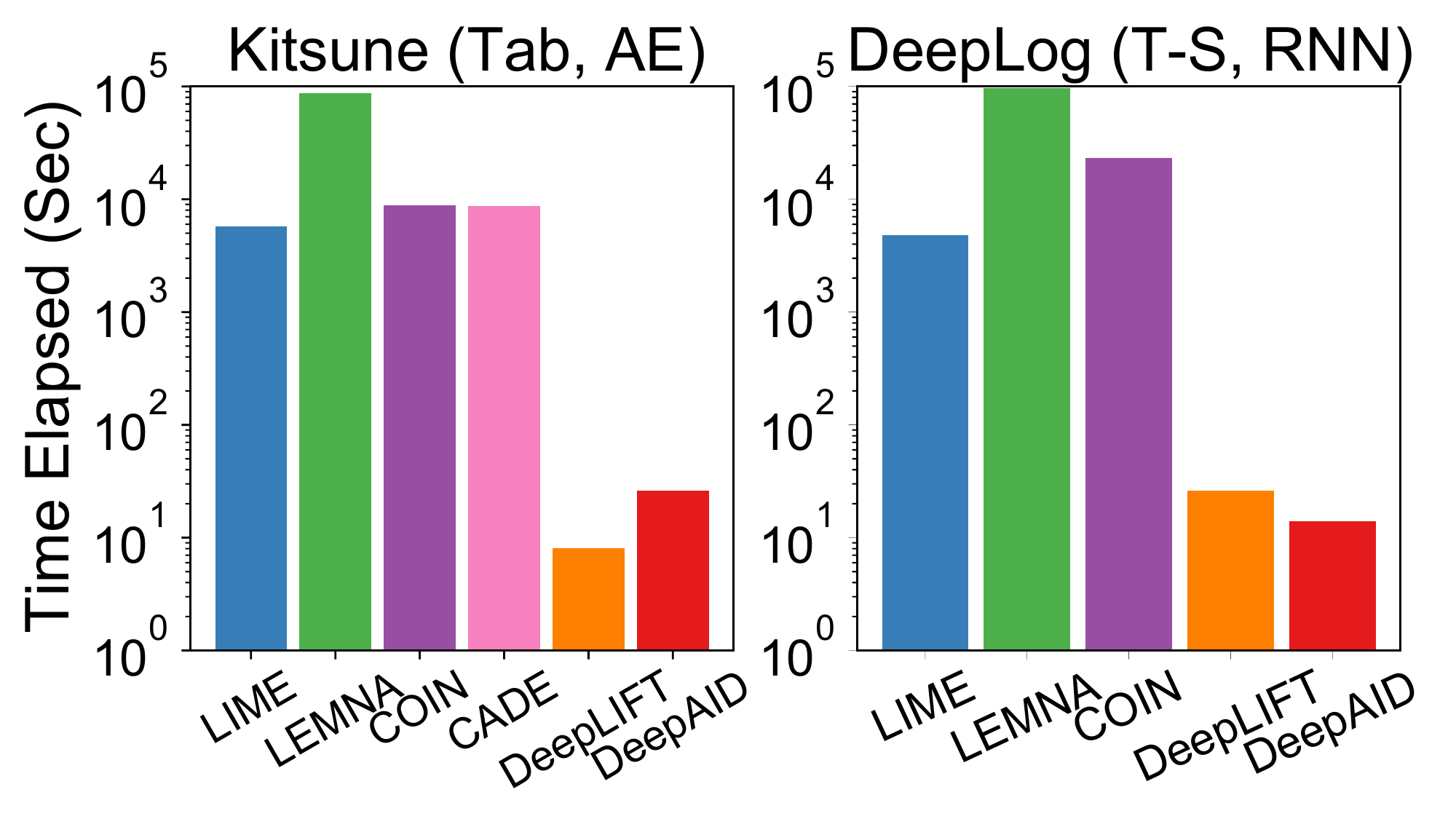}
        \vspace{-5ex}
		\captionof{figure}{\textcolor{black}{Efficiency evaluation of interpreters(\emph{lower is better}).}}
		\label{fig6}
	\end{minipage}

\end{figure*}


\noindent \textbf{Fidelity-Conciseness Evaluation.} To evaluate the fidelity of interpretations, we define an indicator similar to \cite{guo2018lemna} called \emph{Label Flipping Rate} (\textbf{LFR}) as the ratio of abnormal data that becomes normal after being replaced by interpretation results. The intuition is that LFR of high-fidelity interpreters will be higher since they accurately pinpoint important dimensions inducing the abnormality. Obviously, LFR tends to be higher when bringing more dimensions in interpretations, which destroys conciseness on the other hand.
We evaluate this fidelity-conciseness trade-off in Figure \ref{fig3a} under tabular and time-series based scenarios. The x-axis is in decreasing order of used dimensions (increasing order of conciseness). 
From the tabular result, \texttt{CADE} generally outperforms approximation-based interpreters w.r.t. fidelity, while is not good as back propagation based ones.
We can observe that directly selecting ``reference'' from existing in-distribution samples (S.R.T.D.) performs poorly especially with fewer dimensions, which demonstrates the necessity of \emph{searching} reference in \textsf{DeepAID}.
Results also demonstrate that supervised interpreters are not suitable for unsupervised learning due to their low fidelity.
Particularly, \textsf{DeepAID} significantly outperforms other methods,
especially when using fewer dimensions (e.g., \textsf{DeepAID} with 10\% dimensions exceeds others by >50\%).


\noindent \textbf{Stability Evaluation.} Stability of interpreters refers to the similarity of interpretations of the same samples during multiple runs.
Like \cite{warnecke2020evaluating,fan2020can}, we ignore the value in results and keep the index of important feature dimensions. 
Given two interpretation vectors $v_1$ and $v_2$ of important dimension indexes, we leverage \emph{Jaccard Similarity} (\textbf{JS}) to measure the similarity of two results, which is defined as $|\mathsf{set}(v1)\cap\mathsf{set}(v2)|/|\mathsf{set}(v1)\cup\mathsf{set}(v2)|$.
\textcolor{black}{We repeatedly measure the similarity of two interpretation results (from two runs with the same setting) for each anomaly and calculate the average.} The results are shown in Figure \ref{fig3b}. \textcolor{black}{We can observe that approximation/perturbation based methods perform poorly due to their random sampling/perturbation}, while back propagation based methods (\textsf{DeepAID} and \texttt{DeepLIFT}) have strong stability.


Below, we evaluate two kinds of robustness of interpreters: robustness to \emph{random noise} and \emph{adversarial attacks}. Only tabular \texttt{Kitsune} is evaluated since time series in \texttt{DeepLog} is discrete.

\noindent \textbf{Robustness Evaluation (to Noise).}
We measure \textbf{JS} of results interpreting tabular features before and after adding noise sampling from Gaussian $\mathcal{N}(0,\sigma^2)$ with $\sigma=0.01$. As shown in Figure \ref{fig3c}, \textcolor{black}{S.R.T.D. shows a high sensitivity to noise when selecting reference}. \texttt{COIN} and \texttt{DeepLIFT} are relatively robust compare with other baselines (\textcolor{black}{note that those unstable interpreters perform poorly even without noise}), but are still lower than our \textsf{DeepAID}.

\noindent \textcolor{black}{\textbf{Robustness Evaluation (to Attacks).} \textsf{DeepAID} falls into the category of
back propagation (B.P.) based method. Thus, we borrow the ideas from existing adversarial attacks against B.P. based interpreters \cite{ghorbani2019interpretation, zhang2020interpretable} to develop an adaptive attack on \textsf{DeepAID}, called \textbf{optimization-based} attack, which can be defined as adding small perturbation on anomaly $\boldsymbol{x}^\circ$ (denoted with $\boldsymbol{\tilde{x}}^\circ$) that can induce large changes of finally searched reference $\boldsymbol{x}^\ast$. Formally,}
\begin{equation*}
\textcolor{black}{{\rm argmax}_{\boldsymbol{\tilde{x}}^\circ} \|\mathcal{D}_{tab}(\boldsymbol{\tilde{x}}^\circ;\boldsymbol{\tilde{x}}^\circ) - \mathcal{D}_{tab}(\boldsymbol{x}^\circ;\boldsymbol{x}^\circ)\|_{p} \ \  {\rm s.t.} \ \|\boldsymbol{\tilde{x}}^\circ-\boldsymbol{x}^\circ\|_{p} < \delta_{a},}
\end{equation*}
\textcolor{black}{where $\delta_{a}$ limits the perturbation scale. Details of solving $\boldsymbol{\tilde{x}}^\circ$ are in Appendix \ref{appD.1}.
As mentioned in \S \ref{sec4.2}, \textsf{DeepAID} can \textbf{I}nitialize \textbf{R}eference from the \textbf{N}eighborhood of anomaly (called \textbf{I.R.N.} henceforth) to improve the adversarial robustness.
Below, we conduct three parts of experiments. (1) \textbf{RQ-A}: Robustness of \textsf{DeepAID} against the proposed attack without I.R.N. (2) \textbf{RQ-B}:
Impact of I.R.N. against the attack.
(3) \textbf{RQ-C}: Impact of I.R.N. on original stability.
RQ-A or RQ-B is measured by \textbf{JS} of interpretations before and after the attack without (RQ-A) or with (RQ-B) I.R.N., and RQ-C is measured by \textbf{JS} with and without I.R.N. in the absence of attack.}


\textcolor{black}{We extensively evaluate the impact of perturbation scale $\delta_{a}$ of optimization-based attack and neighborhood scale $\sigma_{n}$ of I.R.N.. Results are shown in Figure \ref{fig9}. We first fix $\delta_{a}$ in \ref{fig9a} and observe the impact of $\sigma_{n}$, and vice versa in \ref{fig9b}/\ref{fig9c}.
We find \textsf{DeepAID} without I.R.N. is naturally robust against such attack (\textbf{JS} $>0.91$ when $\delta_{a}=0.2$ and \emph{linearly} decreases as $\delta_{a}$ increases in \ref{fig9b}/\ref{fig9c}),
thanks to the search-based idea analyzed in \S \ref{sec4.2} (\textbf{RQ-A}). Results also demonstrate the effectiveness of I.R.N. In \ref{fig9a}, I.R.N. mitigates most attack effect when $\sigma_{n}\geq0.02$ (\textbf{RQ-B}). We also find I.R.N. has a very small impact on the original stability (\textbf{JS} $>0.95$ even when $\sigma_{n}=0.04$), which can be viewed as a trade-off. Hence, a simple and safe way is to choose a small $\sigma_{n}$, which can also mitigate the attack effect without loss of the original stability, as demonstrated in \ref{fig9c} (\textbf{RQ-C}).}

\textcolor{black}{We also evaluate another type of adversarial attacks that misleads the distance calculations in \textsf{DeepAID} (called \textbf{distance-based} attacks). The results demonstrate the strong robustness of \textsf{DeepAID} against such attacks. For reasons of space, detailed definitions, results, and analysis against such attacks are in Appendix \ref{appD.2}.}   



\begin{table*}[]
	\setstretch{0.75}
	\caption{Example interpretations ($K=5$) for \texttt{Kitsune}.}
	\vspace{-3mm}	
	\label{tab3}
	\centerline{\small \textbf{(a) Ground truth: Remote command execution}}
	\resizebox{0.99\linewidth}{!}{%
		\begin{tabular}{|@{ }c@{ }|@{ }c@{ }|@{ }l@{ }|@{ }l@{}|}
			\Xhline{2\arrayrulewidth}
			\textbf{Feature Name} &
			$\mathbf{x}^\circ<>\mathbf{x}^\ast$ \footnotemark[1]&
			\textbf{Feature Meaning} &
			\textbf{Expert's Understanding} \\ \Xhline{2\arrayrulewidth}
			HpHp\_1.0\_mean &
			90 > 60 &
			The average packet length of the current connection within 1 sec time window &
			Transport layer communication within a single connection \\ \hline
			HpHp\_1.0\_std &
			30 > 10 &
			The standard deviation of pkt length of the current connection within 1 sec &
			The length of the packet payload is not fixed (possibly some commands) \\ \hline
			HH\_1.0\_mean &
			90 > 60 &
			The average length of the all pkts with the same src-dst IPs within 1 sec  &
			No other port is used between src-dst IPs (since 90 is the same as the first feature) \\ \hline
			HpHp\_5.0\_std &
			20 > 10 &
			The standard deviation of pkt length of the current connection within 5 sec &
			This is a rather longer communication (since 5 sec) \\ \hline
			HpHp\_3.0\_mean &
			90 < 1,000 &
			The average packet length of the current connection within 1 sec time window &
			Not a lot of content transferred (possibly short commands) \\ \Xhline{2\arrayrulewidth}
		\end{tabular}%
	}
	
	\vspace{2mm}
	\centerline{\small \textbf{(b) Ground truth: ARP scan}}
	\resizebox{0.99\linewidth}{!}{%
		\begin{tabular}{|@{ }c@{ }|@{ }c@{ }|@{ }l@{ }|@{ }l@{}|}
			\Xhline{2\arrayrulewidth}
			\textbf{Feature Name} &
			\textbf{$\mathbf{x}^\circ<>\mathbf{x}^\ast$ \footnotemark[1]} &
			\textbf{Feature Meaning} &
			\textbf{Expert's understanding} \\ \Xhline{2\arrayrulewidth}
			MI\_dir\_5.0\_weight &
			$10^{5}>10^{4}$
			&
			\# packets send from the same MAC\&IP within 5 sec time window &
			A large number of packets sent from the same host and long-term (5 sec) \\ \hline
			MI\_dir\_0.1\_weight &
			300 > 100 &
			\#  packets send from the same MAC\&IP within 0.1 sec time window &
			A large number of packets sent from the same host (possibly Scan) \\ \hline
			MI\_dir\_5.0\_std &
			0.1 < 10 &
		    std. of packet length of the same MAC\&IP within 5 sec &
			The length of  packets sent from the same source are almost the same (std. is small) \\ \hline
			MI\_dir\_0.1\_std &
			1 < 10 &
			std. of packet length of the same MAC\&IP within 0.1 sec &
			The length of  packets sent from the same source are almost the same (std. is small) \\ \hline
			HH\_1.0\_weight &
			10 < 200 &
			The average length of pkts with the same src-dest IPs within 1 sec &
			The communication is not from the same connection (10 is small, more likely Scan) \\ \Xhline{2\arrayrulewidth}
		\end{tabular}%
	}
	
	\vspace{2mm}
	\centerline{\small \textbf{(c) False positive example}}
	\resizebox{0.99\linewidth}{!}{%
		\begin{tabular}{|@{ }c@{ }|@{ }c@{ }|@{ }l@{ }|l|}
			\Xhline{2\arrayrulewidth}
			\textbf{Feature Name} &
			\textbf{$\mathbf{x}^\circ<>\mathbf{x}^\ast$ \footnotemark[1]} &
			\textbf{Feature Meaning} &
			\textbf{Expert's understanding} \\ \Xhline{2\arrayrulewidth}
			HpHp\_1.0\_covariance &
			$10^{30}>10$ &
			The src-dest covariance of packet length of the current connection within 1 sec time window &
			\multirow{5}{*}{\begin{tabular}[c]{@{}l@{}}There may be bugs or bias in the \\ implementation of feature extraction\end{tabular}} \\ \cline{1-3}
			HpHp\_1.0\_pcc &
			$10^{28}>1$ &
			The src-dest correlation coefficient of packet length of the current connection within 1 sec time window &
			\\ \cline{1-3}
			HH\_0.1\_covariance &
			$10^{28}>10$ &
			The src-dest covariance of packet length with the same src-dest IPs within 0.1 sec time window &
			\\ \cline{1-3}
			HH\_0.1\_pcc &
			$10^{28}>1$ &
			The src-dest correlation coefficien of packet length with the same src-dest IPs within 0.1 sec time window &
			\\ \cline{1-3}
			HpHp\_3.0\_covariance &
			$10^{29}>10$ &
			The src-dest covariance of packet length of the current connection within 3 sec time window &
			\\ \Xhline{2\arrayrulewidth}
		\end{tabular}%
	}
\end{table*}

\noindent \textbf{Efficiency Evaluation.} We evaluate the efficiency by recording runtime for interpreting $2,000$ anomalies for each interpreter. For approximation-based interpreters (\texttt{LIME}, \texttt{LEMNA}, and \texttt{COIN}), the time to train the surrogate model will also be counted, \textcolor{black}{and the same for time to train the encoding model in \texttt{CADE}.} 
The results are shown in Figure \ref{fig6} under tabular and time-series based scenarios. We observe that \textsf{DeepAID} and \texttt{DeepLIFT} are at least two orders of magnitude faster than the other methods. 

\noindent \textbf{Conclusions.} We conclude experiments of \S \ref{sec6.2} here (and also in Table \ref{tab2}) that only our \textsf{DeepAID} can effectively meet all special requirements of security domains and produce high-quality interpretations for unsupervised DL-based anomaly detection. 

Below, we will showcase how \textsf{DeepAID} can help to improving the practicability of security systems from several aspects. Figure \ref{fig4} shows the flowchart of DeepAID usage with the following use cases.

\begin{figure}[]
	\centering
	\includegraphics[width=0.9\linewidth]{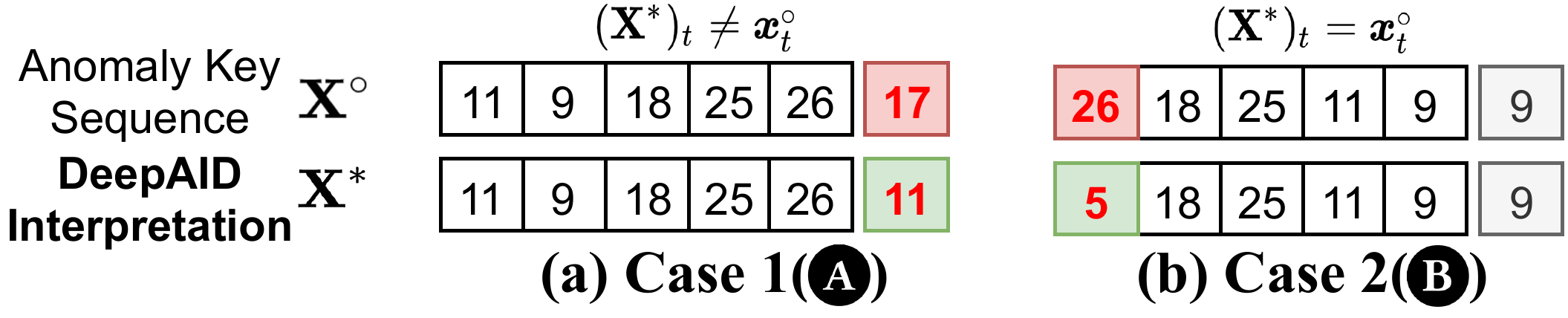}
	
	\vspace{1mm}
	\resizebox{\linewidth}{!}{%
		\begin{tabular}{|c|l||c|l|}
			\hline
			\textbf{Log Key} & \textbf{Meaning}                      & \textbf{Log Key} & \textbf{Meaning}                            \\ \hline
			5                & Receiving block                       & 18               & Starting thread to transfer block           \\ \hline
			9                & Receiving block with size             & 22               & NameSystem(NS).allocateBlock                    \\ \hline
			11               & Responder for block terminating & 25               & Ask to replicate block                      \\ \hline
			17               & Failed to transfer block              & 26               & NS.addStoredBlock: blockMap updated \\ \hline
		\end{tabular}%
	}
	\vspace{-1ex}
	\caption{Example interpretations ($K=1,t=6$) for \texttt{DeepLog}.}
	\label{fig5}
	\vspace{-1ex}
\end{figure}

\subsection{Understanding Model Decisions}
\label{sec6.3}
In this section, we use several cases to demonstrate that \textsf{DeepAID} can capture well-known rules (\ballnumber{A} in Figure \ref{fig4}) as well as discover new knowledge (\ballnumber{B}).

\footnotetext[1]{For ease of illustration, the feature values have been de-normalized and approximated.}

\noindent \textbf{Tabular Data Interpretations.}
We use cases of network intrusion detector \texttt{Kitsune} for illustration. Two representative anomalies in Mirai botnet traffic are interpreted, where the first one is to remotely control the compromised bot to execute malware ``mirai.exe'' and execute a series of commands to collect information, and the second one is to ARP scan of open hosts in its LAN for lateral movement. Suppose that the operator already knows the ground truth (mentioned above) of these two anomalies by analyzing the raw traffic, their goal is to see whether  the DL model has captured well-known rules with the help of \textsf{DeepAID} Interpreter. Here we use $K=5$ important feature dimensions, and their names and meanings are listed in Table \ref{tab3}(a)(b), where the second column is the value of anomaly $\mathbf{x}^\circ$ and reference vector $\mathbf{x}^\ast$ in corresponding dimensions. 
Expert's understandings based on the interpretations are also listed.
Since (a) and (b) in Table \ref{tab3} are both interpretable and reasonable for experts, we can draw the conclusion that DL model has learned the expected well-known rules.    

\noindent \textbf{Time-Series Interpretations.} As shown in Figure \ref{fig5}, we use two simple cases of HDFS log anomaly detector \texttt{DeepLog} for illustration. We set sequence length $t=6$ and use only $K=1$ dimension for interpretation. Recall that time-series Interpreter needs to determine the location of anomalies before solving reference vector $\mathbf{X}^\ast$ through saliency testing (\S \ref{sec4.3}). In case 1, our Interpreter determines that the abnormality occurs at $\boldsymbol{x}^\circ_{t}$ and replaces 17 with 11. Since 17 is originally an abnormal log, we conclude that our Interpreter captures well-known rules. In case 2, Interpreter determines that abnormality occurs at $\boldsymbol{x}^\circ_{1}\boldsymbol{x}^\circ_{2}...\boldsymbol{x}^\circ_{t-1}$ and replaces $\boldsymbol{x}^\circ_{1}$ with 5. The anomaly is probably because blockmap is updated without any preceding block operations. This is not an explicit rule and needs to be analyzed by domain experts, which demonstrates that \textsf{DeepAID} can discover new heuristics beyond explicit knowledge.

\begin{figure}
	\begin{minipage}{.65\linewidth}
		\includegraphics[width=\linewidth]{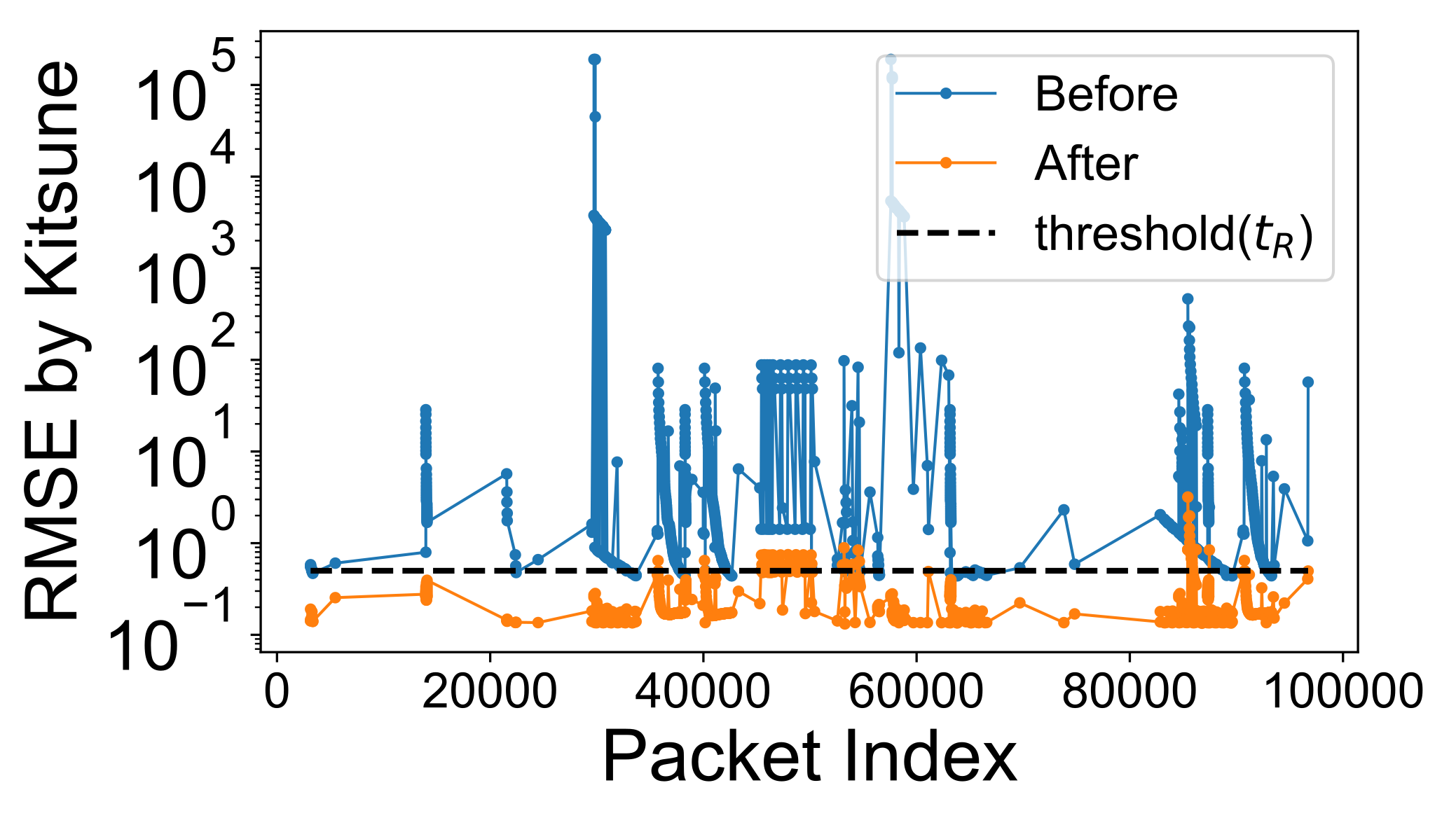}
	\end{minipage}
	\hfill
	\begin{minipage}{.34\linewidth}
		\resizebox{\linewidth}{!}{
			\begin{tabular}{@{ }c@{ }@{ }c@{ }@{ }c@{ }}
				\toprule
				& \textbf{TP} & \textbf{FP}\\
				\midrule
				Before & 66,597 & 1,556 \\
				After & 66,476 & 79 \\ 
				Change & -0.17\%$\downarrow$  & -94.92\%$\downarrow$ \\
				\bottomrule
			\end{tabular}
		}
		\begin{flushleft}
			\setstretch{0.4}
			\small
			\noindent \textbf{*} \emph{The points in the left figure are all normal data (true label)}.
		\end{flushleft} 
	\end{minipage}
	\vspace{-3ex}
	\caption{Case for debugging \texttt{Kitsune}.}
	\label{fig7}
\end{figure}

\begin{table*}
	\parbox{0.7\textwidth}{
		\setstretch{0.95}
		\caption {Performance of reliable detection based on Distiller.}
		\label{tab4}
		\centering
		\resizebox{0.7\textwidth}{!}{%
			\begin{tabular}{|c|c|c|c|c|c|c|c|c|c|c|c|}
				\hline
				\multicolumn{2}{|c|}{\multirow{2}{*}{Method}} & \multicolumn{5}{c|}{Number of Classes = 5} & \multicolumn{5}{c|}{Number of Classes = 10} \\ \cline{3-12} 
				\multicolumn{2}{|c|}{} & f1-micro & f1-macro & f1-micro* & f1-macro* & UACC & f1-micro & f1-macro & f1-micro* & f1-macro* & UACC \\ \hline
				\multicolumn{2}{|c|}{RF} & 0.9980 & 0.9983 & 0.6876 & 0.7351 & N/A(0.) & 0.9836 & 0.9827 & 0.7629 & 0.7848 & N/A(0.) \\ \hline
				\multicolumn{2}{|c|}{MLP} & 0.8207 & 0.8491 & 0.8637 & 0.8399 & N/A(0.) & 0.8732 & 0.8791 & 0.8196 & 0.8101 & N/A(0.) \\ \hline
				\multirow{3}{*}{\begin{tabular}[c]{@{}c@{}}\texttt{Kitsune}+\\ \textsf{DeepAID}\end{tabular}} & K=5 & 0.9782 & 0.9821 & 0.8904 & 0.8444 & 0.7330 & 0.9415 & 0.9408 & 0.8204 & 0.7780 & 0.5217 \\ \cline{2-12} 
				& K=10 & 0.9891 & 0.9911 & 0.9690 & 0.9641 & 0.7944 & 0.9797 & 0.9790 & 0.8822 & 0.8441 & 0.9357 \\ \cline{2-12} 
				& K=15 & \textbf{0.9990} & \textbf{0.9991} & \textbf{0.9779} & \textbf{0.9736} & \textbf{1.0} & \textbf{0.9975} & \textbf{0.9975} & \textbf{0.9001} & \textbf{0.8728} & \textbf{0.9801} \\ \hline
			\end{tabular}%
		}
		
		\begin{flushleft}
			\setstretch{0.7}
			\footnotesize
			\noindent \emph{\textbf{f1-micro} and \textbf{f1-macro} are evaluated under  \textbf{training data}, and \textbf{f1-micro*} and \textbf{f1-macro*} under \textbf{test data};} \\
			\noindent \texttt{Kitsune} (w/o. \textsf{DeepAID}): \textbf{UACC=0.28} \emph{(not evaluated for f1-micro(*) and f1-macro(*) since it cannot classify multi-class data).} \\
			\noindent (\textbf{Higher is better} for all five metrics.)
		\end{flushleft} 
	}
	\hfill
	\parbox{0.29\textwidth}{
		
		\caption{Reducing FPs by Distiller.}	
		\vspace{-3ex}
		\label{tab5}
		\centering
		\resizebox{0.28\textwidth}{!}{%
			\setstretch{0.75}
			\begin{tabular}{|c|c|c|c|c|c|}
				\hline
				\multicolumn{2}{|c|}{\multirow{2}{*}{Method}} & \multicolumn{2}{c|}{\# classes = 5} & \multicolumn{2}{c|}{\# classes = 10} \\ \cline{3-6} 
				\multicolumn{2}{|c|}{} & TPR & FPR & TPR & FPR \\ \hline
				\multicolumn{2}{|c|}{RF} & 0.9989 & 0.9719 & 0.9990 & 0.7504 \\ \hline
				\multicolumn{2}{|c|}{MLP} & 1.0 & 0.0026 & 0.9998 & 0.3319 \\ \hline
				\multicolumn{2}{|c|}{\texttt{Kitsune}} & 0.9755 & 0.0816 & 0.9717 & 0.0845 \\ \hline
				\multirow{3}{*}{\begin{tabular}[c]{@{}c@{}}\texttt{Kitsune}+\\ \textsf{DeepAID}\end{tabular}} & K=5 & \textbf{1.0} & \textbf{0.0} & 0.9999 & \textbf{0.0002} \\ \cline{2-6} 
				& K=10 & 1.0 & 0.0 & \textbf{1.0} & 0.0046 \\ \cline{2-6} 
				& K=15 & 1.0 & 0.0 & 1.0 & 0.0220 \\ \hline
			\end{tabular}%
		}
		\vspace{1ex}
		
		\includegraphics[width=\linewidth]{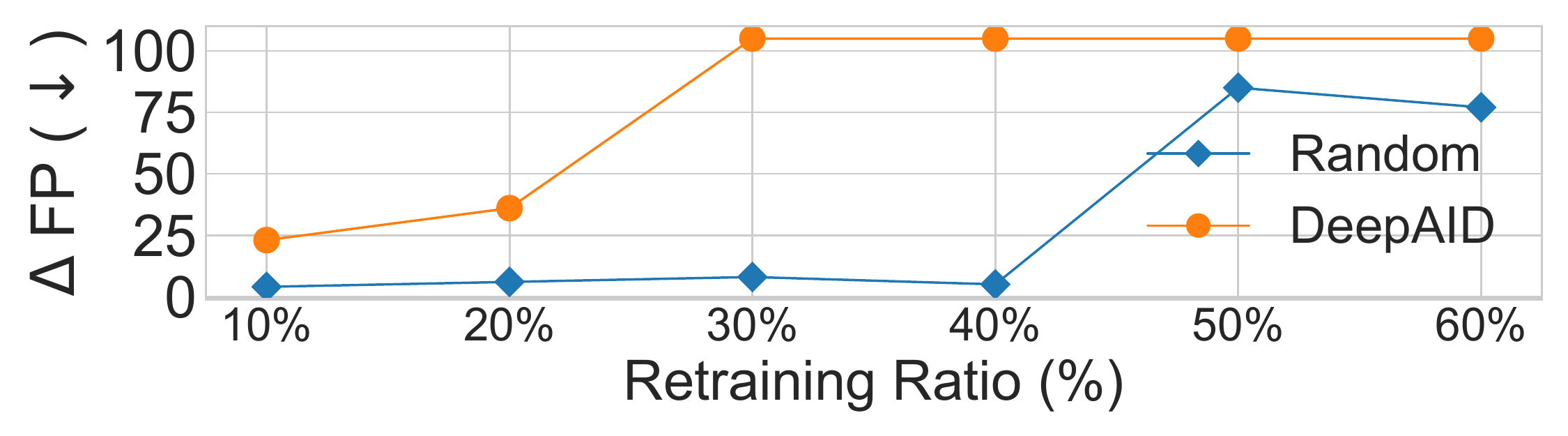}
		\vspace{-6.5ex}
		\captionof{figure}{\small Reducing FPs by retraining.}
		\label{fig8}
	}
\end{table*}

\subsection{Enabling Debuggability}
\label{sec6.4}
Below, we primarily use cases of \texttt{Kitsune} to illustrate how to use \textsf{DeepAID} interpretations to diagnose \emph{explicit} errors in the system (i.e., \ballnumber{C} in Figure \ref{fig4}). Here \emph{explicit} means human-factor errors that are easy to repair. (In contrast to the \emph{implicit} errors caused by deviant model learning, which will be discussed in \S \ref{sec6.6}).

In \texttt{Kitsune}, we find some false positives (FP) with extremely high reconstruction error (RMSE). 
From interpretations (a representative one is listed in Table \ref{tab3}(c)), we find that these FPs are caused by morbidly large features related to the covariance (cov.) or Pearson  correlation coefficient (pcc.). After analyzing the original traffic, we find the relevant features should not be so large. Therefore, we infer that there may be bugs or bias in the implementation of feature extraction algorithm in \texttt{Kitsune}. By examining the open-source implementation of \texttt{Kitsune} \cite{mirsky2018kitsune},
we note that they claimed that they used approximation algorithms to compute related features instead of counting real values. Therefore, we modified the implementation of extracting cov./pcc. related features to calculate their real value. As shown in Figure \ref{fig7} (with Mirai botnet traffic), by fixing this error, we reduce the FPs (94.92\%$\downarrow$) in \texttt{Kitsune} without much affecting the detection of anomalies (only 0.17\%$\downarrow$). 


\subsection{Human-in-the-Loop Reliable Detection}
\label{sec6.5}
Next we evaluate the performance of Distiller as an extension to Interpreter to perform human-in-the-loop detection (\S \ref{sec5}).
Specifically, we design experiments to answer the following questions:
\begin{itemize}[leftmargin=*]
	\item \textbf{RQ1:} Can Distiller accurately match the existing rules?
	
	\item \textbf{RQ2:} Whether Distiller has generalization ability to detect similar threats that do not appear in existing rules?
	
	\item \textbf{RQ3:} Can Distiller preserve the ability to detect unknown threats?
\end{itemize}

\noindent \textbf{Experimental Setting.} 
To evaluate Distiller, expert feedback to anomaly interpretations is required as rules for updating Distiller. However, human feedback is experience-dependent and very subjective.
For fairness and simplicity, we select two well-labeled multi-class datasets, \texttt{Kitsune} dataset \cite{mirsky2018kitsune} and CIC-IDS2017 \cite{sharafaldin_toward_2018}. 
As mentioned in \S \ref{sec5}, we use two ML/DL classifiers as the baseline: Random Forest (RF) and Multi-Layer Perceptron (MLP).
We also evaluate the impact of $K$ for Distiller with $K=5,10,15$.
We respectively use 5 and 10 different classes/feedback states of attack traffic and 200 typical data(/rules) in each class for training/updating (thus totally 1000 and 2000). \textcolor{black}{Note that, we \emph{update} Distiller with the same training set for \textbf{RQ1}/\textbf{RQ2}/\textbf{RQ3} while \emph{test} on training set (\textbf{RQ1}) and test set (\textbf{RQ2}). Particularly for \textbf{RQ3}, we test Distiller on another class of anomalies not in the training set.} As for metrics, we evaluate f1-micro and f1-macro (f1-score in multi-classification) for \textbf{RQ1}/\textbf{RQ2}, and accuracy rate of unknown attack detection (UACC) for \textbf{RQ3}.
Details about the datasets/metrics are in Appendix \ref{app3}.

\noindent \textbf{Results and Conclusions.} Results are shown in Table \ref{tab5}.
From f1-micro/f1-macro, we find
\textsf{DeepAID} can match existing rules very accurately (\textbf{RQ1}). From f1-micro*/f1-macro*, we find \textsf{DeepAID} significantly outperforms baselines on generalization ability (\textbf{RQ2}). Especially for $K$=15, f1-micro*/f1-macro* of \textsf{DeepAID} exceeds RF and MLP by ~10\% and ~20\%. \textcolor{black}{We find MLP is even worse than RF, which may be because the training set is relatively small, while Distiller performs very well with fewer rules.}
The choice of $K$ can be viewed as a trade-off. That is, a larger $K$ ensures that the rules are more detailed (hence more accurate) but will lose the conciseness.
From the results of UACC, we find that \textsf{DeepAID} not only retains the ability to detect unknown threats, but also significantly improves the original 28\% to >98\% (for $K$=15). This is because DL models failed to report unknown threats that are similar to normal data, while Distiller has explicit rules to match known anomalies, thus can judge unknown threats more accurately (\textbf{RQ3}).
This also shows that \textsf{DeepAID} can also reduce false negatives (FN) incidentally.
This experiment confirms \textsf{DeepAID} can help security systems to handle thousands of rules with superior performance on detecting stored (\textbf{RQ1}), similar (\textbf{RQ2}), and unknown (\textbf{RQ3}) anomalies.



\subsection{Reducing FPs}
\label{sec6.6}

We evaluate two methods mentioned in \S \ref{sec5} to reduce FPs in security systems with Distiller (also \ballnumber{D} and \ballnumber{E} in Figure \ref{fig4}).

\noindent \textbf{Modifying Distiller.} We select only 10 rules from 10 FPs to a new feedback state representing normal data, and use the same baselines and datasets in \S \ref{sec6.5}. Here we consider all abnormal classes as one class and use two well-known metrics for evaluation: TPR and FPR (True/False Positive Rate). The results are shown in Table \ref{tab5}. \textsf{DeepAID} demonstrates extremely superior performance with nearly no FP and 100\% TPR. It significantly reduces FPR of \texttt{Kitsune} (from 8\% to 0\%) while improving its TPR. The performance of baselines are poor with respect to FPR. Here we find a small number of FPs appears when $K$=15.
This is because higher $K$ is more likely to induce more ``\emph{conflicts}'', which means that one state in the first FSM transits to multiple feedback states. We provide the detailed theoretical analysis of conflicts in Appendix \ref{app2}.

\noindent \textbf{Retraining the DL model.} Another method to reduce FPs mentioned in \S \ref{sec5} is to retrain DL models with concept drift FPs found by Distiller. Here we use a simple baseline by randomly selecting FPs for retraining, in order to evaluate whether \textsf{DeepAID} can find the more optimal retraining set. As shown in Figure \ref{fig8}, we conclude that \textsf{DeepAID} can help to save the cost of retraining and achieve high performance (e.g., with \textsf{DeepAID}, retraining 30\% samples reduces more FPs than 60\% in the random method).

%% file: end.tex
\section{Discussions}
\label{sec7}
Below, we discuss design choices, limitations, and future works.

\noindent \textbf{Why only Interpret Anomalies?} In practical security applications, operators are more concerned about anomalies. Actually, they cannot even handle all anomalies, let alone the normal data.


\noindent \textbf{Why White-box Interpretation?}
Besides the taxonomy in \S \ref{sec2.1}, local DL interpretations can also be separated as \emph{white-box} and \emph{black-box}. White-box methods require interpreters to have full access to the DL model, while black-box methods treat the DL model as a black box without detailed knowledge. \textsf{DeepAID} and \texttt{DeepLIFT} are white-box for using model gradients, while other baseline interpreters are black-box.
Indeed, black-box methods are more convenient for operators. However, as justified in related works and our experiments (\S \ref{sec6.2}), they are difficult to use in practice due to the 
relatively poor performance. 
On the other hand, the white-box condition makes sense when the one seeking interpretations is the model owner. \cite{guo2018lemna} thinks white-box condition is unreasonable when using pre-trained models, but this is not the case for unsupervised DL models since they do not need pre-training.
In fact, \textsf{DeepAID} only requires the back-propagated gradient information with no need for fully understanding or modifying the original model, which avoids burdening the operators.

\noindent \textcolor{black}{\textbf{Limitations and Future Work.} 
First, the adversarial robustness evaluation and claim of \textsf{DeepAID} are mainly against optimization-based and distance-based attacks (in \S \ref{sec6.2} and Appendix \ref{app6}).
There are other target attacks that may fail \textsf{DeepAID}, such as poisoning the original models (as \textsf{DeepAID} is highly faithful to model decisions), hijacking the search process to generate false references, and crafting anomalies \emph{extremely} or \emph{equally} far from the decision boundary in many dimensions to force interpretations to use more features (spoiling the conciseness).
Future work can investigate the robustness against more attacks.
Second, hyper-parameters of \textsf{DeepAID} are configured empirically. 
We have evaluated the sensitivity of hyper-parameters in Appendix \ref{app5}, and provided brief guidelines about how to configure them in the absence of anomalies. 
Future work can develop more systematic strategies to configure hyper-parameters. 
Third, we implement Distiller only for tabular data. Future work will focus on extending Distiller to other data types.
Fourth, the practical effectiveness of \textsf{DeepAID} may depend on the expert knowledge of operators, since they do not have prior knowledge of anomalies in the unsupervised setting. Future work can investigate the impact of knowledge level and look into approaches to relax the requirement of expert knowledge for operators.
}


\section{Related Work}
\label{sec8}
Most related works have been introduced in \S \ref{sec2}. Below, we briefly discuss other related works from two aspects.


\noindent \textcolor{black}{\textbf{Interpreting Unsupervised Learning \& Anomaly Detection.} Recently, there are a few studies discussing interpretation methods on unsupervised learning or anomaly detection.
First, some studies develop global interpretation through model distillation \cite{song2018exad} and rule extraction \cite{abs-1911-09315} as well as interpretation for clustering models \cite{abs-1906-07633}, which are beyond the scope of this study. 
As for local interpretation of anomaly detection, in addition to \texttt{COIN} \cite{liu2018contextual} and \texttt{CADE} \cite{yang2021cade} which are used as baselines, most of other works simply modify the existing supervised methods to unsupervised learning.
For example, SHAP interpretation \cite{LundbergL17} is adapted to (single) Autoencoder in \cite{antwarg2019explaining}, which cannot be applied to \texttt{Kitsune} (ensemble Autoencoders) or other DL models. 
Deep Taylor Decomposition \cite{MontavonLBSM17} is leveraged for interpreting one-class SVMs in \cite{KauffmannMM20}, which cannot be applied to DNNs. In \cite{carletti2019explainable}, a dedicated interpretation for Isolation Forest is developed which also cannot be applied to DNNs.
 An interpretation method for convolutional models and image analysis is proposed in \cite{KitamuraN19} by visualizing abnormal region, which is unadaptable for other DL models in security domains. 
There are also works converting anomaly detection to supervised tasks and leverage existing supervised interpretations \cite{amarasinghe2018toward, morichetta2019explain}. Other works \cite{BailisGMNRS17,abs-2010-05073} for interpreting anomaly in big-data streaming systems use the knowledge of many other anomalies (time-series intervals), which is hard to be generalized into the unsupervised setting in security domains.
}

\noindent \textbf{Improving DL-based security systems.} Recently, several works have been proposed for improving the practicality of (unsupervised) DL-based security systems, such as improving the robustness of DL-based systems \cite{gehr2018ai2,han2020evaluating}, performing lifelong learning \cite{du2019lifelong}, detecting concept drift \cite{yang2021cade}, \textcolor{black}{learning from low-quality labeled data} \cite{liang2021fare}. They are orthogonal to our work and could be  adopted together for developing more practical security systems. 


\section{Conclusions}
\label{sec9}

In this paper, we propose \textsf{DeepAID}, a general framework to interpret and improve DL-based anomaly detection in security applications.
We design \textsf{DeepAID} Interpreter by optimizing well-designed objective functions with special constraints for security domains to provide high-fidelity, human-readable, stable, robust, and efficient interpretations.
Several applications based on our interpretations and the model-based extension Distiller are proposed to improve the practicality of security systems, including understanding model decisions, discovering new knowledge, diagnosing mistakes, and reducing false positives.
By applying and evaluating \textsf{DeepAID} over three types of security-related anomaly detection systems, we show that \textsf{DeepAID} can provide high-quality interpretations for DL-based anomaly detection in security applications.

\begin{acks}
	We are grateful to the anonymous reviewers for their insightful comments. 
	We also thank for the suggestions from Shize Zhang, Bin Xiong, Kai Wang, as well as all other members from NMGroup and CNPT-Lab in Tsinghua University.
	This work was supported in part by the National Key Research and Development Program of China under Grant 2018YFB1800200. Zhiliang Wang is the corresponding author of this paper.
\end{acks}

%% file: supp.tex
\section*{Appendices}

\appendix

\section{Algorithms of Interpreters}
\label{app1}
We provide the procedure of time-series and graph-data Interpreter.
\subsection{Procedure of Time-Series Interpretation}
The procedure of \textsf{DeepAID} Interpreter for time-series based systems is in Algorithm \ref{alg2}.

\setlength{\textfloatsep}{.5ex}
\begin{algorithm}[!h]
	\small
	\setstretch{0.9}
	\SetKwComment{Comment}{$\triangleright$\ }{}
	\SetKwComment{Commentld}{$\triangledown$\ }{} 
	\SetCommentSty{em}
	\SetKwFunction{Adam}{Adam}
	
	\caption{Interpreting time-series anomalies (univariate)}
	\label{alg2}
	
	\KwIn{Interpreted anomaly $\mathbf{X}^\circ$; $max\_iter$, learning rate $\alpha$; $\mu_1,\mu_2$}
	\KwOut{Interpretation result $\mathbf{X}^\circ-\mathbf{X}^\ast$ with the reference $\mathbf{X}^\ast$ }

	$\boldsymbol{x}^c \leftarrow \operatorname{argmax}_{\boldsymbol{x}^c} \ {\rm Pr}(\boldsymbol{x}^c|\boldsymbol{x}^\circ_1\boldsymbol{x}^\circ_2...\boldsymbol{x}^\circ_{t-1})$;
	
	\eIf(\Comment*[f]{saliency testing}){$ST(\mathbf{X}^\circ;\mu_1,\mu_2) = {\textbf{True}}$ }{
		
		$\mathbf{X}^\ast \leftarrow \boldsymbol{x}^\circ_1\boldsymbol{x}^\circ_2...\boldsymbol{x}^\circ_{t-1}\boldsymbol{x}^c$;
	}{  
	
		Initialize $\mathbf{X}^\ast$ with $\mathbf{X}^\circ$; $t \leftarrow 0$;
		
		\While(\Comment*[f]{iterative optimization}){$t<max\_iter$}
			{	
				$\mathbf{X}^\ast \leftarrow$\Adam{$\mathbf{X}^\ast;\mathcal{D}_{ts},\alpha$} \Comment*{update $\mathbf{X}^\ast$ by Adam optimizer}
				
				$i^\ast \leftarrow \operatorname{argmax}_{i} \big(\nabla \mathcal{D}_{ts}(\mathbf{X}^\ast;\boldsymbol{x}^\circ_t)\big)_{i}$;
				\Comment*{effectness measurement}
				
				\For(\Comment*[f]{replacing ineffective dimensions}){$i = 1$ to $N$}{
					\lIf{$i \neq i^\ast$}{
						$(\mathbf{X}^\ast)_i \leftarrow (\mathbf{X}^\circ)_i$
					}
				}
				
				Discretize $\mathbf{X}^\ast$ into one-hot vectors with only $0$ or $1$;
				
				$t \leftarrow t+1$;
			}
	}

	\Return{$\mathbf{X}^\circ-\mathbf{X}^\ast$}
	
\end{algorithm}

\setlength{\textfloatsep}{.5ex}
\begin{algorithm}[!h]
	\small
	\setstretch{0.9}
	\SetKwComment{Comment}{$\triangleright$\ }{}
	\SetKwComment{Commentld}{$\triangledown$\ }{} 
	\SetCommentSty{em}
	\SetKwFunction{Adam}{Adam}
	\SetKwFunction{POP}{POP}
	\SetKwFunction{PUSH}{PUSH}
	
	\caption{Interpreting graph link anomalies (unattributed)}
	\label{alg3}
	\KwIn{Link anomaly $\mathcal{X}^\circ=(x_a^\circ,x_b^\circ)$; $max\_iter$, learning rate $\alpha$, vertex set $\mathcal{V}$}
	\KwOut{The reference $\mathcal{X}^\ast=(x_a^\ast,x_b^\ast)$ for interpretation}
	
	$\boldsymbol{e}^\circ \leftarrow \mathsf{E}_{G}(\mathcal{X}^\circ)$;
	
	Solve $\boldsymbol{e}^\ast$ through Algorithm (\ref{alg1}) with the input $\boldsymbol{e}^\circ$;
	
	\eIf{$\mathsf{E}_{G}$ is differentiable}{
		
		$\mathcal{X}^\ast \leftarrow$\Adam{$\mathcal{X}^\ast;\mathcal{D}_{gra}(\mathcal{X}^\ast;\boldsymbol{e}^\ast),\alpha$} \Comment*{gradient-based method}
		
		Discretize $\mathcal{X}^\ast$ into two one-hot vectors with only $0$ or $1$;
		
	}(\Comment*[f]{greedy search}){  
		
		\For(\Comment*[f]{Initialize $\mathsf{PQ}$}){each $x \in \{x_a^\circ,x_b^\circ\}$}
		{
			
			\For{each $y \in (\mathcal{N}_G(x)\cap\mathcal{V})$}
			{
				$\mathbf{v}.a \leftarrow x$;\ \
				$\mathbf{v}.b \leftarrow y$;\\
				$\mathbf{v}.w \leftarrow \mathcal{D}_{gra}((\mathbf{v}.a,\mathbf{v}.b);\boldsymbol{e}^\ast)$;\\
				\PUSH{$\mathbf{v}$,$\mathsf{PQ}$};\\
			}
			
			$\mathcal{V} \leftarrow \mathcal{V} - \{x\}$
			\Comment*{delete visited vertex from $\mathcal{V}$}
			
		}
		
		Initialize $\mathcal{X}^\ast$ with $\mathcal{X}^\circ$; $t \leftarrow 0$;

		\While(\Comment*[f]{prioritized BFS}){$t<max\_iter$ and $\mathsf{PQ}\neq\varnothing$}
		{	
			
			$\mathbf{u} \leftarrow$ \POP{$\mathsf{PQ}$};\\
			\uIf{$\mathbf{u}.w < \mathcal{D}_{gra}(\mathcal{X}^\ast;\boldsymbol{e}^\ast)$}{
				
				$\mathcal{X}^\ast \leftarrow (\mathbf{u}.a,\mathbf{u}.b)$
				\Comment*{update reference}
			}
			
			\For{each $y \in (\mathcal{N}_G(\mathbf{u}.b)\cap\mathcal{V})$}
			{
				$\mathbf{v}.a \leftarrow \mathbf{u}.b$;\ \
				$\mathbf{v}.b \leftarrow y$;\\
				$\mathbf{v}.w \leftarrow \mathcal{D}_{gra}((\mathbf{v}.a,\mathbf{v}.b);\boldsymbol{e}^\ast)$;\\
				\PUSH{$\mathbf{v}$,$\mathsf{PQ}$};\\
			}
			
			$\mathcal{V} \leftarrow \mathcal{V} - \{\mathbf{u}.b\}$
			\Comment*{delete visited vertex from $\mathcal{V}$}
			
			$t \leftarrow t+1$;
		}
	}

	\Return{$\mathcal{X}^\ast$}
	
\end{algorithm}

\subsection{Procedure of Graph Data Interpretation}
\label{suppA.2}


Recall in \S \ref{sec4.4}, we claim that problem (\ref{eq12}) cannot be directly solved by gradient-based optimizer if $\mathsf{E}_{G}$ is indifferentiable. To address this problem, we propose an alternative greedy solution as follows:

\noindent \textbf{Greedy Search.} In a nutshell, we start searching two reference nodes $x_a^\ast, x_b^\ast$ from  $x_a^\circ,x_b^\circ$, and gradually search \emph{outward}. In this way, searched reference link will not be too far from the abnormal link, which increases the interpretability. 
Specifically, we denote the objective function in (\ref{eq12}) with  $\mathcal{D}_{gra}(\mathcal{X}^\ast;\boldsymbol{e}^\ast)$, which is used to measure the \emph{priority} of link $\mathcal{X}^\ast$.
Then, the greedy search is conducted by breadth-first search (BFS) with a \emph{priority queue} denoted with $\mathsf{PQ}$.
Let $\mathcal{N}_G(x)$ represents the set of neighbor nodes of $x$.
Initially, push links consisting of $x_a^\circ,x_b^\circ$ and their neighbors (formally, $\{(x,y)|x\in\{x_a^\circ,x_b^\circ\}, y\in\mathcal{N}_G(x)\}$) into the queue $\mathsf{PQ}$. By measuring the priority with $\mathcal{D}_{gra}$, the node popped from $\mathsf{PQ}$ and its (unvisited) neighbor nodes are then pushed into $\mathsf{PQ}$. After a fixed number of iterations, the current optimal link will be used as the reference. 
The whole procedure of graph Interpreter is in Algorithm \ref{alg3}.



\section{Theoretical Analysis of Distiller}
\label{app2}
We provide theoretical analysis about the correctness of Distiller (in \ref{app2.1}) and its complexity (in \ref{app2.2}). 

\subsection{Theoretical Proofs}
\label{app2.1}

\noindent \textbf{Preliminaries (rule, query, result).}
Recall that Distiller introduced in \S \ref{sec5} consists of two FSMs, where the first one is used to store representative interpretation vectors and the second one is used to store feedback by experts.
In update mode of Distiller, we call an interpretation vector containing $K$ dimensions together with a \emph{feedback} state as a ``\emph{rule}''. In test mode of Distiller, we call the given interpretation vector a ``\emph{query}''. The feedback state in the second FSM with the highest probability transferred from this query is called the ``\emph{result}'' of this query, denoted with $r_q$.
Interpretation vectors in \emph{rule} or \emph{query} $\boldsymbol{x}^\circ-\boldsymbol{x}^\ast$ need to firstly be mapped into states $\hat{S}=\hat{s}_1\hat{s}_{2}...\hat{s}_K$ in the first FSM. Then, for query $\hat{S}$, the result is derived by 
$r_q = \operatorname{argmax}_{\hat{r}} \mathcal{P}\left(\hat{r} \mid \hat{S}\right)$ where $\mathcal{P}$ is computed by (\ref{eq14}).

We first provide a theorem of the correctness of Distiller (also the correctness of (\ref{eq14})) as follows:

\noindent \textbf{Theorem 1 (Correctness of Distiller)}. The queried probability is always equal to the probability expectation of all states transferring to the query result according to the queried states. Formally, as (\ref{eq14}):
\begin{equation*}
\mathcal{P}\left(\hat{r} \mid \hat{s}_{1} \hat{s}_{2} ... \hat{s}_{K}\right)  =  \frac{1}{K} \sum_{i=1}^{K}\big(\mathcal{P}_2(\hat{r} \mid \hat{s}_{i}) \prod_{j=1}^{i-1}\mathcal{P}_1(\hat{s}_{j+1} \mid \hat{s}_{j})\big).
\end{equation*}

\noindent \textbf{\emph{Proof} of Theorem 1}. Obviously, $\mathcal{P}_1$ and $\mathcal{P}_2$ are two Markov process, or we say they have Markov property since probability distribution of next states of the process  $\mathcal{P}_1$ or $\mathcal{P}_2$ depends only upon the present state. Meanwhile, $\mathcal{P}_1$ and $\mathcal{P}_2$ are statistically independent since they are updated separately. $\mathcal{P}_1$ captures the probability of $\hat{s}_{i}$ transferring to $\hat{s}_{K}$, while $\mathcal{P}_2$ captures the probability of $\hat{s}_{i}$ transferring to $\hat{r}$.
Thus, we multiply $\mathcal{P}_1$ and $\mathcal{P}_2$ to compute the probability of co-occurrence of the two processes (due to the independence of $\mathcal{P}_1$ and $\mathcal{P}_2$). 
Finally, we calculate the mean for each state in $\hat{S}$ as the expectation of co-occurrence of the two processes. \hfill $\blacksquare$

Next, we further introduce some definitions of terms for sake of analyzing the accuracy bound of Distiller.
	
\noindent \textbf{Definition 1 (State conflict).}
We say that \emph{\textbf{state conflict}} happens when 
two rules indicating different feedback states have the same state(s) in $\hat{S}$. We say that the number of state conflicts in two rules is $i$ if there are $i$ identical states between two rules. For example, there are 2 state conflicts in $\hat{s}_1\hat{s}_2\hat{s}_3\rightarrow \hat{r}_1$ and $\hat{s}_2\hat{s}_4\hat{s}_1\rightarrow \hat{r}_2$, namely $\hat{s}_1$ and $\hat{s}_2$. Note that conflicts happen only when $\hat{r}_1 \neq \hat{r}_2$.

\noindent \textbf{Definition 2 (Transition conflict).} Similar to state conflict, we say that \emph{\textbf{transition conflict}} happens when 
two rules indicating different feedback states have the same transition(s) in $\hat{S}$. For example, there are 2 state conflicts in $\hat{s}_1\hat{s}_2\hat{s}_3\hat{s}_4\rightarrow \hat{r}_1$ and $\hat{s}_3\hat{s}_4\hat{s}_1\hat{s}_2\rightarrow \hat{r}_2$, namely $\hat{s}_1\hat{s}_2$ and $\hat{s}_3\hat{s}_4$. 

\noindent \textbf{Definition 3 (Complete conflict).} We say that \emph{\textbf{complete conflict}} happens when 
there are $K$ state conflicts and $K-1$ transition conflicts between two rules ($K$ is the number of states in $\hat{S}$).
For example, complete conflict happens in  $\hat{s}_1\hat{s}_2\hat{s}_3\hat{s}_4\rightarrow \hat{r}_1$ and $\hat{s}_1\hat{s}_2\hat{s}_3\hat{s}_4\rightarrow \hat{r}_2$, namely only feedback states are different in two rules. 

\noindent \textbf{Definition 4 (Query accuracy).}
Given a Distiller in test mode and a validation query set with true-label feedback states, \emph{\textbf{query accuracy}} can be computed as the ratio of the query result that equals the true label.

Intuitively, the conflicts, especially complete conflicts between rules are the root cause of the decrease in query accuracy. Next, we analyze the relationship between conflicts and query accuracy. 
Here we use $R$ and $F$ to respectively represent the set of all rules and feedback states. Therefore, the number of different rules and feedback are $|R|$ and $|F|$.
Obviously, more rules represented in one Distiller will more likely to induce (more) conflicts. Below, we analyze the relationship between the number of rules, conflicts and query accuracy.
 
\noindent \textbf{Lemma 1.} Suppose that rules are randomly sampled (with uniform probability), the expectation of the number of different rules represented in one state is $\frac{K|R|}{MN}$, where $M$ is the number of feature intervals in Distiller, and $N$ is the total number of dimensions in tabular data, and $K$ is the number of used dimensions in 
the interpretation vector. 

\noindent \textbf{\emph{Proof} of Lemma 1}.
Recall that there are $MN$ states totally in the first FSM.
Because one rule will sample $K$ states, and each state is sampled with equal probability, the probability of each state being selected by a rule is $\frac{K}{MN}$. Suppose that different rules are sampled independent, then for $|R|$ rules, we get the expectation of the number of different rules represented in one state $\frac{K|R|}{MN}$.
\hfill $\blacksquare$

\noindent \textbf{Lemma 2.} Suppose that rules are randomly sampled (with uniform probability), the expectation of the number of states transferring to a certain feedback state in the second FSM is $\frac{K|R|}{MN|F|}$, where $|F|$ is the number of feedback states in the second FSM.

\noindent \textbf{\emph{Proof} of Lemma 2}.
Because rules are randomly sampled with uniform probability, the probability of a certain state to different feedback states is also equal.
Then based on Lemma 1, the $\frac{K|R|}{MN}$ rules that represented by a certain state have equal probability transferring to $|F|$ feedback states, thus we get $\frac{K|R|}{MN|F|}$.
\hfill $\blacksquare$

\noindent \textbf{Theorem 2 (Query accuracy bound)}. The expectation of query accuracy rate is 100\% if the number of rules $|R|$ represented by Distiller satisfies that $|R|<\frac{MN|F|}{K}$.



\begin{figure}[!h]
	\vspace{-.5em}
	\includegraphics[width=0.4\linewidth]{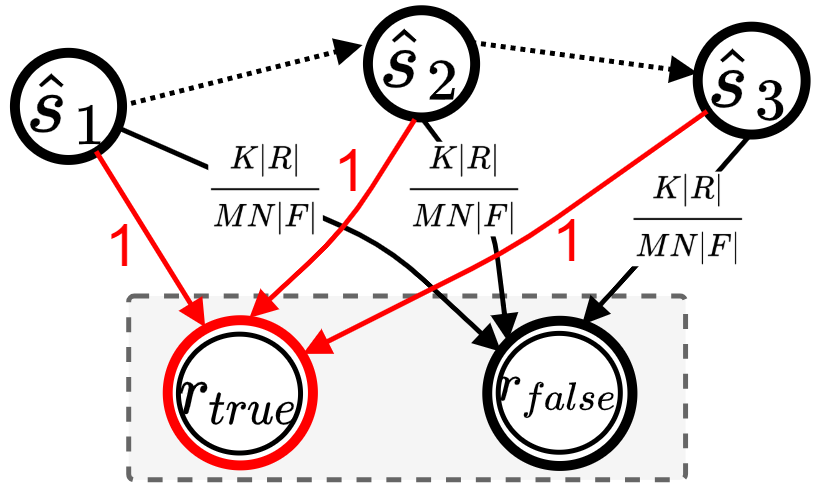}
	\vspace{-1em}
	\caption{Illustration of Proof of Theorem 2.}
	\label{fig_supp1}
	\vspace{-1em}
\end{figure}

\noindent \textbf{\emph{Proof} of Theorem 2}. If $|R|<\frac{MN|F|}{K}$, then $\frac{K|R|}{MN|F|}<1$. Then according to Lemma 2, this means that the expectation of the number of a certain state transferring to a feedback state is less than 1. For a given rule $\hat{s}_1\hat{s}_{2}...\hat{s}_K \rightarrow \hat{r}_{true}$, we consider the worst case that only states $\hat{s}_1\hat{s}_{2}...\hat{s}_K$ in this rule transfer to $\hat{r}_{true}$. This is to say, the number of rules where $\hat{s}_i(i \in \{1,2,...,K\})$  transferring to $\hat{r}_{true}$ is 1. According to Lemma 2, the number of rules where $\hat{s}_i(i \in \{1,2,...,K\})$ transferring to other feedback denoted with $\hat{r}_{false}$ are expected to be $\frac{K|R|}{MN|F|}<1$. In this case, if we have a query $\hat{s}_1\hat{s}_{2}...\hat{s}_K$, we will have the result $\hat{r}_{true}$, according to the probability  transition equation proven in Theorem 1. because:
\vspace{-.5em}
\begin{align}
	\mathcal{P}\left(\hat{r}_{true} \mid \hat{s}_{1} \hat{s}_{2} ... \hat{s}_{K}\right) & =  \frac{1}{K} \sum_{i=1}^{K}\big(\frac{1}{1+\varkappa(|F|-1)} \prod_{j=1}^{i-1}\mathcal{P}_1(\hat{s}_{j+1} \mid \hat{s}_{j})\big) \notag \\  
	& > \frac{1}{K} \sum_{i=1}^{K}\big(\frac{\varkappa}{1+\varkappa(|F|-1)} \prod_{j=1}^{i-1}\mathcal{P}_1(\hat{s}_{j+1} \mid \hat{s}_{j})\big) \notag \\
	& = \mathcal{P}\left(\hat{r}_{false} \mid \hat{s}_{1} \hat{s}_{2} ... \hat{s}_{K}\right),
\end{align}
where $\varkappa=\frac{K|R|}{MN|F|}<1$. We also depict Figure \ref{fig_supp1} for ease of understanding. Since for any query $\hat{S}$, we have $\mathcal{P}\left(\hat{r}_{true} \mid \hat{S}\right)>\mathcal{P}\left(\hat{r}_{false} \mid \hat{S}\right)$, this confirms the expectation of query accuracy rate is 100\% (Note that this is under the worst case). 
\hfill $\blacksquare$

\noindent \textbf{Corollary 1}. Distiller can store $\frac{MN}{K}$ different rules (in the first FSM) at most for the same feedback state (in the second FSM), to ensure that the expectation of query accuracy rate is 100\%.

\noindent \textbf{\emph{Proof} of Corollary 1}.
According to Theorem 2, the expectation of query accuracy rate is 100\% if $|R|<\frac{MN|F|}{K}$, then $\frac{|R|}{|F|}<\frac{MN}{K}$.
\hfill $\blacksquare$

\noindent \textbf{Corollary 2}. The expectation number of complete conflicts is less than 1 if $|R|<\frac{MN|F|}{K}$.

\noindent \textbf{\emph{Proof} of Corollary 2}. 
Assume for the purpose of contradiction that there has complete conflict (namely the number is equal or more than 1) when $|R|<\frac{MN|F|}{K}$, according to Theorem 2, query results related to the conflict will be wrong in the 
worst case. This indicates the query accuracy is not 100\%, which contradicts Theorem 2. Thus, we 
conclude the expectation of the number of complete conflicts is less than 1 if $|R|<\frac{MN|F|}{K}$.
\hfill $\blacksquare$ 

\subsection{Complexity Analysis}
\label{app2.2}
Distiller supports online fashion and can be updated in linear time since only related states are updated. In test mode, the time complexity of Distiller is also linear according to (\ref{eq14}).
As for the space complexity, the primary cost is two FSMs maintained in Distiller. We use 
\emph{probability transition matrix} to model two FSMs, which also referred to as stochastic matrix is commonly used to describe the transitions of a Markov process. Since it is a square matrix, and there are $MN$ states totally, the space complexity of Distiller is $\mathcal{O}{((MN)^2)}$. 
This shows that Distiller can be deployed in security systems 
with a low memory overhead with thousand-scale $MN$.


\section{Detailed Experimental Settings}
\label{app3}
Below, we introduce details of security systems and settings in our experiments which are omitted in the main body of our paper.

\subsection{Datasets}
For sake of fairness and reproducibility, we primarily evaluate 
anomaly detection based security systems with datasets released in their original work. The information of used datasets are listed in Table \ref{tab6}. For \texttt{Kitsune}, we use the
dataset collected from a IoT (Internet-of-Things) network in its work.
We additionally use another open dataset for network intrusion detection called CIC-IDS2017 \cite{sharafaldin_toward_2018}, in order to collect various types of attack traffic for the evaluation of reliable detection by Distiller in \S \ref{sec6.5}. CIC-IDS2017 collected traffic for various common attacks in a large-scale testbed, which covers many common devices and middleboxes in the network. 
Anomalies indicate various types of network attacks.
For \texttt{DeepLog}, we use HDFS dataset \cite{xu2009detecting} which is the same dataset evaluated in its work. 
This dataset consists of key number sequence of Hadoop file system logs, and anomalies indicate malfunction or malicious behavior in the file system.
For \texttt{GLGV}, we also use a dataset evaluated in its
original work called LANL-CMSCSE\cite{akent-2015-enterprise-data}, which collects tons of multi-source events of a real internal computer network in 58 consecutive days. Here we mainly use authentication events collected from individual devices. Anomalies here indicate lateral movement in the APT campaign. 

\begin{table}[!h]
	\vspace{-1ex}
	\caption{Datasets used in this study.}
	\vspace{-2mm}	
	\label{tab6}
	\resizebox{\linewidth}{!}{%
		\begin{tabular}{cccc}
			\Xhline{2\arrayrulewidth}
			\textbf{Security Systems} & \textbf{Datasets} & \textbf{\# Samples} & \textbf{\# Selected Anomaly} \\	
			\Xhline{2\arrayrulewidth}
			\multirow{10}{*}{\begin{tabular}[c]{@{}c@{}}\texttt{Kitsune}\cite{mirsky2018kitsune}\\ (Tabular Data\\ based System)\end{tabular}} & Kitsune-Mirai\cite{mirsky2018kitsune} & 764,137 & 78,739 \\
			& Kitsune-Fuzzing\cite{mirsky2018kitsune} & 2,244,139 & 15,000 \\
			& Kitsune-SSDP Flood\cite{mirsky2018kitsune} & 4,077,266 & 15,000 \\
			& Kitsune-ARP MitM\cite{mirsky2018kitsune} & 2,504,267 & 15,000 \\
			& Kitsune-SYN DoS\cite{mirsky2018kitsune} & 4,077,266 & 15,000 \\ \cline{2-4} 
			& IDS17-DDoS\cite{sharafaldin_toward_2018} & \multirow{6}{*}{44,547,764} & 335,181 \\
			& IDS17-DoS Hulk\cite{sharafaldin_toward_2018} &  & 10,000 \\
			& IDS17-Infiltration\cite{sharafaldin_toward_2018} &  & 10,000 \\
			& IDS17-PortScan\cite{sharafaldin_toward_2018} &  & 10,000 \\
			& IDS17-Botnet\cite{sharafaldin_toward_2018} &  & 10,000 \\
			& IDS17-Slowhttptest\cite{sharafaldin_toward_2018} &  & 5,000 \\ \hline
			\begin{tabular}[c]{@{}c@{}}\texttt{DeepLog}\cite{du2017deeplog}\\ (Time Series\\ based System)\end{tabular} & HDFS\cite{xu2009detecting,du2017deeplog} & 553,366 & 16,838 \\ \hline
			\begin{tabular}[c]{@{}c@{}}\texttt{GLGV}\cite{bowman2020detecting} \\ (Graph Data\\ based System)\end{tabular} & LANL-CMSCSE\cite{akent-2015-enterprise-data} & 1,051,430,459 & 747 \\
			\Xhline{2\arrayrulewidth}
		\end{tabular}%
	}
	\vspace{-2ex}
\end{table}
\begin{table}[!h]
	\caption{Detection performance of interpreted systems.}
	\vspace{-2mm}	
	\label{tab7}
	\resizebox{0.6\linewidth}{!}{%
		\begin{tabular}{cccc}
			\Xhline{2\arrayrulewidth}
			\textbf{Systems} & \textbf{Datasets} & \textbf{TPR} & \textbf{FPR} \\ \Xhline{2\arrayrulewidth}
			\multirow{2}{*}{\texttt{Kitsune}$^*$} & Kitsune-Mirai & 0.8549 & 0.0127 \\ \cline{2-4} 
			& IDS17-DDoS & 0.9120 & 0.0012 \\ \hline
			\texttt{DeepLog} & HDFS & 0.9144 & 0.0124 \\ \hline
			\texttt{GLGV} & LANL-CMSCSE & 0.8099 & 0.0005 \\ \Xhline{2\arrayrulewidth}
		\end{tabular}%
	}
	
	\begin{flushleft}
		\setstretch{0.6}
		\small
		\noindent $*$ \emph{Original implementation instead of the modified version in \S \ref{sec6.4}.}
	\end{flushleft} 
	
\end{table}

\subsection{Implementations}
\noindent \textbf{Security Systems.}
For \texttt{Kitsune}, we primarily use its open-source implementation\footnotemark[1] (in \S \ref{sec6.4}, we modified this implementation to improve the system). 
By default, we use settings claimed in their paper and code. We use 50,000 normal samples for training its unsupervised DL models. 
For \texttt{DeepLog}, we primarily use an open-source implementation\footnotemark[2] and achieve similar performance to the original work. 
For \texttt{GLGV}, we implement it according to the details provided in the paper, and achieved a similar performance to the original paper (TP is slightly lower, but FP is also lower).
The results of original performance of security systems are shown in Table \ref{tab7}.
 
\footnotetext[1]{\texttt{Kitsune}:  \url{https://github.com/ymirsky/Kitsune-py}}

\footnotetext[2]{\texttt{DeepLog}: \url{https://github.com/wuyifan18/DeepLog}}

\noindent \textbf{Baseline Interpreters.} Similarly, we primarily use open-source implementations \footnotemark[3]\footnotemark[4]\footnotemark[5]\footnotemark[6]\footnotemark[7] and default parameters and settings for several baseline interpreters.
For supervised baselines (\texttt{LIME}, \texttt{LEMNA}, and \texttt{DeepLIFT}), we need to first convert the unsupervised anomaly detection models to supervised learning models since the unsupervised DNNs used in security systems are threshold-based but not parametric.
Thus, we use supervised DNNs trained with 5,000 normal data and additionally sampled 5,000 anomalies to approximate the detection boundary, and then evaluate the supervised interpreters on approximated models. 
For \texttt{DeepLIFT}, we randomly select normal training data as the ``reference points'' \cite{shrikumar2017learning}.
\textcolor{black}{For \texttt{CADE}, since the interpreter is originally designed for explaining concept drift samples, we simulate interpreted anomalies in Kitsune Dataset as concept drift samples by training a supervised classifier with 5,000 normal data and 5,000 malicious data from CIC-IDS2017 Dataset. 
After training the concept drift detector (essentially Contrastive Autoencoder \cite{yang2021cade}) with two-class data, we interpret anomalies in Kitsune Dataset with \texttt{CADE} interpreter.}

\footnotetext[3]{\texttt{LIME}: \url{https://github.com/marcotcr/lime}}

\footnotetext[4]{\texttt{LEMNA}: \url{https://github.com/nitishabharathi/LEMNA}(DNN),  \url{https://github.com/Henrygwb/Explaining-DL}(RNN)}

\footnotetext[5]{\texttt{COIN}: \url{https://github.com/ninghaohello/Contextual-Outlier-Interpreter}}

\footnotetext[6]{\texttt{DeepLIFT}: \url{https://github.com/pytorch/captum}}

\footnotetext[7]{\texttt{CADE}: \url{https://github.com/whyisyoung/CADE}}

\subsection{Experimental Settings}
Below, we provide supplementary statements on the settings of experiments in our study.

\noindent \textbf{Hardware.} Our experiments are evaluated on a Dell PowerEdge R720 server using Ubuntu 14.04 LTS with 6 cores Intel Xeon(R) CPU E5-2630 @ 2.60GHz and 94G RAM. All deep learning models are using GPU mode with NVIDIA GeForce GTX 1080ti.

\noindent \textbf{Software.} The software implementation of \textsf{DeepAID} is primarily based on Python 3 with several packages, including scikit-learn for machine learning models and pytorch for deep learning models.


\noindent \textbf{Settings of \S \ref{sec6.2}.} Here we evaluate \texttt{Kitsune} with Kitsune-Mirai dataset. We use the first 50,000 normal data for training and use 10,000 anomalies for evaluating the performance of interpreters. For \texttt{DeepLog}, we use ~5,000 normal data for training and use 5,000 anomalies for evaluating the performance of interpreters. For \texttt{GLGV}, we use 50M normal event logs for training and use 747 anomalies for evaluating Interpreter. 
Note that in the \emph{efficiency} evaluation, we select 2,000 anomalies for tabular and time-series data and 500 anomalies for graph data, and record the runtime of interpreters.

\noindent \textbf{Settings of \S \ref{sec6.3}.} Note that the values in Table \ref{tab3} are de-normalized since only the original value of the feature is interpretable. For ease of illustration, these values have been approximated.

\noindent \textbf{Settings of \S \ref{sec6.4}.} Here we evaluate \texttt{Kitsune} with Kitsune-Mirai dataset. And used the same method (with the first 50,000 normal data) for training. Then we use all 78,739 anomalies and 10,000 normal data for evaluating and comparing the performance before and after fixing the problem of feature extraction, which is discovered with the help of \textsf{DeepAID} interpretations.

\noindent \textbf{Settings of \S \ref{sec6.5}.} Here  \texttt{Kitsune} is trained with the same method again. 
In the evaluation of 5 classes, we use all five Kitsune datasets, while we together use the first five CIC-IDS2017 datasets for the evaluation of 10 classes. The ratio of training set to test set is 1:9. Namely, we have totally 9,000 and 18,000 test data for 5 classes and 10 classes.
This ratio is reasonable, because in practice it is impossible for the operators to update the rules to the Distiller frequently, indicating that the ratio of training data should be relatively small. For \textbf{RQ3}, we choose the last class of CIC-IDS2017 dataset as the ``unknown'' attack. 

\noindent \textbf{Settings of \S \ref{sec6.6}.} Here we have two parts of experiments. The first part (i.e., modifying Distiller) uses the same settings as \S \ref{sec6.5}. 
In the second part (i.e., retraining the DL model), we
train \texttt{Kitsune} with normal traffic from IDS2017, and evaluate it with Kitsune-Mirai traffic. 
Since these two datasets are collected from different testbeds, concept drift may occur during testing.
Indeed, we find there are totally >2,000 FPs when using the modified version of \texttt{Kitsune}. In contrast to 79 FPs (recall in Figure \ref{fig7}), we confirm the occurrence of concept drift. In Figure \ref{fig8}, we choose 1,000 FPs to put into Distiller for concept drift detection, and use another 1,000 FPs for testing.

  
\subsection{Evaluation Metrics}
We introduce the definition of metrics for the evaluation of Distiller omitted in the main body, including f1-micro/f1-macro/UACC in \S \ref{sec6.5} and TPR/FPR in \S \ref{sec6.6}.
The detection performance of binary classification tasks can be measured by firstly counting its true positives (\emph{TP}), true negatives (\emph{TN}), false positives (\emph{FP}), and false negatives (\emph{FN}). In this context, we can define the evaluation metrics:
\begin{equation*}
\begin{aligned}
	Precision = \frac{TP}{TP+FP}, \ & \ \textbf{TPR} = Recall = \frac{TP}{TP+FN}, \\
	\textbf{FPR} = \frac{FP}{FP+TN}, \ & \ \textbf{ACC} = \frac{TP+TN}{TP+FN+TN+FP},
\end{aligned}
\end{equation*}
\begin{equation*}\textbf{F1-Score} =  \frac{2 \times Precision \times Recall}{Precision+Recall}. \end{equation*}

UACC is just ACC of treating the unknown class as positive. As for f1-micro/f1-macro, they are f1-score for multi-class evaluation.
Specifically, f1-micro calculates metrics globally by counting the total TP, FN, and FP, while f1-macro calculates metrics for each label, and finds their unweighted mean. 
 
\section{Adversarial Robustness of Interpreter}
\label{app6}
\textcolor{black}{Below, we provide details of adaptive attacks for adversarial robustness evaluation of the Interpreter in \S \ref{sec6.2}.
We show the strong adversarial robustness of \textsf{DeepAID} and analyze the reasons.
As mentioned in \S \ref{sec6.2}, we evaluate two types of adversarial attacks, including optimization-based attack and distance-based attacks}. 

\subsection{Robustness against optimization-based attack}
\label{appD.1}
\noindent \textbf{Attack Methodology.}
\textcolor{black}{As mentioned in \S \ref{sec6.2}, we borrow the ideas from existing adversarial attacks against B.P. based interpreters \cite{ghorbani2019interpretation, zhang2020interpretable} to develop the following adaptive attack on \textsf{DeepAID}:}
\begin{equation}
\textcolor{black}{{\rm argmax}_{\boldsymbol{\tilde{x}}^\circ} \|\mathcal{D}_{tab}(\boldsymbol{\tilde{x}}^\circ;\boldsymbol{\tilde{x}}^\circ) - \mathcal{D}_{tab}(\boldsymbol{x}^\circ;\boldsymbol{x}^\circ)\|_{p} \ \  {\rm s.t.} \ \|\boldsymbol{\tilde{x}}^\circ-\boldsymbol{x}^\circ\|_{p} < \delta_{a},}
\label{eqa1}
\end{equation}
\textcolor{black}{where $\mathcal{D}_{tab}$ is the objective function of optimization (\ref{eq7}).
As mentioned in \S \ref{sec4.2}, our Interpreter iteratively optimizes (\ref{eq7}) to find a better reference. However, such iterative optimization is hard for attackers to gain the overall gradients. 
From the perspective of attackers, we simplify Interpreter into solving (\ref{eq7}) only once. Hence, the optimization for interpretation is transformed into single-step gradient descent for $\mathcal{D}_{tab}$ with the input of anomaly $\boldsymbol{x}^\circ$. Formally, $\boldsymbol{x}^\ast
\approx \boldsymbol{x}^\circ -\alpha\nabla_{\boldsymbol{x}^\circ } \mathcal{D}_{tab} (\boldsymbol{x}^\circ;\boldsymbol{x}^\circ)$. Then, the interpretation result is  $\boldsymbol{x}^\circ - \boldsymbol{x}^\ast \approx \alpha\nabla_{\boldsymbol{x}^\circ } \mathcal{D}_{tab} (\boldsymbol{x}^\circ;\boldsymbol{x}^\circ)$. In this context, the objective function of optimization-based attack in (\ref{eqa1}) is transformed into:}
\begin{equation}
\textcolor{black}{{\rm argmax}_{\boldsymbol{\tilde{x}}^\circ} \|\alpha\nabla_{\boldsymbol{\tilde{x}}^\circ} \mathcal{D}_{tab} (\boldsymbol{\tilde{x}}^\circ;\boldsymbol{\tilde{x}}^\circ)
- \alpha\nabla_{\boldsymbol{x}^\circ } \mathcal{D}_{tab} (\boldsymbol{x}^\circ;\boldsymbol{x}^\circ) \|_{p}.}
\label{eqa2}
\end{equation}
\textcolor{black}{Here we set $p=2$ for simplicity ($p=1,2$ in \cite{ghorbani2019interpretation}). This problem can be easily solved by gradient-based methods. Like \cite{ghorbani2019interpretation}, to limit $\boldsymbol{\tilde{x}}^\circ$ and $\boldsymbol{x}^\circ$ to be close, we only modify top-K dimensions of $\boldsymbol{\tilde{x}}^\circ$ in each iteration and clip $\boldsymbol{\tilde{x}}^\circ$ to make $\|\boldsymbol{\tilde{x}}^\circ-\boldsymbol{x}^\circ\|_{2} < \delta_{a}$.} 

\begin{figure}[]
	\centering
	\begin{subfigure}{\linewidth}
		\centering
		\includegraphics[width=0.5\linewidth]{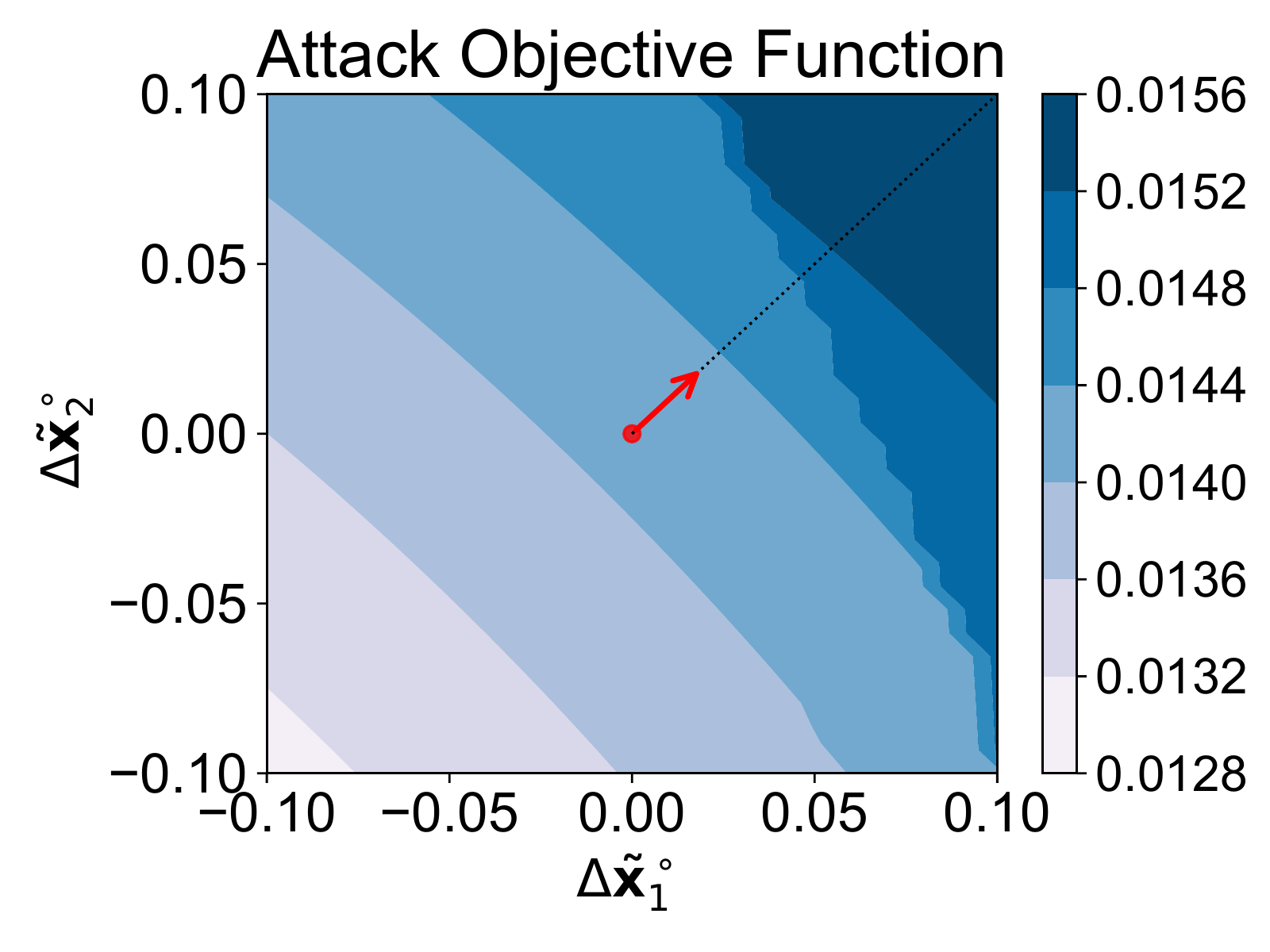}
		\vspace{-1em}
		\caption{Attack effect on attacker's objective function.}
		\label{fig11a}
	\end{subfigure} %
	\vspace{2em}
	\begin{subfigure}{\linewidth}
		\includegraphics[width=\linewidth]{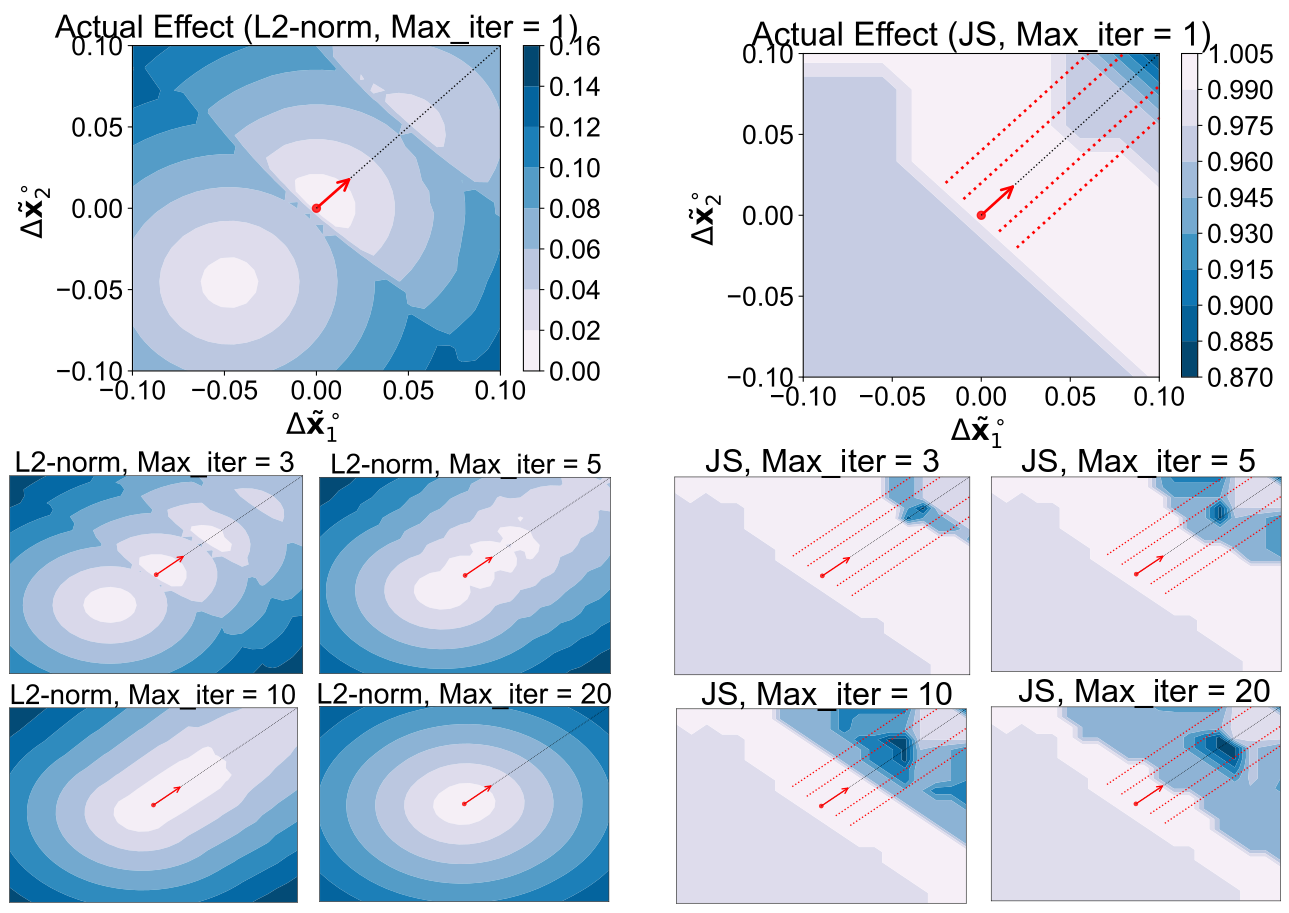}
		\vspace{-1.5em}
		\caption{Attack effect on actual Interpreter's results.}
		\label{fig11b}
	\end{subfigure}
	
	\vspace{-3em}
	\caption{\textcolor{black}{Example of demonstrating robustness against optimization-based attack.}}
	
	\label{fig11}
\end{figure}

\noindent \textbf{Robustness Evaluation and Analysis.} \textcolor{black}{The results of \textsf{DeepAID} Interpreter against optimization-based attack have been shown in the main body (Figure \ref{fig9} in \S \ref{sec6.2}). We conclude that \textsf{DeepAID} Interpreter is naturally robust against such attacks and becomes more robust with I.R.N.. 
In \S \ref{sec4.2}, we have analyzed the reason why \textsf{DeepAID} is naturally robust is due to iterative optimization and search-based idea. And I.R.N. can further improve the robustness by mitigating attack effect on the searching entrance. Specifically, with I.R.N., attackers can only use the perturbation solved from $\boldsymbol{x}^\circ$ to attack the searching process with the \emph{actual} entrance $\boldsymbol{x}^\ast_{(1)} = \boldsymbol{x}^\circ+\mathcal{N}(0,\sigma_{n}^2)$. However, there is a gap between the effect of perturbation over $\boldsymbol{x}^\circ$ and actual attack effect on $\boldsymbol{x}^\ast_{(1)}$. We demonstrate the gap with an intuitive example in Figure \ref{fig11}.}

\textcolor{black}{Here we select an anomaly $\boldsymbol{x}^\circ$ and depict the influence of perturbations in two most important dimensions of $\boldsymbol{\tilde{x}}^\circ$ on the attacker's objective function (\ref{eqa2}) and the actual attack effect (i.e., the interpretation result).
In Figure \ref{fig11a}, attackers choose the direction of perturbation according to the gradient of objective function (\ref{eqa2}), which is the red arrow in the figure.
In Figure \ref{fig11b}, we evaluate two metrics for actual attack effect: the first one is the L2-distance between reference searching from $\boldsymbol{\tilde{x}}^\circ$ and $\boldsymbol{x}^\circ$ (left side of the figure), and the second one is \textbf{JS} between the two references (right side of figure). We evaluate the actual attack effect under different iterations ($max_{iter}$) of Interpreter.
First, the results of L2-distance demonstrate that our Interpreter is naturally robust against such optimization-based attack since attackers cannot accurately simulate the overall gradients in our iterative optimization. As $max_{iter}$ increases, the perturbation effect tends to be smooth, and the attack effect completely disappears when $max_{iter}=20$.
Second, the results of \textbf{JS} demonstrate the large gap between attacker's optimization function and actual effect. The area that improves the actual attack effect can be a very small and transient area in the gradient direction of the attacker's optimizer (See the results of JS when $max_{iter}=3,5$).
In this context, I.R.N. with a small noise on the searching entrance can eliminate the attack effect (see red dotted lines).
Third, we find that the value in the neighborhood of $\boldsymbol{x}^\circ$ usually remains the same (i.e., light-colored near the origin in Figure \ref{fig11b}), which demonstrates that I.R.N. with a small $\sigma_{N}$ has little impact on the original stability.} 

\subsection{Robustness against distance-based attacks}
\label{appD.2}
\textcolor{black}{\noindent \textbf{Attack Methodology.} 
As mentioned in \S \ref{sec6.2}, we evaluate another type of attacks called distance-based attacks specifically designed against the \textsf{DeepAID} by mislead the distance calculations in \textsf{DeepAID}.
The high-level idea of such attacks is to trick the Interpreter to choose \emph{irrelevant} features (i.e., zero dimensions in interpretation $\boldsymbol{x}^\circ-\boldsymbol{x}^\ast$) by subtly perturbing them.
Since there are two distance metrics in our Interpreter ($L_0$ and $L_2$, see Eq.(\ref{eq3})), we evaluate two distance-based attacks: $L_0$ attack adds small perturbations on groups of \emph{irrelevant} features to trick the $L_0$ distance into selecting them, while $L_2$ attack adds a few \emph{irrelevant} features with high values to spoil the computation of the $L_2$.
Specifically, for $L_0$ attack, we choose 50\% irrelevant  
features and add noise sampling from Gaussian $\mathcal{N}(0,\sigma^2)$ with default $\sigma=0.01$. This evaluation is similar to robustness against noise in \S \ref{sec6.2}, while the difference is that $L_0$ attack only perturb 50\% dimensions.
For $L_2$ attack, we randomly choose one irrelevant dimension and change it into  
$A$ times the maximum value in anomaly with default $A=0.8$. 
Note that, the huge change to the selected dimension will
change it from \emph{irrelevant} to \emph{relevant}.
Thus we ignore the selected dimension when calculating JS, and only observe whether it has an impact on \emph{other} irrelevant dimensions.}

\begin{figure}
	\centering
	\begin{subfigure}{\linewidth}
		\centering
		\includegraphics[width=0.9\linewidth]{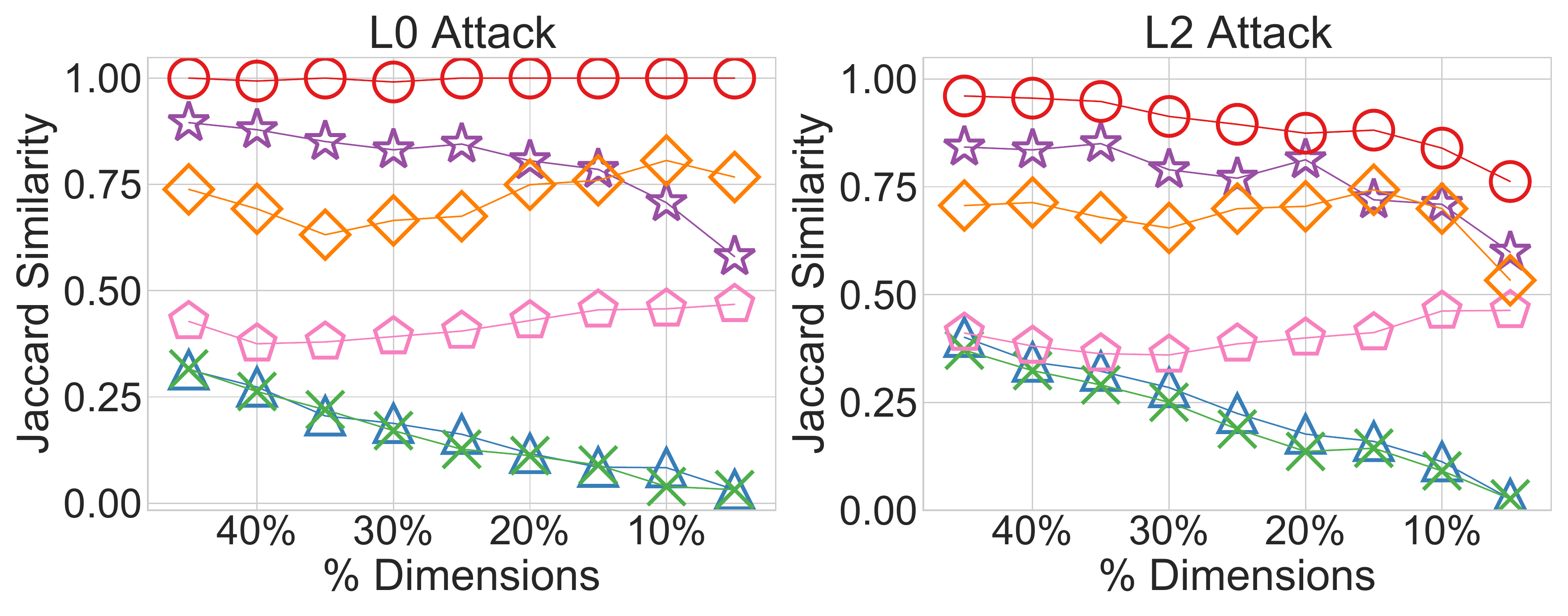}
		\vspace{-1em}
		\caption{Attack scale $\sigma=0.01$, $A=0.8$.}
		\label{fig12a}
	\end{subfigure} %
	\vspace{2.5em}
	\begin{subfigure}{\linewidth}
		\centering
		\includegraphics[width=0.9\linewidth]{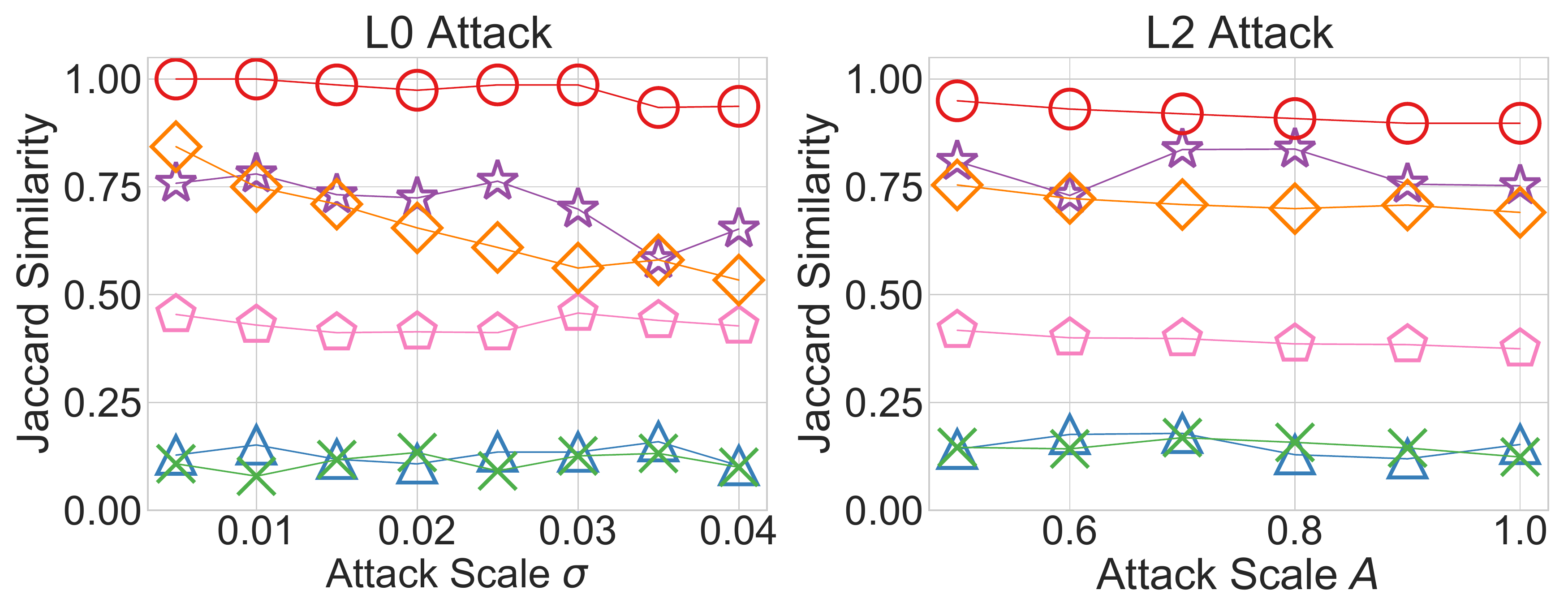}
		\vspace{-.5em}
		\caption{Interp. dimension $K=20$ (20\%).}
		\label{fig12b}
	\end{subfigure}
	
	\begin{subfigure}{\linewidth}
		\centering
		\vspace{-3em}
		\includegraphics[width=\linewidth]{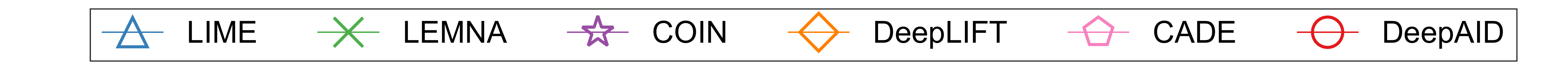}
	\end{subfigure}
	
	\vspace{-2.5em}
	\caption{\textcolor{black}{Robustness evaluation against distance-based attacks (\emph{higher is better}).}}
	\vspace{-1ex}
	\label{fig12}
\end{figure}

\noindent \textbf{Robustness Evaluation and Analysis.} \textcolor{black}{We use the same dataset in \S \ref{sec6.2} and evaluate two distance-based attacks on \textsf{DeepAID} and other baseline Interpreters. We first evaluate with default $\sigma$ or $A$ and also evaluate the impact of two attack scale parameters. The results are shown in Figure \ref{fig12}. We can observe that \textsf{DeepAID} Interpreter is very robust against $L_0$ attack. This is because we use \emph{gradient} $\times$ \emph{input} together instead of only gradients when evaluating the effectiveness of dimensions (as introduced in \S \ref{sec4.2}). This effectively limits the impact of small perturbations on anomalies. As for $L_2$ attack, \textsf{DeepAID} also outperforms other baselines. This is because we only update a small number of features in each iteration to satisfy conciseness (as introduced in \S \ref{sec4.2}). Therefore, \textsf{DeepAID} can update selected features under $L_2$ attack without affecting many other features. This is also why in Figure \ref{fig12a}  \textsf{DeepAID} becomes more robust against $L_2$ attack when the number of dimensions increases. The results in Figure \ref{fig12b} also demonstrate the strong robustness of \textsf{DeepAID}. Particularly, \textsf{DeepAID} is not sensitive to the degree of attack.} 

\begin{figure*}[]
	\includegraphics[width=0.8\textwidth]{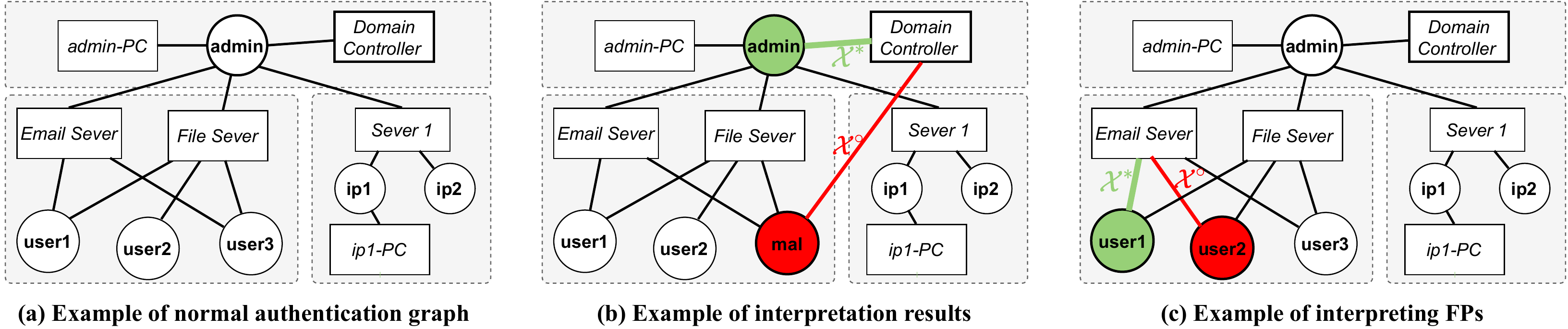}
	\vspace{-1ex}
	\caption{Illustrations of interpreting graph data based system (\texttt{GLGV}).}
	\label{fig_supp2}
	\vspace{-1ex}
\end{figure*}

\section{Graph Data Based Experiments}
\label{app4}
In the main body, we primarily use tabular and time-series based systems for evaluation and illustration due to space limit. 
Another reason is that we have not found any interpreter that can be used as a baseline for unsupervised graph learning tasks (Recall Table \ref{tab2}). Although there are a few methods suitable for graph data such as \cite{meng2020interpreting}, they are difficult to migrate to unsupervised scenarios.
Hence, we only evaluate \textsf{DeepAID} itself for graph data based systems.
Below, we use \texttt{GLGV} to briefly illustrate the adoption of \textsf{DeepAID} on graph data based systems.

\noindent \textbf{Backgrounds.} The backgrounds of \texttt{GLGV} have been introduced in \S \ref{sec2.3}.  \texttt{GLGV} \cite{bowman2020detecting} detects lateral movement in APT campaigns from authentication logs within enterprise networks in an unsupervised (specifically, reconstruction-based) fashion.
It first builds authentication logs into \emph{authentication graph}, where the nodes represent IP/device/users, and the edges represent authentication or access. Figure \ref{fig_supp2}(a) shows an example of a benign authentication graph without any attack (i.e., normal data), which can be divided into three domains According to different permissions or authentication relationships. Only the administrator ``admin'' has the permission to access the Domain Controller (DC).

\noindent \textbf{Settings.} As mentioned in \S \ref{sec4.4},  we implement two versions
of graph data Interpreter according to Algorithm \ref{alg3}, distinguished by whether $\mathsf{E}_G$ is differentiable, denoted with \textsf{DeepAID} and \textsf{DeepAID}$'$. \textsf{DeepAID} means $\mathsf{E}_G$ is differentiable, thus we use gradient-based optimization method in Interpreter, while \textsf{DeepAID}$'$ uses BFS-based method to address the indifferentiable problem. For \textsf{DeepAID}$'$, we use three $max_{iter}$ (Abbreviated as $m$): $m=5,10,20$. $m$ directly reflects how many neighborhoods are viewed in the search process.
We use 10\% normal data with over 10M logs to construct the authentication graph, which consists of 47,319 nodes and 59,105,124 edges.
We use 747 anomalies in the dataset for evaluation.

\begin{figure}
	\centering
	\vspace{-1ex}
	\includegraphics[width=0.5\linewidth]{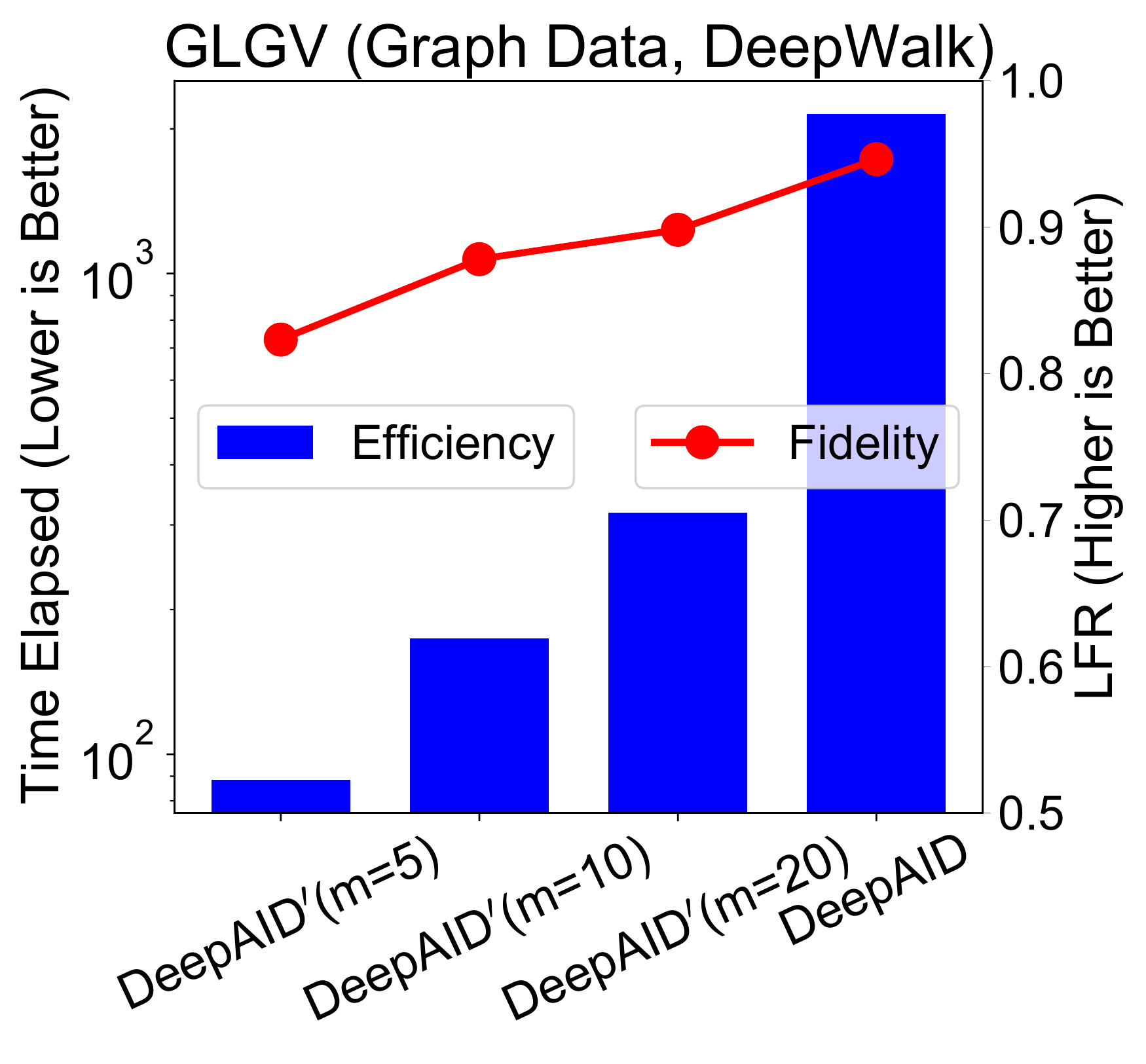}
	\vspace{-2ex}
	\caption{Performance of \textsf{DeepAID} Interpreter for graph data based system.}
	\label{fig_supp3}
\end{figure}

\noindent \textbf{Performance of Interpreter.} We primarily evaluate the fidelity and efficiency of \textsf{DeepAID} Interpreter. The same evaluation methods and indicators as \S \ref{sec6.2} are used here. Note that runtime in efficiency evaluation is per 100 anomalies. The results are shown in Figure \ref{fig_supp3}. We can find that \textsf{DeepAID} (with differentiable $\mathsf{E}_G$) are more accurate (i.e., with high fidelity) but less efficient as the number of parameters of the neural network are too large and the back propagation is very slow. For search-based methods, \textsf{DeepAID} with larger $m$ is more accurate while also more time-consuming (but is acceptable). In summary, for security systems whose $\mathsf{E}_G$ is indifferentiable, we recommend the search-based Interpreter, as \textsf{DeepAID}$'$ needs to re-implement $\mathsf{E}_G$, which will increase the workload.
For systems with differentiable $\mathsf{E}_G$, both methods can be used depending on how the operator treats the fidelity-efficiency trade-off. 

\noindent \textbf{Usage of \textsf{DeepAID}.} 
We provide two simple cases to illustrate how to use \textsf{DeepAID} on \texttt{GLGV} for interpreting anomalies and explaining FPs.
We still use the scenario shown in Figure \ref{fig_supp2}(a).
As shown in Figure \ref{fig_supp2}(b),
suppose that an anomaly is reported by \texttt{GLGV}, which is the link from ``user3'' to DC (denoted by the red line). Now operators would like to find the basis of this decision-making by \texttt{GLGV}.
After using our Interpreter, the reference link denoted by the green line from ``admin'' to DC is provided.
This means that ``user3'' to DC is decided as an anomaly because \texttt{GLGV} thinks (only) ``admin'' can access to (or, authenticate to) DC.
Furthermore, we can also reconstruct possible attack scenarios with the help of \textsf{DeepAID} interpretations: 
``user3'' were compromised by the attacker (e.g., maybe after downloading malware from a phishing email). 
The attacker gaining the initial foothold of user3's devices would like to move laterally in this enterprise network for more harmful attacks. 
Several technicals such as vulnerability exploitation were used by the attackers to pursue a higher permission. Finally, the attacker successfully escalated her privilege to access the DC \cite{bowman2020detecting}.

In Figure \ref{fig_supp2}(c), we provide another case of using \textsf{DeepAID} to analyze FPs in \texttt{GLGV}.
Here \texttt{GLGV} considers the link from ``user2'' to email server is an anomaly, which is actually an FP.
From the interpretation result provided by our \textsf{DeepAID}, we can find that \texttt{GLGV} thinks (only) ``user1'' has the privilege to access the email server. In other words, \texttt{GLGV} thinks ``user1'' and ``user2'' are under different levels of privileges, which is not the case. Therefore, with the help of \textsf{DeepAID}, we can confidently conclude that \texttt{GLGV} make mistakes in its prediction of ``user2''. Through further analysis, we find that this is caused by insufficient representation ability of embedding vectors. Therefore, we solve this FP problem by increasing the length of neighbors traversed by random walk in \texttt{GLGV} and increasing the number of iterations of the embedding process. 

\section{Hyper-parameters Sensitivity}
\label{app5}
\textsf{DeepAID} has some hyper-parameters in Interpreter and Distiller.
Below, we first test the sensitivity of these hyper-parameters and then discuss how to configure them in the absence of anomalies.

\noindent \textbf{Hyper-parameters in \textsf{DeepAID} Interpreter.} We evaluate six hyper-parameters in Interpreter, they are learning rate $\alpha$ in Adam optimizer, number of maximum iteration $max_{iter}$ for optimization, $\lambda$ in the stability term and $\epsilon$ in the fidelity term of objective functions (\ref{eq7}), (\ref{eq8}), (\ref{eq12}). \textcolor{black}{For time-series Interpreter, another two parameters $\mu_{1}$ and $\mu_{2}$ (for saliency testing) are tested.}
We set $K=10$ (10\%) for testing parameters in tabular Interpreter and set  $K=3$ (30\%) for time-series Interpreter, and use the same datasets in \ref{sec6.1}.
We primarily evaluate the sensitivity of fidelity (LFR). The stability and robustness are insensitive to parameters (except for optimization-based attack in \ref{appD.1}, which is more robust when $max_{iter}$ increases).
For efficiency, it is common sense that increasing the learning rate and reducing the number of iterations can reduce runtime, while other parameters are insensitive to efficiency.
The testing method is to change one parameter with a reasonable range while fixing other parameters for each time.
Table \ref{tab8} lists their value range for testing, as well as default values when testing other parameters.
The results of the change and standard deviation of LFR are also listed in Table \ref{tab8}. 



From the results in Table \ref{tab8}, we can find the change of LFR is small (e.g., std. for all parameters is <1\%). 
Thus, the performance of \textsf{DeepAID} Interpreter is basically insensitive to hyper-parameters.

\begin{figure}
	\includegraphics[width=0.6\linewidth]{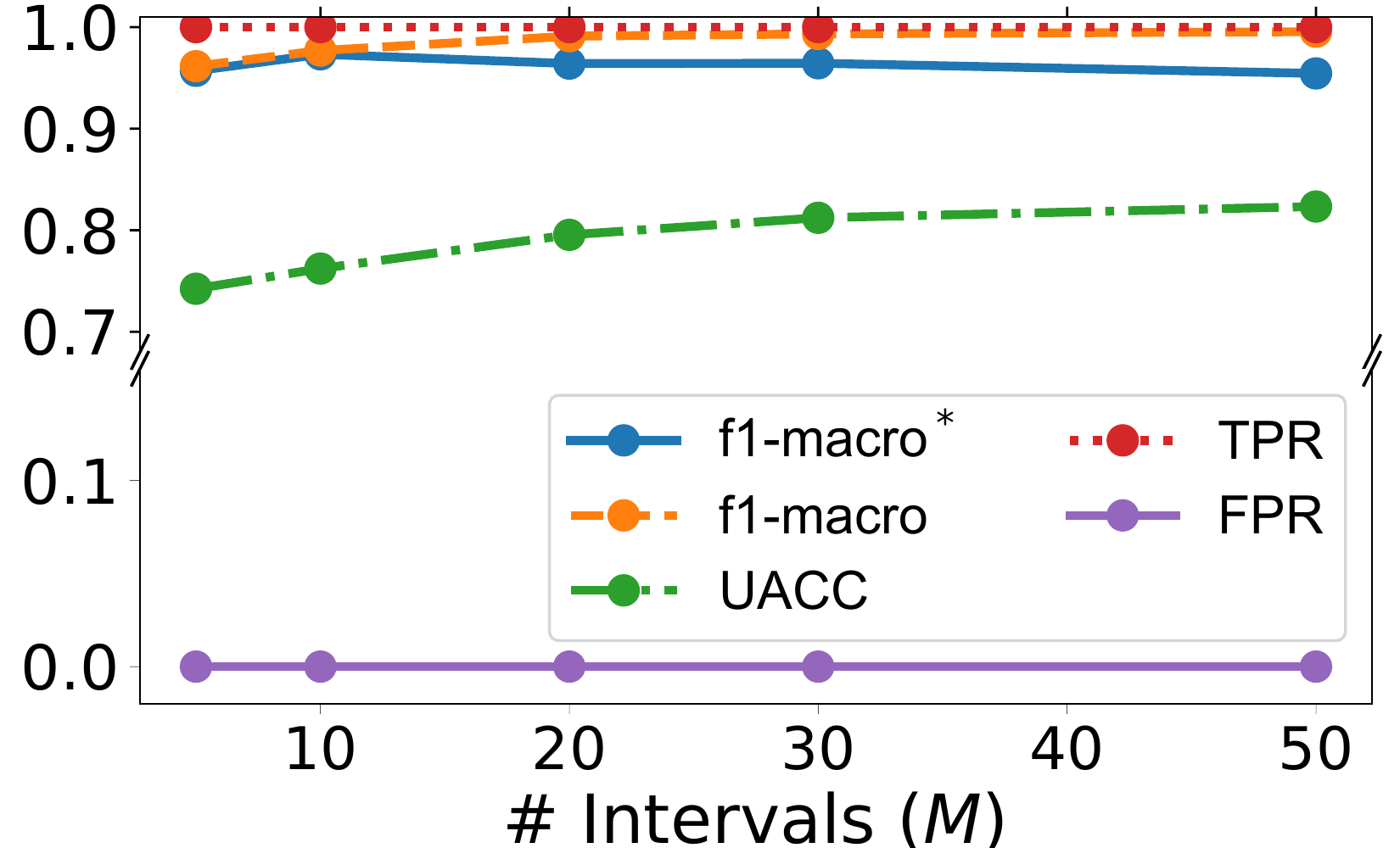}
		\vspace{-1ex}
	\caption{Sensitivity test of hyper-parameter $M$ in Distiller.}
	\label{fig_supp4}
\end{figure}

\noindent \textbf{Hyper-parameters in \textsf{DeepAID} Distiller.}  Distiller introduce only one new parameter $M$, the number of value intervals.
Here we use the same settings in Table \ref{tab4} and
\ref{tab5}, and we set $K=10$ and use 5-class datasets.
We evaluate f1-macro, f1-macro$^*$, TPR, FPR, and UACC under different $M$. The results are shown in Figure \ref{fig_supp4}. We can find that the change of all metrics is very small when $M\geq20$.

\begin{table}
	\caption{\textcolor{black}{Sensitivity test of hyper-parameters in Interpreter.}}
	\vspace{-2ex}	
	\label{tab8}
	\resizebox{\linewidth}{!}{%
		\begin{tabular}{c|cc|cc}
			\hline
			\textbf{Parameters } & \textbf{Default} & \textbf{Testing Range} & \textbf{LFR (range)} & \textbf{LFR (std.)} \\ \hline
			$\alpha$  & 0.5 & {[}0.1, 5.0{]} & {[}0.9150, 0.9215{]} & 0.0031 \\ \hline
			$max_{iter}$  & 20 & {[}10, 100{]} & {[}0.9150, 0.9150{]} & 0.0000 \\ \hline
			$\lambda$  & 0.001 & {[}0.0005,0.005{]} & {[}0.8954, 0.9150{]} & 0.0078 \\ \hline
			$\epsilon$  & 0.01 & {[}0.005, 0.05{]} & {[}0.9019, 0.9281{]} & 0.0098 \\ \hline
			\textcolor{black}{$\mu_{1}$}  & 0.01 & {[}0.001, 0.1{]} & {[}0.9525, 0.9525{]} & 0.0000\\ \hline
			\textcolor{black}{$\mu_{2}$}  & 0.3 & {[}0.2, 0.4{]} & {[}0.9452, 0.9643{]} & 0.0061 \\ \hline
		\end{tabular}%
	}
\end{table}

\noindent \textcolor{black}{\textbf{Hyper-parameters Configuration.} We introduce the guideline of configuring hyper-parameters of \textsf{DeepAID} for operators. Especially in the unsupervised setting, hyper-parameters should be configured in the absence of anomalies (i.e., using normal data only). Note that operators have a high fault tolerance rate of choosing parameters since we have demonstrated their insensitivity.
First, $\alpha$ and $max_{iter}$ are easy to configure since the learning rate of Adam optimizer has been well studied in machine learning community, and $max_{iter}$ should be increased appropriately if $\alpha$ is small. Also considering the adversarial robustness (\ref{appD.1}), we suggest that $max_{iter}$ should not be too small, such as more than 5. $\lambda$ balances the weight between fidelity and stability term. This is a free choice for operators. However, we should ensure that the values of the two terms are at the same order of magnitude. Thus, we can estimate the approximate value of the two terms, and can fine-tune the value after division. Note that, this does not need knowledge of anomalies since we only estimate the magnitude of two terms (i.e., only need to know the value range of normalized features).
For configuring $\epsilon$, we are more concerned about the upper bound, i.e., $\epsilon$ cannot be too large. A feasible method to measure the bound is to measure the value difference between the 99th quantile in normal data and the anomaly threshold ($t_{R},t_{P}$).
On the contrary, we are more concerned about the lower bound of $\mu_{1}$.
A feasible method is to measure the gradient term in (\ref{eq11}) with all normal data, and choose the maximum (or quantile) as the lower bound of $\mu_{1}$.  
For $\mu_{2}$, it can be configured through computing the mean of ${\rm Pr}(x_t|x_1x_2...x_{t-1})$ in normal data. Note that, $\mu_{2}$ must be greater than 
$1/N_{ts}$ where $N_{ts}$ is number of ``classes'' of $x_{t}$, otherwise $\mu_{2}$ cannot capture the class with the largest probability.
}